\documentclass[12pt,a4paper]{article}

\usepackage{graphicx}
\usepackage{amsfonts}
\usepackage{amsmath}
\usepackage[dvipdfmx]{color}
\usepackage{amssymb}

\usepackage{amsthm}
\usepackage[mathscr]{eucal} 
\usepackage{url}

\usepackage[all]{xy}

\def\N{\mathbb{N}}
\def\Z{\mathbb{Z}}
\def\Q{\mathbb{Q}}
\newcommand{\Qbar}{\overline{\Q}}
\def\R{\mathbb{R}}
\def\C{\mathbb{C}}
\def\P{\mathbb{P}}

\begin{document}
\baselineskip 0.6cm
\newcommand{\vev}[1]{ \left\langle {#1} \right\rangle }
\newcommand{\Dsl}{\mbox{\ooalign{\hfil/\hfil\crcr$D$}}}
\newcommand{\Slash}[1]{{\ooalign{\hfil/\hfil\crcr$#1$}}}
\newcommand{\EV}{ {\rm eV} }
\newcommand{\KEV}{ {\rm keV} }
\newcommand{\MEV}{ {\rm MeV} }
\newcommand{\GEV}{ {\rm GeV} }
\newcommand{\TEV}{ {\rm TeV} }

\newcommand{\Thetabar}{\overline{\Theta}}
\newcommand{\epsilonbar}{\overline{\epsilon}}
\newcommand{\Hom}{\mathrm{Hom}}
\newcommand{\Gal}{\mathrm{Gal}}

\newcommand{\KKcomment}[1]{\textsf{\color[HTML]{00008b}[\textit{KK: #1}]}}
\newcommand{\I}{\mathrm{I}}
\newcommand{\II}{\mathrm{II}}
\newcommand{\III}{\mathrm{III}}
\newcommand{\IV}{\mathrm{IV}}
\newcommand{\Sp}{\mathrm{Sp}}
\newcommand{\USp}{\mathrm{USp}}

\def\tr{\mathop{\rm tr}}
\def\Spin{\mathop{\rm Spin}}
\def\SO{\mathop{\rm SO}}
\def\O{\mathop{\rm O}}
\def\SU{\mathop{\rm SU}}
\def\U{\mathop{\rm U}}
\def\Sp{\mathop{\rm Sp}}
\def\SL{\mathop{\rm SL}}

\theoremstyle{definition}
\newtheorem{thm}{Theorem}[subsection]
\newtheorem{defn}[thm]{Definition}
\newtheorem{exmpl}[thm]{Example}
\newtheorem{props}[thm]{Proposition}
\newtheorem{cor}[thm]{Corollary}
\newtheorem{lemma}[thm]{Lemma}
\newtheorem{rmk}[thm]{Remark}
\newtheorem{anythng}[thm]{}

% from KKcommand

\newcommand{\rank}{\mathrm{rank}}
\newcommand{\dfn}[1]{\textbf{#1}}
\newcommand{\p}[1]{\left(#1\right)}
\newcommand{\B}[1]{\left\{#1\right\}}
\renewcommand{\b}[1]{\left[#1\right]}

\newcommand{\ubar}{{\overline{u}}}
\newcommand{\vbar}{{\overline{v}}}
\newcommand{\wbar}{{\overline{w}}}
\newcommand{\omegabar}{\overline{\omega}}
\newcommand{\rhobar}{\overline{\rho}}
\newcommand{\NS}{\mathrm{NS}}

\newcommand{\Omegabar}{\overline{\Omega}}
\newcommand{\Pibar}{\overline{\Pi}}
\newcommand{\Sigmabar}{\overline{\Sigma}}
\newcommand{\todo}[1]{\textcolor{red}{\large \raise0.2ex\hbox{\fbox{\small todo}}\footnote{\textcolor{red}{#1}}}}
\newcommand{\cA}{\mathcal{A}}
\newcommand{\cB}{\mathcal{B}}
\newcommand{\cM}{\mathcal{M}}
\newcommand{\cN}{\mathcal{N}}
\newcommand{\cO}{\mathcal{O}}
\newcommand{\ord}{\mathrm{ord}}
\newcommand{\fref}[1]{Fig.\ref{fig:#1}}
\newcommand{\eref}[1]{(\ref{#1})}
\newcommand{\Aut}{\mathrm{Aut}}
\newcommand{\Isom}{\mathrm{Isom}}
\newcommand{\hor}{\mathrm{hor}}
\newcommand{\ver}{\mathrm{ver}}
\newcommand{\GL}{\mathrm{GL}}

\begin{titlepage}
  
\begin{flushright}
  IPMU20-0127 \\
% % Version \today
\end{flushright}
 
  \vskip 1cm
  \begin{center}
  
  {\large \bf $W=0$ Complex Structure Moduli Stabilization \\
    on CM-type K3 x K3 Orbifolds:} \\
{\bf --- Arithmetic, Geometry and Particle Physics --- } 
  
  \vskip 1.2cm
 
 Keita Kanno and Taizan Watari
 
 \vskip 0.4cm
  
  {\it 
    Kavli Institute for the Physics and Mathematics of the Universe (WPI), 
    UTIAS, The University of Tokyo, Kashiwa-no-ha 5-1-5, 277-8583, Japan
   }

 \vskip 1.5cm
   
\abstract{
It is an important question in string compactification whether 
complex structure moduli stabilization inevitably ends up with a 
vacuum expectation value of the superpotential $\vev{W}$ of the order 
of the Planck scale cubed. Any thoughts on volume stabilization and 
inflation in string theory, as well as on phenomenology of supersymmetric 
Standard Models, will be affected by the answer to this question. 
In this work, we follow an idea for making $\vev{W} \simeq 0$ where 
the internal manifold has a vacuum complex structure with arithmetic 
characterization, and address Calabi--Yau fourfold compactification 
of F-theory. 
The moduli space of K3 $\times$ K3 orbifolds contain infinitely many 
such vacua. Arithmetic conditions for a $\vev{W} =0$ flux are worked out, 
and then all the K3 moduli have supersymmetric mass. Possible 
gauge groups, matter representations and discrete symmetries are studied 
for the case of $\Z_2$-orbifolds. 
}
  
\end{center}
\end{titlepage}

% \tableofcontents

%  \begin{flushright}
%  IPMU18-xxxx
%  \end{flushright}
%   \vskip 1cm
%   \begin{center}
%   
%   {\large \bf F-theory on $\text{K3}\times\text{K3}/\Z_n$ and Flux Vacua
% at CM points}
%   
%   \vskip 1.2cm
%  
% Keita Kanno and Taizan Watari
%  
%  \vskip 0.4cm
  
%   {
%     Kavli Institute for the Physics and Mathematics of the Universe, \\
%     the University of Tokyo, Kashiwa-no-ha 5-1-5, 277-8583, Japan
%
%Kavli Institute for the Physics and Mathematics of the Universe (WPI),\\
%The University of Tokyo Institutes for Advanced Study, The University of Tokyo,\\
%Kashiwa, Chiba 277-8583, Japan
%
%Kavli IPMU (WPI), UTIAS, The University of Tokyo, Kashiwa, Chiba 277-8583, Japan
%http://www.ipmu.jp/ja/node/1819
% }

%  \vskip 1.5cm
   
%  \abstract{In this paper, ...} 
% %   
%  \end{center}
\tableofcontents
% \maketitle

 \clearpage

%%%%%%%%%%%%%%%%%%%%%%%%%%%%%%%%%%%%%%%%%%%%%%%%%%
\section{Introduction}
%%%%%%%%%%%%%%%%%%%%%%%%%%%%%%%%%%%%%%%%%%%%%%%%%%

In an F-theory compactification on a Calabi--Yau fourfold $Y$, 
the effective theory in 3+1-dimensions has the superpotential 
\begin{align}
  W = W_{\rm cpx~str} \propto \int_Y G \wedge \Omega_Y; 
 \label{eq:GVW}
\end{align}
$G$ is a four-form flux in the M-theory formulation of F-theory 
taking value in \cite{flux-qntztn} 
\begin{align}
  \frac{1}{2} c_2(TY) + H^4(Y;\Z), 
  \label{eq:space-of-G}
\end{align}
and $\Omega_Y$ a holomorphic $(4,0)$-form on $Y$. 
$\Omega_Y$ varies relatively to $H^4(Y;\Z)$ over the moduli space 
${\cal M}_{\rm cpx~str}^{[Y]}$ of complex structure of $Y$, so the 
superpotential $W$ is regarded as a function (a section of 
an appropriate line bundle in fact) on ${\cal M}_{\rm cpx~str}^{[Y]}$.
When a topological flux $G$ is fixed, the $F$-term condition 
determines the vacuum expectation value (vev) of the complex structure 
parameters $\vev{z} \in {\cal M}_{\rm cpx~str}^{[Y]}$, and consequently the vev
of $W$ proportional to $\int_Y G \wedge \vev{\Omega_Y}$, 
where $\vev{\Omega_Y} := \Omega_{Y_{\vev{z}}}$.

The vev of $W$ determined in this way are quite often 
of the order of $M_{\rm Pl}^3$, where $M_{\rm Pl}$ is the Planck scale
in (3+1)-dimensions \cite{DDFK}. 
This means that the vacuum\footnote{
In this article, we do not discuss stabilization of K\"{a}hler moduli. 
It makes sense to focus on stabilization of complex structure moduli 
and pose a question if it is possible to achieve $|\vev{W}_{\rm cpx~str}| 
\ll M_{\rm Pl}^3$, 
when one ignores a possibility that $|\vev{W}| \ll M_{\rm Pl}^3$ as a 
result of cancellation between $\vev{W}_{\rm cpx~str} \sim M_{\rm Pl}^3$ 
and $\vev{W}_{\rm Kahler} \sim M_{\rm Pl}^3$ in 
a non-geometric/non-perturbative stabilization of K\"{a}hler moduli. 
} % 
has AdS supersymmetry with the cosmological 
constant of the order of $-M_{\rm Pl}^4$, and the gravitino mass is 
of the order of $M_{\rm Pl}$. Once a topological flux $G$ is chosen, 
there is no chance of continuous tuning of compactification 
parameters (because the complex structure moduli fields are expected to have 
large masses). 
Certainly such a large negative cosmological constant\footnote{
Such a large negative cosmological constant is not an immediate 
consequence of $\vev{W}\sim M_{\rm Pl}^3$, because of cancellation 
that takes place for a special kinds of K\"{a}hler potential. 
One has to make sure, though, that such a special form is maintained 
even after Kaluza--Klein / string / quantum / non-perturbative corrections are 
taken into account. 
} %
 is not a good 
approximation to the vacuum we live in. Large gravitino mass 
and its anomaly mediation to gauginos are a fatal blow to 
the electroweak-ino dark matter scenario, and also to supersymmetric 
grand unification. 

If there is a dynamics, mechanism, theoretical principle or anything else 
that renders the vev of the superpotential much smaller than $M_{\rm Pl}^3$, 
therefore, it is worth investigating it further. In this article, 
we pick up an idea of \cite{MooreTASI, enumerate}, and elaborate more on it. 
The idea is to focus on the subset ${\cal M}_{\rm alg}^{[Y]} \subset 
{\cal M}_{\rm cpx~str}^{[Y]}$ of the complex structure moduli space; it is the set 
of points $\vev{z}$ in ${\cal M}_{\rm cpx~str}^{[Y]}$ where 
all the Hodge components $H^{p,4-p}(Y_{\vev{z}};\C)$ (with $p = 0,1,2,3,4$) 
have basis elements in\footnote{
Here, the field $\overline{\Q} \subset \C$ consists of all the algebraic 
numbers. 
} %
 $H^4(Y_{\vev{z}};\overline{\Q})$. 
Now that the period integrals form a finite dimensional vector space over 
$\Q$, it is not unthinkable any more that their linear combination by an 
appropriately chosen set of flux quanta $G$ in (\ref{eq:space-of-G}) 
vanishes. As an attractive phenomenology idea for the small value of 
the cosmological constant in this local universe, therefore, three 
directions of further investigation will be motivated;
(a) to look for a dynamics or theoretical principle that will favor a choice 
of $\vev{z}$ from the subset ${\cal M}_{\rm alg}^{[Y]}$ than from 
${\cal M}_{\rm cpx~str}^{[Y]}$, (b) to derive physics consequences of the idea 
other than the original input $\vev{W} \simeq 0$, and (c) to elaborate 
more---as string-phenomenology---on how/when a choice 
$\vev{z} \in {\cal M}_{\rm alg}^{[Y]}$ renders $\vev{W}\simeq 0$. 

In Ref. \cite{prev.paper.PRD}, the authors pursued the direction (c) 
for choices of $\vev{z}$ from an even smaller subset 
${\cal M}_{\rm CM}^{[Y]} \subset {\cal M}_{\rm alg}^{[Y]}$. 
The subset ${\cal M}_{\rm CM}^{[Y]}$ consists of choices of complex structure 
$\vev{z}$ where the compactification manifold $Y_{\vev{z}}$ is of CM-type, 
a notion generalizing the complex multiplication on elliptic curves. 
While we wait until section \ref{sssec:math-preparation-tensor} in this 
article to spell out what the CM-type means in math language (a more 
systematic review is found in \cite{prev.paper.PRD, prev.paper.preprint}), 
there are two characterizations of the CM nature of $z \in 
{\cal M}_{\rm cpx~str}^{[Y]}$ that can be stated in string theory. 
First, in a Type IIB Calabi--Yau orientifold case, the CM nature 
of a Calabi--Yau threefold $M_z$ for compactification has been 
conjectured \cite{GV} to be a half of the necessary and sufficient 
condition for the $N=(2,2)$ worldsheet superconformal field theory 
to be described by a rational CFT.\footnote{
The phenomenology idea of assuming $\vev{z} \in {\cal M}_{\rm CM}^{[Y]}$
(with an extra assumption on K\"{a}hler moduli vev)  
therefore attributes the small value of the cosmological constant 
not to a symmetry of the field theory on the space-time $\R^{3,1}$ 
(or $\R^{3,1} \times M_z$), but to an immensely large symmetry (chiral 
algebra) of the conformal field theory on the worldsheet. 
} % 
This observation may (or may not) shed a bit of light in the research 
direction (a) above. 
The other characterization of a CM-type complex structure $z$ is that 
the basis elements of $H^{p,4-p}(Y_z;\C)$ are not just in 
$H^4(Y_z;\overline{\Q})$, but are subject to stronger control under 
the Galois group action. 
Due to this nature, it turns out \cite{prev.paper.PRD} that 
the F-term (supersymmetry) conditions on the flux quanta for a given 
CM-type complex structure $z$ are highly degenerate as a consequence, 
and are satisfied by a space of flux quanta of higher dimension 
(relatively to the estimation for $\vev{z} \in {\cal M}_{\rm alg}^{[Y]}$ 
in \cite{enumerate}).

Study in this article has two motivations. One is to continue on the 
research direction (c) for $\vev{z} \in {\cal M}_{\rm CM}^{[Y]}$ for 
fourfolds $Y$ in the context of M-theory/F-theory compactification; 
the study of \cite{prev.paper.PRD} was only in the context of Type IIB 
Calabi--Yau orientifolds,\footnote{
Neither did we study stabilization of the moduli of 
D7-brane configurations, nor particle-physics consequences of such 
compactifications in \cite{prev.paper.PRD}.
} % 
so $Y$ were of the form $(E_\phi \times M)/\Z_2$ for some Calabi--Yau 
threefold $M$ and an elliptic curve $E_\phi$, both of CM type.
The other is to extract consequences on particle physics / complex 
structure moduli stabilization in F-theory, which is in the research 
direction (b). 

We do not attempt at making a progress in the direction (a) in this 
article. It is worth reminding ourselves, though, that a CM-type K3 
surface has a defining equation with all the coefficients being 
algebraic numbers \cite{PSS-arth, Rizov}. So, if a fourfold $Y$ 
is a K3 x K3 orbifold of a pair of CM-type K3 surfaces, $Y$ 
has defining equations with all the coefficients in $\overline{\Q}$. 
The $L$-function can be defined for each one of such arithmetic models of $Y$. 
Recent articles \cite{Candelas, Kach20, Schimm, Kach20-2} 
suggest---under certain assumptions---that 
Calabi--Yau threefolds $M_z$ for certain $z \in {\cal M}_{\rm alg}$ 
have a simple rational Hodge substructure in $H^3(M_z;\Q)$ whose 
$L$-function is modular. It will be exciting, if such a research direction 
(and its extension to F-theory) manages to elevate such an arithmetic aspect 
into a necessary theoretical principle in string/M-theory in the future. 

Here is what we do in this article.
We work exclusively on fourfolds $Y$ obtained in the form of K3 x K3 
orbifolds.\footnote{
This class of fourfolds includes (modulo birational transformation)
orbifolds of (an elliptic curve) x (a Borcea--Voisin Calabi--Yau threefold).
} % 
This is because CM-type complex structure is known to exist in a 
most systematic way for this class of fourfolds (brief review 
in section \ref{ssec:CM-BV}).\footnote{
A more extensive review is found in \S 2.2 and appendix B.1 
of \cite{prev.paper.preprint}. 
} % 
We deal with a simplest class of $\Z_2$-orbifolds of K3 x K3 
in sections \ref{sec:flux-analysis} and \ref{sec:particle}, while 
section \ref{sec:non-Z2-orbifold} deals with more general orbifolds of 
K3 x K3. M-theory compactification on such fourfolds down to $\R^{2,1}$ 
is studied in sections \ref{sec:flux-analysis} 
and \ref{sec:non-Z2-orbifold}; we work out the conditions 
for a non-trivial supersymmetric flux with $\vev{W}=0$ to exist, and 
also examine the mass terms, interactions and symmetries of the 
complex structure moduli fields.  
Section \ref{sec:particle} is devoted to F-theory compactification
down to $\R^{3,1}$. Some attempts are made in finding fourfolds $Y$ 
birational to a K3 x K3 orbifold so that $Y$ have flat elliptic fibrations.  
Results of section \ref{sec:flux-analysis} are recycled (with a bit 
of care), and we see that the complex structure moduli of the pair 
of K3 surfaces can be given large masses by a flux satisfying $DW=W=0$. 
Sections \ref{ssec:on-fibre} and \ref{ssec:on-base} also derive constraints 
on possible choices of non-Abelian gauge group and matter curve 
configuration.\footnote{
Here is a cautionary remark: we presented in section \ref{sec:particle}
only the F-theory geometry construction in which we have confidence; 
we have a sense of feeling that there will be more constructions for 
F-theory geometry with CM-type Hodge (sub)structure,  
even within the simplest class of $\Z_2$-orbifold of K3 x K3 
(see footnotes \ref{fn:F-geometry-1}, \ref{fn:F-geometry-2} 
and \ref{fn:F-geometry-5}). 
So, it is too early to take those constraints as a final statement,  
or to take them out of the context. 
} %  

This study can be seen as an example that a phenomenological idea 
for small $\vev{W}$ may have particle physics consequences apparently 
totally unrelated to the cosmological constant: discrete gauge symmetry 
(section \ref{sssec:bonus}), approximate accidental symmetry 
in the effective theory (sections \ref{sssec:mass-TX=T0}, 
\ref{ssec:Tx-neq-T0} and \ref{ssec:mass-genOrbfld}) and constraints 
on choices of non-Abelian gauge groups and matter curve configuration.  

%%%%%%%%%%%%%%%%%%%%%%%%%%%%%%%%%%%%%%%%%%%%%%%%%%%%%%
\section{Supersymmetric Flux Vacua on CM-type 
$({\rm K3} \times {\rm K3})/\Z_2$ Orbifolds}
\label{sec:flux-analysis}
%%%%%%%%%%%%%%%%%%%%%%%%%%%%%%%%%%%%%%%%%%%%%%%%%%%%%%

%%%%%%%%%%%%%%%%%%%%%%%%%%%%%%%%%%%%%%%%%%%%%%%%%
\subsection{CM-type Calabi--Yau Fourfolds and Borcea--Voisin Orbifolds}
\label{ssec:CM-BV}
%%%%%%%%%%%%%%%%%%%%%%%%%%%%%%%%%%%%%%%%%%%%%%%%%

In the case $Y=E$ is an elliptic curve, 
a one-dimensional Calabi--Yau manifold, the complex structure of $E$ 
is of CM-type, by definition, if $E$ has complex multiplication 
(see \cite[\S 2.1, B.1.5, and B.1.6]{prev.paper.preprint} 
or footnote \ref{fn:CM-ell} in this article, for example, 
for more background information). The set of CM points 
${\cal M}_{\rm CM}^{[E]}$ in the moduli space of complex structure 
of elliptic curves ${\cal M}_{\rm cpx~str}^{[E]} \cong {\cal H}/{\rm SL}(2;\Z)$
is completely understood; CM points in the upper complex half plane 
${\cal H}$ are the set of the roots of any quadratic polynomial equation of 
one variable with coefficients in $\Q$. 
They are labeled by the imaginary quadratic fields $K$; the CM points 
sharing the same imaginary quadratic field forms an orbit 
under the action of ${\rm GL}(2;\Q) = \mathbb{G}Sp(2;\Q)$. 
In the case $Y=X$ is a K3 surface (a two-dimensional Calabi--Yau 
manifold) with a transcendental lattice $T_X$, 
it is also known that any CM point in the moduli space 
${\cal M}_{CM}^{[X(T_X)]}$ is associated with a CM field\footnote{
See \S A.2 of \cite{prev.paper.preprint} (and a passage before 
section \ref{sssec:FUP} in this article) for the definition of 
a {\bf CM field}.
} %
 $K$ of degree $[K:\Q] = {\rm rank}(T_X)$; the CM points 
sharing the same CM field $K$ form orbits under the action of the 
group\footnote{
The groups $\mathbb{G}Sp$ and $\mathbb{G}O$ consist of linear 
transformations that preserve skew-symmetric and symmetric bilinear forms, 
respectively, up to overall scalar multiplications. 
} %
 $\mathbb{G}O(T_X;\Q)$ 
on ${\cal M}_{CM}^{[X(T_X)]} \subset {\cal M}_{\rm cpx~str}^{[X(T_X)]} 
= {\rm Isom}(T_X) \backslash D(T_X)$; here, $D(T_X)$ is the period domain 
of the signature $(2,{\rm rank}(T_X)-2)$ lattice $T_X$ and ${\rm Isom}(T_X)$
the group of integral isometries of $T_X$. 
In particular, we know that there are 
infinitely many CM points in the moduli space of complex structure 
of elliptic curves and K3 surfaces. 

When it comes to the case $Y=M$ is a 
Calabi--Yau threefold, or a Calabi--Yau fourfold $Y$, however, much less 
is known. It is believed that the Calabi--Yau threefolds $M$ realized by 
rational CFT's have complex structure of CM type \cite{GV}, but they are 
nothing more than a small number of isolated points in the moduli space. 
Although the group ${\rm Sp}(b_3(M))$ is a symmetry of some of the 
relations that the Hodge structure of a Calabi--Yau threefold $M$ satisfies, 
yet the action of the group takes a point in ${\cal M}_{\rm cpx~str}^{[M]}$
outside of ${\cal M}_{\rm cpx~str}^{[M]}$ in general; the latter observation 
also holds true in the case $Y$ is a Calabi--Yau fourfold, when the group 
${\rm Sp}(b_3(M))$ is replaced by the isometry group of the lattice 
$H^4(Y;\Z)$. So, in particular, we do not have an argument in the case 
$Y$ is a threefold or a fourfold that infinitely many CM points 
${\cal M}_{\rm CM}^{[Y]}$ show up in the form of orbits of $\mathbb{G}Sp(b_3)$
or $\mathbb{G}O(b_4(Y))$.\footnote{
This argument still does not rule out infinitely many CM points; 
in fact infinitely many CM points are contained in the 101-dimensional 
moduli space of the quintic Calabi--Yau threefolds 
(e.g., see \cite[footnote 18]{prev.paper.preprint} for references). 
The Fermat sextic fourfold \cite{BV} is CM-type (e.g., \cite{YL}). 
} %
Indeed, the Andr\'e--Oort conjecture hints that 
there are not so many CM points available in ${\cal M}_{\rm cpx~str}^{[Y]}$ 
in those cases. 
For more information, see \cite[\S2.2]{prev.paper.preprint}. 

For a special class of topological types of Calabi--Yau threefolds $[Y=M]$ 
or of fourfolds $[Y]$, however, it is possible to identify systematically 
a set of points ($\vev{z}$'s) of ${\cal M}_{\rm cpx~str}^{[Y]}$ where 
$H^3(Y_{\vev{z}};\Q)$ or $H^4(Y_{\vev{z}};\Q)$ has a CM-type rational Hodge 
substructure.\footnote{
There is a review material on rational Hodge structure in 
section \ref{ssec:cond-flux-MorF} in this article. 
}\raisebox{4pt}{,}\footnote{
\label{fn:CM-CY-CM-substr}
It is a stronger condition for a rational Hodge structure on $H^4(Y;\Q)$
to be of CM-type than for it to have a rational Hodge substructure that 
is of CM-type. See the discussion at the end of 
section \ref{ssec:generic-noFlux}. Whether the Coleman--Oort conjecture 
is relevant in the current context (whether supersymmetric flux is available 
for moduli stabilization) should also be reconsidered along this line. 
}
An idea, originally in \cite{Borcea, voisinhodge},
is to take a product of a CM-type elliptic curve $E$ and a CM-type K3 surface, 
or of a pair of CM-type K3 surfaces, first, and then to take an orbifold 
that preserves the Calabi--Yau condition. Not all the topological types 
available for a Calabi--Yau three/four-fold will be realized in this 
construction. 
The moduli space ${\cal M}_{\rm cpx~str}^{[Y]}$ of a three/four-fold $Y$ 
constructed in that way contains an orbifold locus 
${\cal M}_{\rm cpx~str}^{[Y]BV}$ where the orbifold 
singularity of $Y_{\vev{z}}$ is not deformed in complex structure;\footnote{
We do not talk about choice of K\"{a}hler moduli in this article. 
Whether the orbifold singularity is resolved or not, discussions 
in this article are valid. 
} % 
as long as the building block $E$ or K3 surfaces are of CM-type, 
and the vacuum choice $\vev{z}$ of the complex structure of $Y_{z}$ 
is in the orbifold locus ${\cal M}_{\rm cpx~str}^{[Y]BV}$, then 
$H^3(Y_{\vev{z}};\Q)$ or $H^4(Y_{\vev{z}};\Q)$ has a rational Hodge 
substructure of CM-type indeed. 

The simplest class of Calabi--Yau fourfolds $Y$ as K3 x K3 orbifolds
is of the form $Y=(X^{(1)} \times X^{(2)})/\Z_2$. Both of the K3 surfaces 
$X^{(1)}$ and $X^{(2)}$ are assumed to have a non-symplectic automorphism 
of order two, $\sigma_{(1)}$ and $\sigma_{(2)}$, respectively;   
the holomorphic (2,0)-forms $\Omega_{X^{(1)}}$ and $\Omega_{X^{(2)}}$ get 
transformed as $\sigma_{(i)}^* (\Omega_{X^{(i)}}) = 
- \Omega_{X^{(i)}}$ for $i=1,2$. By choosing the generator $\sigma$ 
of the orbifold group $\Z_2$ to be $(\sigma_{(1)}, \sigma_{(2)})$, 
the orbifold $Y$ becomes Calabi--Yau because 
$(\Omega_{X^{(1)}} \wedge \Omega_{X^{(2)}})$ is invariant under the generator 
$\sigma$, yet $\Omega_{X^{(i)}}$'s are not. 
We call a subclass of those fourfolds---those where 
both $\sigma_{(1)}$ and $\sigma_{(2)}$ act purely non-symplectically 
(more explanations in the next paragraph and also in 
section \ref{ssec:K3-nonsymp-Aut} (footnotes \ref{fn:def-purely-nonsymp} 
and \ref{fn:more-purely-nonsymp} in particular))---as Borcea--Voisin 
K3 x K3 orbifolds in this article; 
more general orbifolds ((a) where $\sigma_{(1)}$ and/or $\sigma_{(2)}$ are 
non-symplectic but not purely non-symplectic, or (b) where the orbifold 
group is not $\Z_2$) as generalized Borcea--Voisin K3 x K3 orbifolds.   
Until the end of this section \ref{sec:flux-analysis}, 
we deal with M-theory compactifications on a Borcea--Voisin fourfold. 

Reference \cite{Nikulin-factor-long} 
provides a theory of topological classification of 
a pair $(X, \sigma)$ of a K3 surface $X$ and an automorphism 
$\sigma \in {\rm Aut}(X)$ of order two ($\sigma^2 = {\bf Id}_X$) acting 
non-symplectically ($\sigma^*\Omega_X \neq \Omega_X$) on the holomorphic 
(2, 0)-form $\Omega_X$. To be more precise, it classifies 
$(S_0, T_0, \sigma)$ modulo isometry of $H^2(X;\Z)$, where 
$S_0$ and $T_0$ are mutually orthogonal primitive sublattices of $H^2(X;\Z)$
of signature $(1,r-1)$ and $(2,20-r)$, respectively, and $\sigma$
an isometry of $H^2(X;\Z)$ that acts trivially on $S_0$ and as $(-1)\times$ 
on $T_0$.  This lattice-theory classification of $(S_0, T_0, \sigma)$ 
is regarded as that of non-symplectic automorphisms of order two, 
because one may choose $\C \Omega_X$ from $D(T_0)/{\rm Isom}(T_0)$. 
For such a complex structure, 
the transcendental lattice $T_X$ is contained within $T_0$, and 
$\sigma^*\Omega_X = - \Omega_X$; the N\'eron--Severi lattice $S_X$ 
contains $S_0$. The list 
of \cite{Nikulin-factor-long} consists 
of 75 choices of $(S_0, T_0, \sigma)$. 
So, we have 75 choices of $(S_0^{(i)}, T_0^{(i)},\sigma_{(i)})$ for each 
one of $i=1,2$; for a given choice, a topological family of Borcea--Voisin 
orbifolds is available for M-theory compactification. 
In the rest of this section, supersymmetric flux configuration is studied 
for a vacuum complex structure in 
\begin{align}
 {\cal M}_{\rm CM}^{[X(T_0^{(1)})]} \times {\cal M}_{\rm CM}^{[X(T_0^{(2)})]} \subset 
 {\cal M}_{\rm cpx~str}^{[X(T_0^{(1)})]} \times {\cal M}_{\rm cpx~str}^{[X(T_0^{(2)})]}
    = {\cal M}_{\rm cpx~str}^{[Y]BV}. 
  \label{eq:CM-X1-X2-subspace}
\end{align}

Here is a remark before moving on.
One may also construct a Calabi--Yau fourfold as an orbifold 
of two elliptic curves $E_\phi$, $E_\tau$, and a K3 surface $X^{(2)}$, 
instead of a pair of K3 surfaces:\footnote{
In sections \ref{sec:flux-analysis} and \ref{sec:non-Z2-orbifold}, we do not 
distinguish a pair of fourfolds that are mutually birational and have 
the same number of complex structure and K\"{a}hler deformations.
That is enough for the purpose of analysing supersymmetric flux 
configuration and complex structure moduli effective field theory. 
 
For example, an orbifold $(E_\phi \times E_\tau \times X^{(2)})/
(\Z_2 \times \Z_2)$
has $\C^3/(\Z_2 \times \Z_2)$ singularity along a curve 
$Z_{(2)} \subset X^{(2)}$; the fourfold 
$(E_\phi \times [(E_\tau \times X^{(2)})/\Z_2])/\Z_2$ in the first line 
and $([(E_\phi \times E_\tau)/\Z_2]\times X^{(2)})/\Z_2$ in the second line
are regarded as different resolutions of the $\C^3/(\Z_2 \times \Z_2)$ 
singularity (cf \cite{DDFK}). Two flops convert one to the other. 
For this reason, we do not even make a clear distinction between 
an orbifold with singularity and a non-singular manifold obtained 
as a crepant resolution of the orbifold in sections \ref{sec:flux-analysis} 
and \ref{sec:non-Z2-orbifold}.
} % 
\begin{align}
   Y & \; 
    = \left( E_\phi \times \left(  E_\tau \times X^{(2)} \right)/\Z_2 \right)
      /\Z_2
    =: \left( E_\phi \times M \right)/\Z_2,  \label{eq:BV-IIB-as-Kummer-K3} \\
 & \; = \left( \left( E_\phi \times E_\tau \right)/\Z_2 \times X^{(2)} \right)
        /\Z_2
   =: \left( X^{(1)} \times X^{(2)} \right) /\Z_2 . \nonumber
\end{align}
This is for Type IIB Calabi--Yau orientifold compactification, where 
the Calabi--Yau threefold is $M = (E_\tau \times X^{(2)})/\Z_2$. 
This construction is nothing more than a special case of the 
Borcea--Voisin K3 x K3 orbifolds; we can see the combination 
$X^{(1)} = (E_\phi \times E_\tau)/\Z_2 =: {\rm Km}(E_\phi \times E_\tau)$ 
as the K3 surface $X^{(1)}$; along with an involution $\sigma_{(1)}$ that 
multiplies $(-1)$ to $E_\tau$, the pair $(X^{(1)}, \sigma_{(1)})$ becomes 
one of the 75 topological types classified by Nikulin (the one\footnote{
$U$ stands for the hyperbolic plane lattice, or equivalently the 
even unimodular lattice of signature (1,1), ${\rm II}_{1,1}$.
See also footnote \ref{fn:lattice-conv}.  
} %
 with 
$T_0 = U[2]U[2]$). 
For this reason, we do not loose generality at all by thinking 
only of K3 x K3 orbifolds. 

%%%%%%%%%%%%%%%%%%%%%%%%%%%%%%%%%%%%%%%%%%%%%%%%%
\subsection{$H^4((X^{(1)}\times X^{(2)})/\Z_2;\Q)$ and Complex Structure 
Deformations}
\label{ssec:generic-noFlux}
%%%%%%%%%%%%%%%%%%%%%%%%%%%%%%%%%%%%%%%%%%%%%%%%%

In an M-theory compactification on a fourfold $Y$ and its F-theory limit, 
a flux is in $H^4(Y;\Q)$, and the complex structure moduli in $H^4(Y;\C)$. 
Before starting analysis for conditions for supersymmetric flux vacua with 
$\vev{W}=0$ to exist, we need to remind ourselves of a bit of math 
of the cohomology groups of a Borcea--Voisin orbifold 
$Y = (X^{(1)}\times X^{(2)})/\Z_2$ with both $X^{(1)}$ and $X^{(2)}$ 
being a K3 surface.

%%%%%%%%%%%%%%%%%%%%%%%%%%%%%%%%%%%%%%%%%%%%%%%%

%%%%%%%%%%%%%%%%%%%%%%%%%%%%%%%%%%%%%%%%%%%%%%%%

The fourfold $Y=(X^{(1)} \times X^{(2)})/\Z_2$ would remain singular, if 
it stays precisely at the orbifold locus without complex structure 
deformation or K\"{a}hler parameter resolution. Because we do not assume 
anything about the vacuum value of the K\"{a}hler parameter, we do not 
need to think that $Y$ is singular, and moreover, we can always take a limit 
from non-zero resolution to the orbifold limit, if we wish. So, topology 
of the fourfold $Y$ is well-defined.\footnote{
In section \ref{sec:particle}, we will use $Y^{BV}$ for the non-singular 
fourfold after resolution, and $Y_0$ the orbifold without deformation or 
resolution of the $\C^2/\Z_2$ singularity.
}
 
To describe the topology of $Y$, we need one more preparation. 
The non-symplectic automorphism $\sigma_{(i)}: X^{(i)} \rightarrow X^{(i)}$ 
may have fixed points (for $i=1,2$ individually), and the locus of 
fixed points are denoted by $Z_{(i)}$ for $i=1,2$. The set $Z_{(i)}$ 
of fixed points in $X^{(i)}$ consists of curves whose irreducible 
components are disjoint from one another, when $\sigma_{(i)}$ acts 
non-symplectically and is order 2 \cite{Nikulin-factor-long}. 
Among the 75 choices of $(S_0, T_0, \sigma)$ in \cite{Nikulin-factor-long}, 
this subset $Z$ of fixed points is empty in just one choice, 
where\footnote{
\label{fn:lattice-conv}
In this article, negative definite root lattices of $A_n$, $D_n$ and $E_n$
type 
are denoted by $A_n$, $D_n$ and $E_n$. For a lattice $L$, $L[n]$ stands 
for a lattice where $L \cong L[n]$ as free abelian groups, and the 
intersection form of the latter is $n$ times that of the former. 
The lattice $L_1 \oplus L_2$ for lattices $L_1$ and $L_2$ are often 
denoted by $L_1L_2$ by dropping ``$\oplus$'' in this article. 
} %
$S_0 = U[2]E_8[2]$. The subset $Z$ consists of two disjoint elliptic 
curves in the choice with $S_0 = UE_8[2]$. For all other 73 
choices,\footnote{
The discriminant group $G_0 := T_0^\vee/T_0 \cong S_0^\vee/S_0$ is 
always isomorphic to $(\Z_2)^{\oplus a}$ for some $a \in \Z_{\geq 0}$, 
because the order-2 non-symplectic automorphism $\sigma$ is assumed to 
act trivially on $S_0$ in \cite{Nikulin-factor-long}. 
The pair of integers $a$ and $r = {\rm rank}(S_0)$ capture the geometry 
of the set $Z$ of $(X, \sigma)$ associated with $(S_0, T_0, \sigma)$ 
\cite{Nikulin-factor-long}.
In this article, the values of $a$, $r$, $k$ and $g$ for $i=1,2$ are 
denoted by $a_{(i)}$, $r_{(i)}$, $k_i$ and $g_{(i)}$, respectively. 
} % 
the set $Z$ consists of one curve $C_{(g)}$ of genus $g = (22-r-a)/2$ 
in addition to $k=(r-a)/2$ rational curves 
$\P^1$ \cite{Nikulin-factor-long}:
\begin{align}
 Z = C_{(g)} \amalg \cup_{p=1}^{k} L_{p}; \qquad 
    g(C_{(g)}) = (22 - r - a)/2, \qquad 
    L_p \simeq \P^1.
  \label{eq:Z2-fixed-curves-Nikulin}
\end{align}
The subset $Z_{(4)} \subset X^{(1)} \times X^{(2)}$ of 
fixed points under the action of $\sigma = (\sigma_{(1)}, \sigma_{(2)})$
is $Z_{(4)} = Z_{(1)} \times Z_{(2)}$.

The topological cohomology group $H^4(Y;\Q)$ of $[Y]$ is
\cite[Thm. 7.31]{voisinhodge}, as an abelian group,  
\begin{equation}
H^4(Y;\Q)\simeq \left[H^4(X^{(1)} \times X^{(2)};\Q)\right]^\sigma 
    \oplus H^2(Z_{(4)};\Q);
   \label{eq:H4-full-inv-case}
\end{equation}
the superscript ${}^\sigma$ in the first term extracts the part invariant 
under the action of $\sigma$. A two-form on $Z_{(4)}$ has a corresponding 
four-form in $Y$; the two-form on $Z_{(4)}$ is pulled back to the exceptional 
divisor of the resolved $Y$, and then is taken a wedge product 
with the Poincar\'e dual of the exceptional divisor ( = mapped by the 
Gysin homomorphism). 

In the family of fourfolds $[Y]$, the horizontal component of 
$H^4(Y;\Q)$ is 
\begin{align}
 H^4_H(Y;\Q) =  \left( T_0^{(1)}\otimes T_0^{(2)}\right) \otimes \Q \oplus 
   H^1(Z_{(1)}; \Q) \otimes H^1(Z_{(2)};\Q),
    \label{eq:H4-Hpart-inv-case}
\end{align}
where the first term is from $[H^4(X^{(1)} \times X^{(2)};\Q)]^\sigma$, 
and the second term from $H^2(Z_{(4)};\Q)$. The vertical component is 
\begin{align}
  H^4_V(Y;\Q) & \; = \; \left(S_0^{(1)} \otimes S_0^{(2)}\right)\otimes \Q
 \oplus H^4(X^{(1)}; \Q) \otimes H^0(X^{(2)}; \Q)
 \oplus H^0(X^{(1)}; \Q) \otimes H^4(X^{(2)}; \Q) \nonumber \\
  & \oplus H^2(Z_{(1)};\Q) \otimes H^0(Z_{(2)};\Q)
  \oplus H^0(Z_{(1)}; \Q) \otimes H^2(Z_{(2)};\Q), 
\end{align}
where the first line and the second line come from 
$[H^4(X^{(1)}\times X^{(2)})]^\sigma$ and $H^2(Z_{(4)})$, respectively.
The entire 4th cohomology group $H^4(Y;\Q)$ is covered by the direct 
sum of the horizontal component and the vertical component in the case 
of the family of $[Y]$ over ${\cal M}_{\rm cpx~str}^{[Y]}$.
The holomorphic 4-form $\Omega_{Y_z}$ varies for 
$z \in {\cal M}_{\rm cpx~str}^{[Y]}$, but it does so only within 
$H^4_H(Y;\Q) \otimes_\Q \C$. When the point $z$ is in the subvariety 
${\cal M}_{\rm cpx~str}^{[Y]BV} \subset {\cal M}_{\rm cpx~str}^{[Y]}$, 
$\Omega_{Y_z}$ remains within $(T_0^{(1)} \otimes T_0^{(2)})\otimes_\Z \C$.

At any point $z \in {\cal M}_{\rm cpx~str}^{[Y]}$, a ($z$-dependent) Hodge 
structure is introduced in the vector space $H^4_H(Y;\Q)$; the vertical 
subspace $H^4_V(Y;\Q)$ contains only\footnote{
The notions of a rational Hodge substructure and its level will be
introduced in section \ref{sssec:math-suppl}.
} %
 the level-0 Hodge structure. 
For a vacuum complex structure $\vev{z}$ within ${\cal M}_{\rm cpx~str}^{[Y]BV}$, 
the vector subspace $H^1(Z_{(1)};\Q) \otimes H^{1}(Z_{(2)};\Q)$ supports 
a rational Hodge substructure of level 2, and 
$(T_0^{(1)}\otimes T_0^{(2)})\otimes \Q$ a rational Hodge substructure of 
level-4. Linear fluctuation $\delta z$ in the complex structure from 
$\vev{z}$
are in the Hodge (3, 1) component of $(T_0^{(1)}\otimes T_0^{(2)})\otimes \C$
(there are $(20-r_{(1)})+(20-r_{(2)})$ such deformations) 
and also in the vector space 
$H^{1,0}(Z_{(1)};\C) \otimes H^{1,0}(Z_{(2)};\C)$ (there are $g_{(1)} g_{(2)}$ 
of them); 
the former group of fluctuations
are within ${\cal M}_{\rm cpx~str}^{[Y]BV}$ and the latter group ventures out 
from ${\cal M}_{\rm cpx~str}^{[Y]BV}$ into ${\cal M}_{\rm cpx~str}^{[Y]}$ by deforming 
the $\C^2/\Z_2$ singularity of $Y_{\vev{z}}$. 
At the quadratic order in the deformation of complex structure, 
$\Omega_{Y_z} \simeq \Omega_{Y_{\vev{z}}} + (\delta z)^a \psi_a + 
(\delta z)^a(\delta z)^b \psi_{ab}$ for $z = \vev{z} + \delta z$, 
the quadrature of the $(40-r_{(1)}-r_{(2)})$ complex structure deformations 
do not bring $\Omega_{Y_z}$ out of $(T_0^{(1)} \otimes T_0^{(2)}) \otimes \C$. 
The quadrature involving $g_{(1)}g_{(2)}$ complex structure deformations, 
however, may be in the entire $H^{4}_H(Y_{\vev{z}};\C)$.

The observation above on the Hodge substructures on $H^4(Y_{\vev{z}};\Q)$ 
and finitely perturbed $\Omega_{Y_z}$ on them indicates that 
a non-trivial flux is necessary at least in the 
$(T_0^{(1)} \otimes T_0^{(2)}) \otimes \Q$ component in order to generate  
mass terms of the $(40-r_{(1)}-r_{(2)})$ moduli fields.\footnote{
Comments on the $r_{(1)}=r_{(2)}=20$ case will be found later in this article. 
} %
 The $g_{(1)}g_{(2)}$ moduli 
fields may also acquire mass terms from a flux in 
$(T_0^{(1)}\otimes T_0^{(2)})\otimes \Q$, or they may not.\footnote{
\label{fn:heuristic}
Here is a heuristic argument. Think of a case that a Borcea--Voisin 
orbifold $Y=(X^{(1)}\times X^{(2)})/\Z_2$ has a mirror $Y^{\circ} = 
(X^{(1)}_\circ \times X^{(2)}_\circ)/\Z_2$ using the mirror $X^{(1)}_\circ$ 
and $X^{(2)}_\circ$ of $X^{(1)}$ and $X^{(2)}$; suppose that $X^{(i)}_\circ$ can 
be chosen from Nikulin's list, where 
$r^\circ_{(i)} = (20-r_{(i)})$ and $a^\circ_{(i)} = a_{(i)}$ for $i=1,2$. 
The intersection ring of the mirror $Y^\circ$ can be used to infer 
finite perturbation of the complex structure $\Omega_Y$ of $Y$. 
The fourfold $Y^\circ$ has three groups of divisors; $D^{(1)}$ that 
originate from the divisors of $X^{(1)}_\circ$, $D^{(2)}$ that 
originate from the divisors of $X^{(2)}_\circ$, and the exceptional 
divisors $D_\sigma$ associated with the $\C^2/\Z_2$ orbifold singularity.
The intersection numbers of the form $(D_\sigma)^2 \cdot D^{(1)} \cdot D^{(2)}$ 
are determined by the intersection numbers of the curve of the 
involution-fixed points in $X^{(i)}_\circ$ with $D^{(i)}$, and are 
non-zero when the curves of fixed points are non-empty. This observation
hints that there is a good chance that a quadratic order perturbation 
of $\Omega_Y$ deforming the $\C^2/\Z_2$ singularity turns into a mass term 
in the presence of a flux in $(T_0^{(1)}\otimes T_0^{(2)})\otimes \Q$. 

There are logical gaps to fill, however. 
One is that the classical intersection 
ring in $Y^\circ$ has an immediate information on the $\Omega_Y$ in the 
large complex structure limit of $Y$, not in the zero deformation limit 
(orbifold limit) of $\Omega_Y$. The other is that a flux needs to be
in $W_{(20|02)}$ component within $(T_0^{(1)}\otimes T_0^{(2)})\otimes \Q$ 
as we will see in section \ref{ssec:Tx=T0}. 
} % 
We take it out of the scope of this article to study $\Omega_{Y_z}$ 
at the quadratic order in $\delta z$ including those $g_{(1)}g_{(2)}$ moduli. 
So it is not a necessary condition---at this moment---for all the 
complex structure moduli stabilization that 
$H^1(Z_{(1)};\Q)\otimes H^1(Z_{(2)};\Q)$ contains a level-0 rational 
Hodge substructure.\footnote{
\label{fn:Z1xZ2-correspondence}
This condition is equivalent to existence of an algebraic curve
in $Z_{(1)} \times Z_{(2)}$ other than a copy of $Z_{(1)} \times {\rm pt}$ or 
${\rm pt}\times Z_{(2)}$. 
} %
In this article, therefore, we assume that the Hodge structure on 
$(T_0^{(1)}\otimes T_0^{(2)})\otimes \Q$ is of CM-type (starting in 
section \ref{ssec:cond-flux-MorF}), and study when and how 
supersymmetric flux is admitted in this component; 
we do not ask whether the Hodge structure on 
$H^1(Z_{(1)};\Q)\otimes H^1(Z_{(2)};\Q)$ is CM-type, or has a level-0 
Hodge substructure. 

%%%%%%%%%%%%%%%%%%%%%%%%%%%%%%%%%%%%%%%%%%%%%%%%%
\subsection{The Conditions of Supersymmetric Fluxes in M/F-theory: \\
Cases with CM-type Calabi--Yau Compactifications}
\label{ssec:cond-flux-MorF}
%%%%%%%%%%%%%%%%%%%%%%%%%%%%%%%%%%%%%%%%%%%%%%%%

%%%%%%%%%%%%%%%%%%%%%%%%%%%%%%%%%%%%%%%%%%%%%%%%%%%
\subsubsection{Two Perspectives}
\label{sssec:2-perspectives}
%%%%%%%%%%%%%%%%%%%%%%%%%%%%%%%%%%%%%%%%%%%%%%%%%%%

As is well-known, there are two different perspectives in 
describing the way a topological flux $G \in H^4(Y;\Q)$ in a 
Calabi--Yau fourfold $Y$ stabilizes the complex structure 
moduli of $Y$. One is more physical, and the other more mathematical, 
as we repeat them shortly. Either way, the condition for supersymmetry 
is stated concisely by the F-term condition\footnote{
Here, we use the superpotential (\ref{eq:GVW}) and 
the K\"{a}hler potential obtained by dimensional reduction. 
All kinds of corrections expected in an effective theory of four supersymmetry 
charges are not taken into account. 
} %
\begin{align}
   DW = 0: \qquad G^{(1,3)}=0    \label{eq:cond-DW=0}
\end{align}
and the additional condition for the Minkowski spacetime and $m_{3/2}=0$ 
after compactification,  
\begin{align}
  W=0: \qquad G^{(0,4)} = 0.    \label{eq:cond-W=0}
\end{align}

In the more physical perspective, we think that a topological flux $G$ is 
specified as a part of data of compactification first, and then 
the superpotential (\ref{eq:GVW}) gives rise to non-trivial scalar 
potential of the complex structure moduli fields of $Y$; the 
expectation value of those fields adjust themselves in the early 
period of time in the universe to arrive at a potential minimum, 
where the resulting complex structure of $Y$ is such that the Hodge 
$(1,3)$ component of the topological $G \in H^4(Y;\Q)$ must be absent 
when measured in that complex structure. For such a topological $G$ 
and the complex structure of $Y$ so determined, it is a non-trivial 
question whether the Hodge $(0,4)$ component of $G$ vanishes (the 
condition (\ref{eq:cond-W=0}) is satisfied) or not. 

In the more mathematical perspective, on the other hand, we pose questions 
that are concerned about classification of flux vacua, forgetting about 
cosmological time evolution before the complex structure moduli fields  
come down the potential to their vacuum value. We pick up one point in the 
complex structure moduli space $z \in {\cal M}_{\rm cpx~str}^{[Y]}$, and ask 
if there is any topological flux $G \in H^4(Y_z;\Q)$ whose $H^{1,3}(Y_z; \C)$
component vanishes; here, $Y_z=Y$ is the fourfold of the topological type 
$[Y]$ with the complex structure corresponding to the point 
$z \in {\cal M}_{\rm cpx~str}^{[Y]}$, emphasizing the $z$-dependence.
The condition (\ref{eq:cond-W=0}) can also be phrased in the same way. 
At a generic point $z \in {\cal M}_{\rm cpx~str}^{[Y]}$, only the trivial 
purely horizontal flux $G=0 \in H^4(Y_z;\Q)$ satisfy the conditions 
(\ref{eq:cond-DW=0}).
Points in ${\cal M}_{\rm cpx~str}^{[Y]}$ where non-trivial fluxes 
$G \in H^4(Y;\Q)$ satisfy the condition (\ref{eq:cond-DW=0})
form a special sub-locus of ${\cal M}_{\rm cpx~str}^{[Y]}$. This is a 
Noether--Lefschetz problem in a Calabi--Yau fourfold $[Y]$.
In this article, we exploit the latter perspective. 

The D-term condition, or equivalently the primitivity, also needs to be 
satisfied for a flux on $Y$ to be supersymmetric. The F-term condition 
(\ref{eq:cond-DW=0}) and the D-term condition are almost\footnote{
We treat fluxes in this article as elements in the $\Q$-coefficient 
cohomology, not in the $\Z$-coefficient, and the upper bound on the D3-brane 
charge is not imposed. At this level of analysis, fluxes in the purely 
vertical part and purely horizontal part can be regarded completely 
independent. 
} %
independent, 
however, because the F-term [resp. D-term] condition constrains the 
purely horizontal [resp. purely vertical] part of the flux 
(cf \cite{mir-HD, flux-Sakura, BW-vhr}).
In sections \ref{sec:flux-analysis} and \ref{sec:non-Z2-orbifold}, 
we do not deal with the purely vertical part of the flux (or the D-term 
condition), because they are not relevant to the gravitino mass. 

% For this reason, one does 
% not have to pay close attention to whether we should use a fourfold $Y$ 
% or another fourfold birational to $Y$ at this moment. 

In the rest of this section \ref{ssec:cond-flux-MorF}, we will see 
that the condition for existence of a non-trivial and purely horizontal 
supersymmetric flux can be stated 
in a very concise way---(\ref{eq:cond-DW=0-smpl-Hdge}, 
\ref{eq:cond-DW=W=0-smpl-Hdge})---when the vacuum complex structure of 
a Calabi--Yau fourfold $Y$ is of {\it CM-type}.\footnote{
cf see footnote \ref{fn:CM-CY-CM-substr} 
} %
It is done by using the notion of {\it simple components of the Hodge 
structure}. This is a preparation for the analysis in 
sections \ref{ssec:Tx=T0} and \ref{ssec:Tx-neq-T0}.  
%

%%%%%%%%%%%%%%%%%%%%%%%%%%%%%%%%%%%%%%%%%%%%%%%%%%%%%%%%
\subsubsection{Math Supplementary Notes: Simple Hodge Substructure, and CM-type}
\label{sssec:math-suppl}
%%%%%%%%%%%%%%%%%%%%%%%%%%%%%%%%%%%%%%%%%%%%%%%%%%%%%%%%

Before doing anything else, we should have a definition of 
simple Hodge structure and a minimum account for what CM-type stands for. 

(Definition) 
Let $V_\Q$ be a vector space over $\Q$. A decomposition of the 
vector space $V_\Q \otimes_\Q \C$ over $\C$ into the form of 
\begin{align}
  V_\Q \otimes_\Q \C \cong \oplus_{p+q=n} V_\C^{p,q} \qquad \quad 
  \left(\overline{V_\C^{p,q}} = V_\C^{q,p}\right) 
\end{align}
of vector subspaces $V_\C^{p,q}$ for non-negative integers $p$, $q$, and $n$, 
is called a {\bf rational Hodge structure of weight $n$}. For a smooth 
compact K\"{a}hler manifold $M$, the cohomology group $H^n(M;\Q)$ has 
a rational Hodge structure of weight $n$ given by the complex structure of 
the K\"{a}hler manifold $M$, for example.

A rational Hodge structure on a vector space $V_\Q$ is said to be 
{\bf simple}, if there is no vector proper subspace $W_\Q \subset V_\Q$ 
over $\Q$ so that $\oplus_{p,q} (V_\C^{p,q} \cap (W_\Q \otimes \C))$
reproduces $(W_\Q \otimes \C)$. When such a proper subspace $W_\Q$ exists, 
$V_\Q$ [resp. $W_\Q$] is said to have [resp. to support] 
a {\bf rational Hodge substructure}. An example of rational Hodge structure 
that is not simple is the Hodge structure on $H^2(X;\Q)$ of an algebraic K3 
surface $X$; both $S_X \otimes \Q$ and $T_X \otimes \Q$ support a 
rational Hodge substructure of $H^2(X;\Q)$. When a vector space $V_\Q$ with 
a rational Hodge structure is decomposed into vector subspaces over $\Q$, 
$V_\Q \cong \oplus_{a \in A} W_a$, and each $W_a$ supports a rational Hodge 
substructure that is simple, we say that it is a 
{\bf simple component decomposition of the rational Hodge 
structure}.
% 
% \footnote{
% %
% When there is a {\bf polarized} rational 
% Hodge structure on a vector space $V_\Q$, there is just one 
% decomposition into the simple components, as we can require that 
% different simple components are orthogonal.
% {\bf true?  (within the level-0 components)}
% } %

A simple component $W_a$ in such a decomposition is said to be 
{\bf level-$\ell$}, when 
$\ell := {\rm Max}(|p-q|; V^{p,q} \cap (W_a \otimes \C) \neq \phi )$. 
For example, $T_X \otimes \Q$ of an algebraic K3 surface $X$ is a simple 
component of level-2, and $S_X \otimes \Q$ contains only level-0 simple 
components. 
(the end of the Definition)

\vspace{5mm}

(Definition) 
CM-type\footnote{
As the present authors have already included a pedagogical review on this 
in \cite{prev.paper.preprint}, a brief explanation in the following 
is kept to the minimum. 
} %  
 is a special property---see the next paragraph---that 
a rational Hodge structure on a vector space over $\Q$ may have. 
For a complex $m$-dimensional K\"{a}hler manifold $M$, its complex structure 
introduces a rational Hodge 
structure on $H^m(M;\Q)$ and one may ask whether the rational Hodge 
structure is of CM-type or not. A choice of complex structure of $M$ 
[resp. a point in ${\cal M}_{\rm cpx~str}^{[M]}$] is said to be {\bf CM-type} 
[resp. a {\bf CM point}] when that is the case. 

Suppose that a vector space $V_\Q$ over $\Q$ is given a rational Hodge 
structure. It is {\bf of CM-type} when the algebra of 
Hodge-structure-preserving $\Q$-linear maps from $V_\Q$ to 
itself---${\rm End}_{\rm Hdg}(V_\Q)$---is abelian, and has 
$\dim_\Q({\rm End}_{\rm Hdg}(V_\Q))$ equal to $\dim_\Q V_\Q$.  
(the end of Definition)

This is a property of the Hodge structure on $H^1(T^2;\Q)$ for 
elliptic curves $T^2$ with complex multiplication.\footnote{
\label{fn:CM-ell}
For an elliptic curve $T^2 = \C/(\Z \oplus \tau \Z)$ with $\tau^2 + 1 = 0$, 
for example, not necessarily one-to-one holomorphic maps such as 
$[(i)\times]: T^2 \rightarrow T^2$ and $[(1+2i)\times]: T^2 \rightarrow T^2$ 
that multiply complex numbers are examples of complex multiplication 
operations. More generally, an elliptic curve $T^2 = \C/(\Z \oplus \tau \Z)$
has non-trivial complex multiplication operations if and only if 
there is a set of non-trivial integers $(a,b,c)$ satisfying 
$a\tau^2+b\tau + c=0$. 
Each complex multiplication operation induces a $\Q$-linear map 
$H^1(T^2;\Q) \rightarrow H^1(T^2;\Q)$ that preserves the Hodge decomposition 
of $H^1(T^2;\Q)\otimes \C$. The field ${\rm End}_{\rm Hdg}(H^1(T^2;\Q))$ 
of an elliptic curve with complex multiplication is an imaginary quadratic 
field (the easiest class of CM fields) whose extension degree is 
equal to $\dim_\Q(H^1(T^2;\Q)) = 2$.
} % 
CM type is a notion that generalizes the complex multiplication on elliptic 
curves to more general complex manifolds. When a CM-type rational Hodge 
structure on $V_\Q$ is simple, then the algebra ${\rm End}_{\rm Hdg}(V_\Q)$ 
is always a field with a special property; this class of fields is called 
a CM field; a brief review on the properties of CM fields is found, 
for example, in \cite[\S A.2]{prev.paper.preprint}.

%%%%%%%%%%%%%%%%%%%%%%%%%%%%%%%%%%%%%%%%%%%%%%%%%%%%%%
\subsubsection{A Frequently Used Property}
\label{sssec:FUP}
%%%%%%%%%%%%%%%%%%%%%%%%%%%%%%%%%%%%%%%%%%%%%%%%%%%%%%

The following is a textbook-level material in math, but is 
a powerful tool frequently used in this article. So, we include 
its statement in this article for the convenience of readers 
(a little more explanation is found in \cite[\S B.2]{prev.paper.preprint}).

%
% \begin{props}\label{prp:embd-vect-corresp}
Let $V_\Q$ be a vector space over $\Q$, and $F$ a number field of degree 
$[F:\Q] = \dim_\Q V_\Q$; suppose that $F$ acts on $V_\Q$ through 
$\phi: F\hookrightarrow {\rm End}_\Q(V_\Q)$. Let us choose an (arbitrary) 
isomorphism $\imath: F \cong V_\Q$ as a vector space over $\Q$. Then 
the action of $\phi(F) \subset {\rm End}_\Q(V_\Q)$ on $V_\Q \otimes_\Q 
F^{\rm nc}$ can be diagonalized simultaneously; to be more specific,  
$V_\Q \otimes_\Q F^{\rm nc}$ has a diagonalization basis\footnote{
The superscript ``nc'' stands for the normal closure.
} %  
\begin{align}
V_\Q \otimes_\Q F^{\rm nc} = {\rm Span}_{F^{\rm nc}} \{ v_a \; 
| \; a=1,\cdots, \dim_\Q V_\Q\}, 
\end{align}
there is a 1-to-1 correspondence between those $\dim_\Q V_\Q$ basis elements 
and the set of all the $[F:\Q]$ embeddings $\Phi_F^{\rm full} := 
{\rm Hom}_\Q(F,\overline{\Q})$, and  
\begin{align}
  x \cdot v_a = v_a \rho_a(x), \qquad \qquad \rho_a \in \Phi_F^{\rm full}, \quad 
{}^\forall x \in \phi(F). 
\end{align}
Moreover, when we express the eigenvectors $v_a$ as $F^{\rm nc}$-coefficient 
linear combinations of a $\Q$-basis $\{ e_i \; | \; i=1,\cdots, \dim_\Q V_\Q\}$ 
of $V_\Q$, $v_a = \sum_i e_i c^i_a$, there exists a basis 
$\{ y_i \; | \; i=1,\cdots, [F:\Q]\}$ of the vector space $F$ over $\Q$ so that 
\begin{align}
  c^i_a = \rho_a(y_i).
  \label{eq:c-y-egvct-relation}
\end{align}

In the context of this article, we wish to use the property above 
for $V_\Q$ as $T_X \otimes \Q$ of a CM-type K3 surface $X$, and also 
as one of simple components of $H^4(Y;\Q)$ with a CM-type rational Hodge 
structure. The role of the field $F$ above is played by the CM field 
${\rm End}_{\rm Hdg}(V_\Q)$. In that context, each one of the eigenvectors, 
say, $v_a$, belongs to a definite Hodge $(p_a,q_a)$ component.  
% \end{props}

%%%%%%%%%%%%%%%%%%%%%%%%%%%%%%%%%%%%%%%%%%%%%%%%%%%%%%%%%%%
\subsubsection{The Conditions}
%%%%%%%%%%%%%%%%%%%%%%%%%%%%%%%%%%%%%%%%%%%%%%%%%%%%%%%%%%%

%
With the jargons prepared in section \ref{sssec:math-suppl}, 
and the special property of CM-type Hodge structure explained in 
section \ref{sssec:FUP}, we can now translate the conditions on 
supersymmetric fluxes as follows. First, let 
\begin{align}
  H^4(Y_z;\Q) \cong \oplus_{a \in A} \left(H^4(Y_z;\Q) \right)_a 
\end{align}
be a simple component decomposition of the rational Hodge structure of 
$H^4(Y_z; \Q)$ at $z \in {\cal M}_{\rm cpx~str}^{[Y]}$. 
We already take the second (math) perspective in 
section \ref{sssec:2-perspectives}. 
A flux $G$ in $H^4(Y;\Q)$---if there is any---can be decomposed into 
$\sum_{a\in A} G_a$ with $G_a \in \left( H^4(Y_z;\Q)\right)_a$ accordingly. 
Whether $G$ satisfies the $DW=0$ condition or the $DW=W=0$ condition 
for the complex structure $z$ can be discussed separately for individual 
$G_a$'s. 

Now, the property stated in section \ref{sssec:FUP} implies for a 
CM-type simple Hodge structure in $\left(H^4(Y_z;\Q)\right)_a$ that 
there is a basis $\{ v^{(a)}_b \}$ in $\left( H^4(Y_z;\C) \right)_a$ 
over which the action of the CM field 
${\rm End}_{\rm Hdg}((H^4(Y_z;\Q))_a)$ is diagonalized; 
$b = 1,\cdots, \dim_\Q(  (H^4(Y_z;\Q))_a )$.  
An intriguing property of the coefficients $c^i_b$'s 
in (\ref{eq:c-y-egvct-relation}) and hence that of the basis elements 
$v_b$'s is that the Galois group 
${\rm Gal}(  {\rm End}_{\rm Hdg}((H^4(Y_z;\Q))_a)/\Q  )$ acts on them as 
permutation and transitively; $\sigma \cdot \rho_b =: \rho_{\sigma(b)}$, and 
$\sigma(v^{(a)}_b) = v^{(a)}_{\sigma(b)}$. 
 When a rational element $G_a \in ( H^4(Y_z;\Q) )_a$ is decomposed 
in this basis, $G_a = \sum_b v^{(a)}_b g^{(a)}_b$ with coefficients $g^{(a)}_b$'s 
in $\overline{\Q}$, $G_a$ should be invariant 
under the Galois group action, so we obtain 
$\sigma(g^{(a)}_b) = g^{(a)}_{\sigma(b)}$. In particular, if $G_a$ is such that 
$g^{(a)}_b \neq 0$ for some $b$, then $g^{(a)}_b \neq 0$ for all $b = 1,\cdots, 
\dim_\Q((H^4(Y_z;\Q))_a)$. 

Therefore, the condition that a topological flux $G \in H^4(Y_z;\Q)$ does 
not have the (1,3) Hodge component is translated as follows:  
\begin{align}
  {}^\forall G_a \in \left( H^4(Y_z;\Q) \right)_a \quad 
 & \; {\rm if~} \left(\left( H^4(Y_z:\Q) \right)_a \! \otimes_\Q \C \right) 
       \cap H^{1,3} = \phi,     \nonumber \\
  G_a = 0 \quad & \; {\rm if~} \left(\left( H^4(Y_z:\Q) \right)_a \! 
      \otimes_\Q \C \right) \cap H^{1,3} \neq \phi.  
  \label{eq:cond-DW=0-smpl-Hdge}
\end{align}
In particular, if all the simple components have non-empty Hodge $(1,3)$ 
components, then only the trivial flux $G=0$ is consistent with the 
$DW=0$ condition at $z \in {\cal M}_{\rm cpx~str}^{[Y]}$. Similarly, 
the condition that the topological flux $G \in H^4(Y_z;\Q)$ has neither 
the $(1,3)$-component nor $(0, 4)$ component is translated as follows:
\begin{align}
  {}^\forall G_a \in \left( H^4(Y_z;\Q) \right)_a \quad 
 & \; {\rm if~} \left(\left( H^4(Y_z:\Q) \right)_a \otimes_\Q \C \right) \cap 
  \left( H^{1,3} \oplus H^{0,4} \right) = \phi, 
    \nonumber \\
  G_a = 0 \quad & \; {\rm if~} \left(\left( H^4(Y_z:\Q) \right)_a \otimes_\Q 
     \C \right) \cap 
   \left( H^{1,3} \oplus H^{0,4} \right) \neq \phi. 
   \label{eq:cond-DW=W=0-smpl-Hdge}
\end{align}
The $DW=W=0$ condition, and hence this last condition is further translated 
as follows: $G_a=0$ in all the simple components with the level $\ell >0$.

%%%%%%%%%%%%%%%%%%%%%%%%%%%%%%%%%%%%%%%%%%%%%%%%%%%%%%%
\subsection{Cases with a Generic CM Point in $D(T_0)$}
\label{ssec:Tx=T0}
%%%%%%%%%%%%%%%%%%%%%%%%%%%%%%%%%%%%%%%%%%%%%%%%%%%%%%%%%

In sections \ref{ssec:Tx=T0} and \ref{ssec:Tx-neq-T0}, we work out the 
conditions (\ref{eq:cond-DW=0-smpl-Hdge}, \ref{eq:cond-DW=W=0-smpl-Hdge}) 
for existence of a non-trivial supersymmetric flux in the 
$(T_0^{(1)}\otimes T^{(2)}_0)\otimes \Q$ component for a vacuum complex 
structure in (\ref{eq:CM-X1-X2-subspace}). It is done by translating the 
conditions (\ref{eq:cond-DW=0-smpl-Hdge}, \ref{eq:cond-DW=W=0-smpl-Hdge}) 
into arithmetic characterizations on the vacuum complex structure.  
Before doing anything, however,

In section \ref{ssec:Tx=T0}, we deal with the cases where 
complex structure of $X^{(1)}$ and $X^{(2)}$ are CM-type but otherwise 
generic in the period domains $D(T_0^{(1)})$ and $D(T_0^{(2)})$; this means
that $T_X^{(i)}=T_0^{(i)}$. 
Analysis in sections \ref{sssec:math-preparation-tensor} 
and \ref{sssec:flux-tx-is-t0} reveals that 
the condition (\ref{eq:cond-DW=0-smpl-Hdge}) for a $DW=0$ flux 
is translated to (\ref{eq:flux-cond-p1}), and 
the condition (\ref{eq:cond-DW=W=0-smpl-Hdge}) for a $DW=W=0$ flux to 
(\ref{eq:flux-cond-math2}); busy readers might choose 
to skip the analysis and proceed to a recap in 
p. \pageref{pg:phys-recap} at the end of section \ref{sssec:flux-tx-is-t0}.  
The effective field theory (including mass matrices and symmetries) 
of complex structure moduli fields is studied in 
section \ref{sssec:mass-TX=T0}.  
%

%%%%%%%%%%%%%%%%%%%%%%%%%%%%%%%%%%%%%%%%%%%%%%%%%%%%%
\subsubsection{Tensor Product of a Pair of CM-type Hodge Structures}
\label{sssec:math-preparation-tensor}
%%%%%%%%%%%%%%%%%%%%%%%%%%%%%%%%%%%%%%%%%%%%%%%%%%%%%

For a complex structure in (\ref{eq:CM-X1-X2-subspace}) generic enough 
to have $T_0^{(i)}=T_X^{(i)}$, the rational Hodge structure on 
$V_i := T_0^{(i)} \otimes \Q$ is simple and CM-type (by assumption) for both 
$i=1,2$; let $K^{(i)}$ denote their CM fields. It is then known 
\cite[Prop. 1.2]{Borcea} that the rational Hodge structure on 
$V_1 \otimes V_2 = (T_0^{(1)}\otimes T_0^{(2)})\otimes \Q$ is also of CM type. 
The rational Hodge structure on $(T_0^{(1)}\otimes T_0^{(2)})\otimes \Q$ is 
not necessarily simple, however. 

In fact, the non-simple nature of a rational Hodge structure of 
$H^4(Y;\Q)$ (or of $H^3(M;\Q)$) is an essential ingredient for $\vev{W}=0$
\cite{DeWolfe}. In Ref. \cite{prev.paper.PRD}, for example, 
$M = (E_\tau \times X^{(2)})/\Z_2$ with a CM elliptic curve $E_\tau$ 
and a CM-type K3 surface $X^{(2)}$ is used for a Type IIB orientifold; 
the authors found $DW=W=0$ fluxes by exploiting a case the rational 
Hodge structure is not simple on $V_1 \otimes V_2$ with 
$V_1 = H^1(E;\Q)$ and $V_2 = T_X^{(2)} \otimes \Q$. 
We will also do so on $V_1 \otimes V_2 = 
(T_0^{(1)}\otimes T_0^{(2)})\otimes \Q$ in this article. 

It was not difficult to work out the simple component 
decomposition of $(V_1 \otimes V_2)$ in \cite{prev.paper.PRD}, 
when $V_1 = H^1(T^2;\Q)$ is of 
just two-dimensions, and we know that $K^{(1)}$ is an imaginary quadratic 
field. For a general $V_1 = T_X^{(1)} \otimes \Q$ and $K^{(1)}$ of a CM-type 
K3 surface $X^{(1)}$, however, we need to be equipped with an understanding on 
general structure of the simple component decomposition of $V_1 \otimes V_2$. 
That is what we do in section \ref{sssec:math-preparation-tensor} 
(by exploiting \cite[\S5]{ST}), and we will arrive 
at (\ref{eq:K1-K2-Chinese}, \ref{eq:V1-V2-Chinese}, \ref{eq:Hdg-isom-smpl-dcmp}, \ref{eq:temp-temp}). 

\vspace{1cm}

Step 1: The algebras of endomorphisms $K^{(1)} \subset {\rm End}_{\rm Hdg}(V_1)$ 
and $K^{(2)} \subset {\rm End}_{\rm Hdg}(V_2)$ give rise to an algebra of Hodge 
endomorphisms of the vector space $(V_1 \otimes_\Q V_2)$; 
$K^{(1)} \otimes_\Q K^{(2)} \hookrightarrow {\rm End}_{\rm Hdg}(V_1 \otimes_\Q V_2)$.
Similarly to the fact that the rational Hodge structure on 
$(V_1 \otimes_\Q V_2)$ is not necessarily simple, the algebra 
$(K^{(1)} \otimes_\Q K^{(2)})$ of endomorphisms of $(V_1 \otimes_\Q V_2)$ 
is not necessarily a field. The first step is to look at the structure 
of the algebra $K^{(1)}\otimes_\Q K^{(2)}$. 

First, let us explain\footnote{
Any introductory textbook on field theory (such as \cite{Fujisaki, Roman}) 
will be useful in following the discussions in 
sections \ref{sssec:math-preparation-tensor} 
and \ref{sssec:flux-tx-is-t0}.
} %
 the decomposition (\ref{eq:K1-K2-Chinese}).
The field extension $K^{(1)}$
over $\Q$ is always expressed in the form of $K^{(1)} = \Q(\alpha)$ 
for some $\alpha \in K^{(1)}$. Let $f_{\alpha/\Q} \in \Q[x]$ be a minimal 
polynomial of $\alpha \in K^{(1)}$ over $\Q$, which means that 
$K^{(1)} \cong \Q(\alpha) \cong \Q[x]/(f_{\alpha/\Q})$, and 
\begin{align}
  K^{(2)} \otimes_\Q K^{(1)} \cong K^{(2)} \otimes_\Q \Q[x]/(f_{\alpha/\Q}) 
   = K^{(2)}[x]/(f_{\alpha/\Q}).
\end{align}
Although the minimal polynomial $f_{\alpha/\Q}$ is irreducible in the ring 
$\Q[x]$, it may in principle be factorizable in the ring $K^{(2)}[x]$; 
let $f_{\alpha/\Q} (x) = \prod_{i=1}^r g_i(x)$ be an irreducible factorization, 
where $g_i(x) \in K^{(2)}[x]$. The Chinese remainder theorem is used to obtain 
\begin{align}
  (K^{(1)} \otimes_\Q K^{(2)}) \cong K^{(2)}[x]/(f_{\alpha/\Q}) \cong \oplus_{i=1}^r \; 
    K^{(2)}[x]/(g_i) =: \oplus_{i=1}^r L_i.
  \label{eq:K1-K2-Chinese}
\end{align}
The algebra $K^{(1)} \otimes_\Q K^{(2)}$ is decomposed into a direct sum 
of number fields $K^{(2)}[x]/(g_i)$; each component is a degree 
$[L_i:K^{(2)}] = {\rm deg}(g_i)$ extension field over $K^{(2)}$. 

Second, let us spell out the relation between the sets of embeddings 
of $K^{(1)}$ and $K^{(2)}$, 
\begin{align}
 \Phi_{K^{(1)}}^{\rm full} := {\rm Hom}_\Q(K^{(1)}, \overline{\Q})
     \quad {\rm and} \quad
 \Phi_{K^{(2)}}^{\rm full} := {\rm Hom}_\Q(K^{(2)}, \overline{\Q}), 
\end{align}
respectively, and those of the number fields $L_i$; 
remember that the set of embeddings of the CM fields play an important role in describing a Hodge structure of CM-type (section \ref{sssec:FUP}). 
The set of embeddings $\Phi_{K^{(1)}}^{\rm full} \times \Phi_{K^{(2)}}^{\rm full}$ 
of the algebra $K^{(1)} \otimes_\Q K^{(2)}$ is decomposed into 
\begin{align}
  \Phi_{K^{(1)}}^{\rm full} \times \Phi_{K^{(2)}}^{\rm full} = \amalg_{i=1}^r \Phi_{L_i}^{\rm full}, 
\qquad \Phi_{L_i}^{\rm full} = \left\{ (\rho^{(1)}, \rho^{(2)}) \; | \; \rho^{(1)}(\alpha) {\rm ~is~a~root~of~} (\rho^{(2)}(g_i))(x)=0 \right\}; 
  \label{eq:decmp-PhiK1-PhiK2-PhiLs}
\end{align}
obviously individual $\Phi_{L_i}^{\rm full}$'s consist of 
${\rm deg}(g_i) \times [K^{(2)}:\Q]$ distinct embeddings of the number field 
$L_i$ (so the notation $\Phi^{\rm full}_{L_i}$ is appropriate), and the subsets 
$\Phi_{L_i}^{\rm full}$ for $i=1,\cdots, r$ are mutually exclusive in 
$\Phi_{K^{(1)}}^{\rm full} \times \Phi_{K^{(2)}}^{\rm full}$, 
because the polynomial $f_{\alpha/\Q}$ is separable. 
Now, both the algebra $K^{(1)}\otimes_\Q K^{(2)}$ and its set of embeddings 
$\Phi^{\rm full}_{K^{(1)}} \times \Phi^{\rm full}_{K^{(2)}}$ have decompositions, 
(\ref{eq:K1-K2-Chinese}) and (\ref{eq:decmp-PhiK1-PhiK2-PhiLs}), respectively.
The two decompositions are compatible in fact, in that the 
embeddings in $\Phi_{L_i}^{\rm full}$ are trivial on the other direct sum 
components, $L_j$'s with $j \neq i$, as follows. As a part of the Chinese 
remainder theorem, there exists 
$a_i \in K^{(2)}[x]/(g_i)$ for $i=1,\cdots, r$ so that 
\begin{align}
  1 = \sum_i a_i f'_i \in K^{(2)}[x]/(f_{\alpha/\Q}), \qquad \qquad 
     f'_i := \prod_{j \neq i} g_j, 
\end{align}
in line with\footnote{
memo: 
The Chinese remainder theorem is valid for a PID $R$, and its ideals $P_i$ 
that are mutually prime. Let $P_i = (p_i)_R$. 
Then the element $a_i \in R/P_i$ satisfies 
$a_i \cdot (\prod_{j\neq i} p_j) = 1 \in R/(p_i)$. 
The homomorphism from $R/P$ to $\oplus_i (R/P_i)$ is obtained by 
just dividing further; divide a residue mod 
$\prod_i p_i$ by $p_i$ to find its residue mod $p_i$. Under the homomorphism 
from $\oplus_i (R/P_i)$ to $R/P$, on the other hand, 
$(y_1, \cdots, y_r) \in \oplus R/(p_i)$ is assigned 
$\sum_i y_i a_i (\prod_{j \neq i} p_j) \in R/(\prod_k p_k)$. 
} %  
the decomposition $K^{(2)}[x]/(f_{\alpha/\Q}) \cong 
\oplus_i K^{(2)}[x]/(g_i)$.  So, an element in $L_j$ can be regarded as 
a polynomial in $K^{(2)}[x]$ times $a_j f'_j$ (mod $f_{\alpha/\Q}$), whose 
image by any embedding in $\Phi_{L_i}^{\rm full}$ with $i\neq j$ vanishes 
because $f'_j$ contains the factor $g_i$.

It is useful to note that the Galois group 
${\rm Gal}((K^{(1)}K^{(2)})^{\rm nc}/\Q)$ acts on the set 
$\Phi_{K^{(1)}}^{\rm full} \times \Phi_{K^{(2)}}^{\rm full}$; a Galois 
transformation $\sigma \in {\rm Gal}((K^{(1)}K^{(2)})^{\rm nc}/\Q)$ converts 
an embedding $\rho^{(1)} \otimes \rho^{(2)} \in \Phi_{K^{(1)}}^{\rm full} \times 
\Phi_{K^{(2)}}^{\rm full}$ to another embedding given by 
$\sigma \cdot (\rho^{(1)} \otimes \rho^{(2)}): K^{(1)} \otimes_\Q K^{(2)} 
\rightarrow \overline{\Q} \rightarrow \overline{\Q}$. 
\label{pg:Gal-act-on-PhiPhi}
The decomposition (\ref{eq:decmp-PhiK1-PhiK2-PhiLs}) can be 
regarded as the orbit decomposition under this group action. 
So, this observation further indicates that the 
decomposition (\ref{eq:decmp-PhiK1-PhiK2-PhiLs}) is independent 
of the choice of the primitive element $\alpha$ of $K^{(1)} \cong \Q(\alpha)$. 

Instead of exploiting the structure of $K^{(1)}$ as $\Q({}^\exists \alpha)$, 
we could have exploited the same structure of $K^{(2)}$; 
$K^{(2)}$ is regarded as $K^{(2)} \cong \Q(\alpha')$ for an appropriate choice 
of $\alpha' \in K^{(2)}$; 
find its minimal polynomial over $\Q$, and factorize the polynomial 
over $K^{(1)}$ to find another decomposition of $K^{(1)} \otimes_\Q K^{(2)}$ 
into a direct sum of number fields. So, yet another 
decomposition of the set $\Phi_{K^{(1)}}^{\rm full} \times \Phi_{K^{(2)}}^{\rm full}$ 
also follows. This decomposition must be the orbit decomposition of the 
action of ${\rm Gal}((K^{(1)}K^{(2)})^{\rm nc}/\Q)$ on 
$\Phi_{K^{(1)}}^{\rm full} \times \Phi_{K^{(2)}}^{\rm full}$, where the same group 
acts on the same set precisely in the same way as before. 
So, the decomposition of the embeddings should be independent of whether 
we exploit $K^{(1)} \cong \Q(\alpha)$ or $K^{(2)} = \Q(\alpha')$, and so is  
the decomposition of the algebra $K^{(1)} \otimes_\Q K^{(2)} \cong 
\oplus_{i=1}^r L_i$.
It also follows that $[L_i:\Q]$ is divisible by both $[K^{(2)}:\Q]$ and 
$[K^{(1)}:\Q]$.

Step 2: 
Remember that one can find a non-canonical isomorphism 
$i_1: K^{(1)} \cong V_1$ and $i_2: K^{(2)} \cong V_2$ as vector spaces over $\Q$. 
Then an isomorphism 
\begin{align}
  i_1 \otimes i_2: (K^{(1)} \otimes_\Q K^{(2)}) \cong V_1 \otimes_\Q V_2
\end{align}
combined with (\ref{eq:K1-K2-Chinese}) introduces a decomposition 
of the vector space 
\begin{align}
  V_1 \otimes_\Q V_2 \cong \oplus_{i=1}^r W_i.
  \label{eq:V1-V2-Chinese}
\end{align}
Individual $W_i$'s in $V_1 \otimes V_2$ are vector subspaces over $\Q$;
the number field $L_i$ acts on $W_i$, 
$[L_i:\Q] = \dim_\Q W_i$, and each one of the simultaneous 
eigenstates $v_a \in W_i \otimes_\Q (K^{(1)}K^{(2)})^{\rm nc}$ of the action 
of $L_i$ is in a definite Hodge $(p,q)$ component 
(section \ref{sssec:FUP}), so all the elements in $L_i$ are 
in ${\rm End}_{\rm Hdg}(W_i)$. So, the decomposition (\ref{eq:V1-V2-Chinese}) 
over $\Q$ is compatible with the rational Hodge substructure, and 
each $W_i$ has a rational Hodge structure of CM type. 

Step 3: 
Independently from the decomposition (\ref{eq:V1-V2-Chinese}) 
of $V_1 \otimes V_2$ that follows from the structure (\ref{eq:K1-K2-Chinese}),
one may also think of a simple component decomposition of a not necessarily 
CM-type rational Hodge structure of $V_\Q$:
\begin{align}
  V_\Q  \cong \oplus_{a \in A} V_a . 
  \label{eq:Hdg-smpl-rat-dcmp}
\end{align}
Combining this structure (\ref{eq:Hdg-smpl-rat-dcmp}) and 
the structure theorem of semi-simple algebras, one can state---as we 
do in the following---the structure of the entire algebra 
${\rm End}_{\rm Hdg}(V_\Q)$; as a reminder, 
$K^{(1)}\otimes_\Q K^{(2)}$ is a part of ${\rm End}_{\rm Hdg}(V_\Q)$. 

For any pair of simple components $V_a$ and $V_b$ 
in (\ref{eq:Hdg-smpl-rat-dcmp}), any $\phi \in {\rm Hom}_{\rm Hdg}(V_a, V_b)$ is 
either a zero map or an invertible Hodge morphism.\footnote{
If $\phi$ is 
not surjective, then $V_b$ has a rational Hodge substructure, which 
contradicts against the assumption that the rational Hodge structure 
on $V_b$ is simple. If $\phi$ has a non-trivial kernel, that implies 
that $V_a$ has a rational Hodge substructure, which is a contradiction 
once again. So, $\phi$ must be an isomorphism between the vector spaces 
$V_a$ and $V_b$ over $\Q$.
} % 
One can think of grouping the simple components 
$\{ V_a \; | \; a \in A\}$ into Hodge-isomorphism classes based on whether 
the set ${\rm Hom}_{\rm Hdg}(V_a, V_b)$ is empty or non-empty (i.e., a Hodge 
isomorphism exists). The set of Hodge isomorphism classes of the simple 
components in $V_\Q$ is denoted by ${\cal A}$, and one can think of the 
decomposition 
\begin{align}
  V_\Q \cong \oplus_{\alpha \in {\cal A}} \left( \oplus_{a \in A; [a]=\alpha}
    V_a \right) =: \oplus_{\alpha \in {\cal A}} V_\alpha .
  \label{eq:Hdg-isom-smpl-dcmp}
\end{align}
The algebra of Hodge endomorphisms of $V_\Q$ has the structure 
\begin{align}
  {\rm End}_{\rm Hdg}(V_\Q) \cong \oplus_{\alpha \in {\cal A}} 
    M_{n_\alpha \times n_\alpha}(D_\alpha), \qquad 
  D_{\alpha=[a]} = {\rm End}_{\rm Hdg}(V_a), 
  \label{eq:smpl-subalg-dcmp}
\end{align}
where $n_\alpha$ is the number of simple components ($V_a$'s) that fall into 
a given Hodge-isomorphism class $\alpha$, $[a] = \alpha$. $D_\alpha$ is a division 
algebra over $\Q$ (because all the non-zero element is invertible). Therefore, 
${\rm End}_{\rm Hdg}(V_\Q)$ is a semi-simple algebra over $\Q$, and the factor 
$M_{n_\alpha \times n_\alpha}(D_\alpha)$ for a Hodge-isomorphism class $\alpha \in {\cal A}$ 
is a simple algebra.\footnote{
It is a simple algebra in the sense that it does not 
have a non-trivial two-sided ideal.
} % 

Now, we can invoke a few known facts about semi-simple algebras. 
One is that $V_a$ is regarded as an irreducible module of 
$D_{\alpha = [a]}$, and further 
\begin{align}
   \dim_\Q V_a = \dim_\Q D_\alpha.
  \label{eq:temp-D}
\end{align}
As another fact (\cite[Thm. II.4.11]{TsushimaNagao} or 
\cite[Cor.2.2.3]{ssAlg}), 
\begin{align}
  \dim_\Q D_{\alpha}  = q_\alpha^2 [K_\alpha : \Q]
   \label{eq:temp-E}
\end{align}
for some $q_\alpha \in \mathbb{N}_{>0}$, where $K_\alpha$ is the centre of 
the division algebra $D_\alpha$. 

Step 4: The general structure of ${\rm End}_{\rm Hdg}(V_\Q)$ in Step 3
is for a general rational Hodge structure not necessarily of CM type, 
whereas the CM-type nature of the Hodge structure on $V_1 \otimes V_2$ 
has been exploited in Steps 1 and 2.  
Let us see in the following (by following \cite[\S5]{ST}) 
how the decomposition (\ref{eq:V1-V2-Chinese}) is related 
to (\ref{eq:Hdg-isom-smpl-dcmp}) in Step 3, and how 
$K^{(1)}\otimes K^{(2)}$ with the structure (\ref{eq:K1-K2-Chinese}) 
fits into the general structure (\ref{eq:smpl-subalg-dcmp}) 
of ${\rm End}_{\rm Hdg}(V_\Q)$ in Step 3, when $V_\Q = V_1 \otimes_\Q V_2$.  

First observation is that one $\alpha \in {\cal A}$ is assigned to 
each label $i \in \{ 1,\cdots, r\}$ in the decomposition 
(\ref{eq:K1-K2-Chinese}, \ref{eq:V1-V2-Chinese}); the corresponding $\alpha$ 
is denoted by $\alpha(i)$. To see this correspondence, think of 
\begin{align}
L_i \hookrightarrow (K^{(1)}\otimes K^{(2)}) \hookrightarrow 
{\rm End}_{\rm Hdg}(V_\Q) \rightarrow M_{n_\alpha \times n_\alpha}(D_\alpha) 
\end{align}
for a given $i \in \{ 1,\cdots, r\}$ and an arbitrary $\alpha \in {\cal A}$.
The image of $L_i$ must be non-trivial at least for one $\alpha \in {\cal A}$;  
now we wish to see that that is the case for only one Hodge isomorphism 
class $\alpha$ in ${\cal A}$. 
 
Suppose that the image of $L_i$ is non-zero for $\alpha_0 \in {\cal A}$. 
Then the vector space $V_{\alpha_0}$ contains a vector subspace isomorphic 
to $L_i$,
and the algebra $L_i \hookrightarrow M_{n_{\alpha_0}\times n_{\alpha_0}}(D_{\alpha_0})$
is represented on this copy of the vector space $L_i$ as a full set of 
$\Phi^{\rm full}_{L_i}$. 
If there were distinct $\alpha_0, \alpha'_0 \in {\cal A}$ where 
$L_i$ is embedded non-trivially, then the set of representations 
$\Phi^{\rm full}_{L_i}$ would appear more than once in 
$V_{\alpha_0} \oplus V_{\alpha'_0} \subset V_\Q = (V_1 \otimes V_2)$; 
that contradicts against 
the fact that all the representations in $\Phi^{\rm full}_{K^{(1)}} \times 
\Phi^{\rm full}_{K^{(2)}}$ appear just once on $V_\Q$. We have thus 
established a claim that there is just one $\alpha \in {\cal A}$ 
where the image of $L_i$ in $M_{n_\alpha \times n_\alpha}(D_\alpha)$ is non-trivial. 

Second, we will see how $L_i$ fits into the algebra 
$M_{n_\alpha \times n_\alpha}(D_\alpha)$ with $\alpha = \alpha(i)$ by exploiting 
the CM nature of the Hodge structure on $V_\alpha$. 
The following argument (built on Step 3) is almost\footnote{
The original version \cite[\S5]{ST} is for $V_\Q = H^1(A;\Q)$ for an 
abelian variety $A$. cf \cite{K3-EndHdg, huybrechts2016lectures} 
for $V_\Q = T_X$ of a K3 surface $X$.
} %
 a copy of the logic of \S 5 of \cite{ST}. 

For a given $\alpha \in {\cal A}$, now consider a set of the label $i$'s 
in $\{1,\cdots, r\}$ with $\alpha(i) = \alpha$. Due to the CM nature, 
the relation 
\begin{align}
  \sum_{i \; {\rm s.t.}}^{(\alpha(i)=\alpha)} [L_i: \Q] = \dim_\Q V_\alpha
    \label{eq:temp-A}
\end{align}
holds for individual $\alpha$'s in ${\cal A}$. Furthermore, 
general arguments in Step 3---(\ref{eq:temp-D}) and (\ref{eq:temp-E})---
implies that 
\begin{align}
  \dim_\Q V_\alpha = n_\alpha \dim_\Q V_{a \; ([a]=\alpha)}
  = n_\alpha q_\alpha^2 [K_\alpha: \Q].
   \label{eq:temp-B}
\end{align}
On the other hand, the algebra 
\begin{align}
  L' = \left( \oplus_{i \; {\rm s.t.}}^{(\alpha(i)=\alpha)} L_i \right) \cdot 
    \left( K_\alpha {\bf 1}_{n_\alpha \times n_\alpha}\right) \subset 
   M_{n_\alpha \times n_\alpha}(D_\alpha)
\end{align}
remains to be a commutative subalgebra, and any commutative subalgebra 
of a central simple algebra $M_{n_\alpha \times n_\alpha}(D_\alpha)$ is bounded 
in its dimension by 
\begin{align} 
  \sum_{i \; {\rm s.t.}}^{(\alpha(i)=\alpha)} [L_i:\Q] \leq \dim_\Q L'  \leq 
       n_\alpha \times q_\alpha \times [K_\alpha:\Q]. 
   \label{eq:temp-C}
\end{align}
So, by combining (\ref{eq:temp-A}, \ref{eq:temp-B}) against (\ref{eq:temp-C}), 
we can see that $q_\alpha =1$ (which means that $D_\alpha = K_\alpha$), and also 
that $K_\alpha {\bf 1}_{n_\alpha \times n_\alpha}$ is contained in 
$\oplus_{i; \alpha(i)=\alpha} L_i$. The latter statement 
further indicates\footnote{
This is because 
$K_\alpha {\bf 1} \ni 1 \cdot {\bf 1} = \sum_{i} \epsilon_i \in \oplus_{i}L_i$; 
$K_\alpha {\bf 1} \cdot (0,\cdots, \epsilon_i, \cdots , 0) \subset L_i \subset 
\oplus_j L_j$ is a subfield of $L_i$ and is isomorphic to $K_\alpha$.
} %  
that those $L_i$'s can be regarded as an extension of $K_{\alpha(i)}$.
For the field $L_i$ to be a non-trivial extension of $K_{\alpha(i)}$, 
at least some of the endomorphisms in $L_i \subset 
M_{n_\alpha \times n_\alpha}(K_\alpha)$ 
must mix multiple different simple components $V_{a}$ with $[a]=\alpha$. 

To summarize, 
\begin{align}
   V_\alpha \cong \oplus_{i \; {\rm s.t.}}^{(\alpha(i)=\alpha)} W_i. 
  \label{eq:temp-temp}
\end{align}
$K_{\alpha}$ is the endomorphism field of the CM-type simple rational Hodge 
structure of $V_a$ (such that $[a]=\alpha$), $L_i$ is an extension of 
$K_{\alpha(i)}$, and $\oplus_{i \; {\rm s.t.}}^{(\alpha(i) = \alpha)} L_i 
\hookrightarrow M_{n_\alpha \times n_\alpha}(K_{\alpha}) = {\rm End}_{\rm Hdg}(V_\alpha)$.   

(Remark) 
% \label{rmk:embd-vect-corresp}
The vector space $W_i \otimes \overline{\Q}$ has a well-motivated basis;
basis elements are in one-to-one with the embeddings  
$\rho^{(1)} \otimes \rho^{(2)}$ in $\Phi_{L_i}^{\rm full}$. This is just 
a special case of section \ref{sssec:FUP} with $F=L_i$ 
and $V_\Q=W_i$. Each one of the basis elements are also associated 
with a particular Hodge (p,q) type, so each embedding 
$\rho^{(1)}\otimes \rho^{(2)}$ of $L_i$ has its corresponding Hodge 
type (p, q).
This correspondence will be exploited in the following analysis.

%%%%%%%%%%%%%%%%%%%%%%%%%%%%%%%%%%%%%%%%%%%%%%%%%%%%%%%%%%%%%
\subsubsection{$DW=0$ Flux and $DW=W=0$ Flux, Assuming $T_X = T_0$}
\label{sssec:flux-tx-is-t0}
%%%%%%%%%%%%%%%%%%%%%%%%%%%%%%%%%%%%%%%%%%%%%%%%%%%%%%%%%%%%%

Toward the end of section \ref{ssec:cond-flux-MorF}, we used the 
language of the simple Hodge component decomposition to write down the 
conditions for the presence of a non-trivial supersymmetric flux. 
Whether a non-trivial flux with those conditions exists or not can 
be studied for individual simple rational Hodge components.  
For simple Hodge components that are mutually Hodge-isomorphic, say, 
$\phi: V_a \cong V_{b}$, $[a]=[b] = \alpha \in {\cal A}$,
\begin{align}
  h^{p,q}[V_a] = h^{p,q}[V_b]
\end{align}
holds for all (p,q). We can thus talk of the level of individual 
Hodge-isomorphism classes, $\alpha \in {\cal A}$, and we can also study 
whether fluxes with $DW=0$ and/or $W=0$ exists for individual 
Hodge-isomorphism classes. 
\begin{itemize}
\item In a $V_\alpha$ of level 0, there are only Hodge (2,2) components 
(by definition). Any rational flux here satisfies both of the 
$DW=0$ and $W=0$ conditions. 
\item In a $V_\alpha$ of level 2, any non-zero rational flux breaks the 
$DW=0$ condition (although $W=0$ would be satisfied).
\item There is just one level-4 simple rational Hodge component 
of $H^4(Y;\Q)$ of a Calabi--Yau fourfold $Y$, so this simple component alone 
forms one Hodge-isomorphism class of the simple components in $H^4(Y;\Q)$.
This simple component admits a rational flux with $DW=0$ if and 
only if $h^{3,1}=0$ holds in this simple component. Let us say that 
a simple component is \dfn{$(3,1)$-free} if the component 
has $h^{3,1}=0$. Even when this condition is satisfied, 
such a flux does not satisfy the $W=0$ condition.
\end{itemize}

Let us continue to focus on a Borcea--Voisin orbifold $Y = (X^{(1)} \times X^{(2)})/\Z_2$ of a pair of CM-type K3 surfaces with $T^{(1)}_X=T^{(1)}_0$ and $T^{(2)}_X=T^{(2)}_0$. We have seen in section \ref{sssec:math-preparation-tensor} that the Hodge structure on $V_1 \otimes V_2$, where $V_1 \cong T^{(1)}_X \otimes \Q$ and $V_2 \cong T^{(2)}_X \otimes \Q$, has the decomposition (\ref{eq:V1-V2-Chinese}), which is compatible with the Hodge-isomorphism-class decomposition (\ref{eq:Hdg-isom-smpl-dcmp}), although (\ref{eq:V1-V2-Chinese}) may be a finer classification\footnote{
because of (\ref{eq:temp-temp}) } %
 than (\ref{eq:Hdg-isom-smpl-dcmp}). Therefore, we can rephrase the criteria for the existence of non-trivial supersymmetric fluxes, which is stated above, by simply replacing Hodge-isomorphism classes of simple components by individual components $W_i$ in (\ref{eq:V1-V2-Chinese}).

In the decomposition \eref{eq:V1-V2-Chinese} of the Hodge structure on $V_1 \otimes_\Q V_2$, the individual components $W_i$ are either level-4, level-2, or level-0. We will see, first, that there are at most only two $W_i$'s that are not level-2 (so, a $DW=0$ flux is possible only in those at most two $W_i$'s); this is the Step 1 below.  In Step 2, we work out the conditions on the CM fields 
$K^{(1)}$ and $K^{(2)}$ for those one or two component(s) to be (3,1) free, 
so that a $DW=0$ flux is indeed available. A physics recap (Step 3) comes 
at the end of this section \ref{sssec:flux-tx-is-t0}. 

Step 1: To show that there are at most two $W_i$'s, let us introduce some notations. We denote the extension degrees of $K^{(1)}$ and $K^{(2)}$ over $\Q$ by $n_1:=[K^{(1)}:\Q]$ and $n_2:=[K^{(2)}:\Q]$, respectively. The embeddings of $K^{(i)}$ with $i=1,2$ are denoted by $\Hom(K^{(i)},\Qbar)=\B{\rho^{(i)}_{(20)},\rho^{(i)}_{(02)},\rho^{(i)}_3,\dots,\rho^{(i)}_{n_i}}$, where $\rho^{(i)}_{(20)}$ and $\rho^{(i)}_{(02)}$ correspond to the $(2,0)$ component and $(0,2)$ component of $H^2(X^{(i)})$, respectively, in the sense of a remark at the end of section \ref{sssec:math-preparation-tensor}; 
    % Remark \ref{rmk:embd-vect-corresp}; 
the action of $x\in K^{(i)}$ on the $(2,0)$-form $\Omega^{(i)}_X$ of $X^{(i)}$ is $x:\Omega^{(i)}_X\mapsto \rho^{(i)}_{(20)}(x) \cdot \Omega^{(i)}_X$ for any $x\in K^{(i)}$.
Let us denote by $L_{(20|20)}$ the number field $L_i$ for which $\Phi^\mathrm{full}_{L_i}$ contains $\rho^{(1)}_{(20)}\otimes\rho^{(2)}_{(20)}$, and by $L_{(20|02)}$ the 
number field $L_j$ for which $\Phi^\mathrm{full}_{L_j}$ contains $\rho^{(1)}_{(20)}\otimes\rho^{(2)}_{(02)}$. For those $i$ and $j$, the vector spaces $W_i$ and $W_j$ 
are denoted by $W_{(20|20)}$ and $W_{(20|02)}$, respectively. 
Note that both $i\neq j$ and $i=j$ are possible. 
We claim that these (at most) two components have a chance to be different 
from level-2, and that all other $W_k$'s in (\ref{eq:V1-V2-Chinese}) 
are level-2. 

Obviously, $W_{(20|20)}$ is always the unique level-4 component.
$\Phi^{\rm full}_{L_{(20|20)}}$ contains $\rho^{(1)}_{(20)}\otimes\rho^{(2)}_{(20)}$. 

To see that all other $W_k$'s except $W_{(20|02)}$ are level-2, note first 
that every Hodge $(3,1)$ component in $(V_1\otimes V_2)\otimes\C$ corresponds 
to an embedding of the form 
\begin{align}
    \rho^{(1)}_{(20)}\otimes\rho^{(2)}_b&\text{ with }3\leq b \leq n_2 \qquad 
     {\rm or} \label{eq:31-embed-1}\\
    \rho^{(1)}_a\otimes\rho^{(2)}_{(20)}&\text{ with }3\leq a \leq n_1,\label{eq:31-embed-2} 
\end{align}
because a $(3,1)$-form in $(V_1\otimes V_2)\otimes \C$ is always a product of a $(2,0)$-form in $V_1\otimes\C$ and a $(1,1)$-form in  $V_2\otimes \C$, or vice versa.
On the other hand, each set of embeddings $\Phi_{L_k}^{\rm full}$ contains 
at least one element of the form $\rho^{(1)}_{(20)}\otimes\rho^{(2)}_\beta$ 
for some $\beta$ in $\{(20),(02),3,\dots,n_2\}$, because 
$\Phi^{\rm full}_{L_k}$ forms an orbit under the Galois group action. 
So, $\Phi^\mathrm{full}_{L_k}$ for $k \neq (20|20), (20|02)$ contains 
$\rho^{(1)}_{(20)}\otimes\rho^{(2)}_\beta$ with $\beta \in \{3, \cdots, n_2\}$, 
and the corresponding $W_k$ is of level 2. 
We conclude that a $DW = 0$ flux is not impossible only within 
$W_{(20|20)}$ and $W_{(20|02)}$. 

Step 2: 
Now let us work out the conditions for non-trivial fluxes to exist 
in $W_{(20|20)}$ and $W_{(20|02)}$ in terms of the CM fields $K^{(1)}$, $K^{(2)}$, 
and their actions on $T_X^{(1)}$ and $T_X^{(2)}$.  
The analysis will take several pages, but the conclusion can be summarized quite simply; a non-trivial flux with $DW=0$ exists if and only if 
equation \eref{eq:flux-cond-p1} is satisfied. A stronger condition 
(\ref{eq:flux-cond-math2}) is necessary and sufficient 
for a non-trivial $DW=W=0$ flux. 

We first study the level-4 component $W_{(20|20)}$. Recall that a non-trivial flux in a level-4 component preserves the $DW=0$ condition if and only if the component is $(3,1)$-free, i.e. free of Hodge $(3,1)$ components.\footnote{Note that any non-trivial flux in a level-4 component violates the $W=0$ condition.} We are thus interested in when the component is $(3,1)$-free.
Since we know all the elements in $\Phi^{\rm full}_{K^{(1)}}\times \Phi^{\rm full}_{K^{(2)}}$ that correspond to Hodge $(3,1)$ components, \eref{eq:31-embed-1} and \eref{eq:31-embed-2}, our task reduces to finding out whether or not $\Phi^\mathrm{full}_{L_{(20|20)}}$ contains such embeddings. This is equivalent to working out whether or not there exists an action of $\Gal((K^{(1)}K^{(2)})^{\rm nc}/\Q)$ that maps $\rho^{(1)}_{(20)}\otimes\rho^{(2)}_{(20)}$ to one of such embeddings that correspond to $(3,1)$ components, since $\Phi^\mathrm{full}_{L_{(20|20)}}$ is generated by $\Gal((K^{(1)}K^{(2)})^{\rm nc}/\Q)$ acting
on $\rho^{(1)}_{(20)}\otimes\rho^{(2)}_{(20)}$ 
(see p. \pageref{pg:Gal-act-on-PhiPhi}). Such a map must be contained in $G^{(1)}_{(20)}:=\Gal((K^{(1)}K^{(2)})^{\rm nc}/\rho^{(1)}_{(20)}(K^{(1)}))$ or $G^{(2)}_{(20)}:=\Gal((K^{(1)}K^{(2)})^{\rm nc}/\rho^{(2)}_{(20)}(K^{(2)}))$, since either $\rho^{(1)}_{(20)}$ or $\rho^{(2)}_{(20)}$ must be held fixed by the map. Thus the component $W_{(20|20)}$ is $(3,1)$-free, if and only if the following two conditions are satisfied simultaneously:
\begin{enumerate}
    \item[(i)] 
    There is no element $\sigma^{(1)}\in G^{(1)}_{(20)}$ such that $\sigma^{(1)}\circ\p{\rho^{(1)}_{(20)}\otimes\rho^{(2)}_{(20)}}=\rho^{(1)}_{(20)}\otimes\rho^{(2)}_b$, for any $3\leq b\leq n_2$.
    \item[(ii)] 
     There is no element $\sigma^{(2)}\in G^{(2)}_{(20)}$ such that $\sigma^{(2)}\circ\p{\rho^{(1)}_{(20)}\otimes\rho^{(2)}_{(20)}}=\rho^{(1)}_a\otimes\rho^{(2)}_{(20)}$, for any $3 \leq a\leq n_1$.
\end{enumerate}

Let us work out in turn when each one of these conditions is satisfied.
We first focus on the condition (i). We define $N_1$ to be the extension degree\footnote{Strictly speaking, $L_{(20|20)}$ is defined to be an abstract extension field of $K^{(2)}$, $L_{(20|20)}=K^{(2)}[x]/g(x)$ for some $g\in K^{(2)}[x]$ and the endomorphism field $K^{(1)}$ is not a subfield of it, so the extension degree $[L_{(20|20)}:K^{(1)}]$ does not make sense. However, since we know that $L_{(20|20)}$ is isomorphic to $K^{(1)}[x]/h(x)$ with some $h\in K^{(1)}[x]$, we abuse the notation and define $[L_{(20|20)}:K^{(1)}]:=[\varphi(L_{(20|20)}):K^{(1)}]$ with an isomorphism $\varphi:K^{(2)}[x]/g(x)\to K^{(1)}[x]/h(x)$.} $N_1:=[L_{(20|20)}:K^{(1)}]$.
There are $N_1-1$ non-trivial actions in $G^{(1)}_{(20)}$, which map $\rho^{(1)}_{(20)}\otimes\rho^{(2)}_{(20)}$ to $\rho^{(1)}_{(20)}\otimes\rho^{(2)}_\beta$ for some $\beta=(02),3,\dots,n_2$.
Except for the one element that maps to $\beta=(02)$, which may or may not exist, each one of them will violate the condition (i).
We thus immediately conclude that the condition (i) is violated whenever $N_1>2$.

There are two ways to satisfy the condition (i) when $N_1\leq 2$, 
i.e. $W_{(20|20)}$ does not contain Hodge $(3,1)$ components 
of the form \eref{eq:31-embed-1}:
\begin{itemize}
    \item[(i-1)] When $N_1=1$, the condition (i) is always satisfied. This is because there are no non-trivial action in $G^{(1)}_{(20)}$.
    Note that this also means that $W_{(20|20)}\neq W_{(20|02)}$ in this case, because $\Phi^\mathrm{full}_{L_{(20|20)}}$ does not contain $\rho^{(1)}_{(20)}\otimes\rho^{(2)}_{(02)}$.
    \item[(i-2)] When $N_1=2$ and $\Phi^\mathrm{full}_{L_{(20|20)}}$ contains $\rho^{(1)}_{(20)}\otimes\rho^{(2)}_{(02)}$, then the condition (i) is satisfied. This is because the only non-trivial action of $G^{(1)}_{(20)}$ maps $\rho^{(1)}_{(20)}\otimes\rho^{(2)}_{(20)}$ to $\rho^{(1)}_{(20)}\otimes\rho^{(2)}_{(02)}$. At the same time this means that $W_{(20|20)}=W_{(20|02)}$.
\end{itemize}
Note that, even when $N_1=2$, if $\Phi^\mathrm{full}_{L_{(20|20)}}$ does not contain $\rho^{(1)}_{(20)}\otimes\rho^{(2)}_{(02)}$, the condition (i) is violated; the only non-trivial element in $G^{(1)}_{(20)}$ will map $\rho^{(1)}_{(20)}\otimes\rho^{(2)}_{(20)}$ to an embedding of the form \eref{eq:31-embed-1}.

Similarly, defining $N_2:=[L_{(20|20)}:K^{(2)}]$, one can argue that there are only two ways to satisfy the condition (ii), i.e. $W_{(20|20)}$ does not contain Hodge $(3,1)$ components of the form \eref{eq:31-embed-2}:
\begin{itemize}
    \item[(ii-1)] When $N_2=1$, the condition (ii) is satisfied, since there are no non-trivial action in $G^{(2)}_{(20)}$.
    This means that $W_{(20|20)}\neq W_{(20|02)}$ in this case, because $\Phi^\mathrm{full}_{L_{(20|20)}}$ does not contain $\rho^{(1)}_{(02)}\otimes\rho^{(2)}_{(20)}$, which is the complex conjugate of $\rho^{(1)}_{(20)}\otimes\rho^{(2)}_{(02)}$ and must be contained in $\Phi^\mathrm{full}_{L_{(20|02)}}$.
    
    \item[(ii-2)] When $N_2=2$ and $\Phi^\mathrm{full}_{L_{(20|20)}}$ contains $\rho^{(1)}_{(02)}\otimes\rho^{(2)}_{(20)}$, then the condition (ii) is again satisfied. At the same time this means that $W_{(20|20)}=W_{(20|02)}$.
\end{itemize}

Now we are ready to see when the conditions (i) and (ii) are simultaneously satisfied. In order to satisfy both (i) and (ii), there seems to be four choices for not having a Hodge $(3,1)$ component in $W_{(20|20)}$, i.e. two choices for the condition (i) and another pair of choices for the condition (ii). However, two of them, (i-1)-(ii-2) and (i-2)-(ii-1) cannot happen; (i-1) or (ii-1) imply $W_{(20|20)}\neq W_{(20|02)}$, whereas (i-2) or (ii-2) imply $W_{(20|20)}=W_{(20|02)}$, thus contradiction.

In summary, there are two cases, (i-1)-(ii-1) and (i-2)-(ii-2), where the level-4 component $W_{(20|20)}$ is $(3,1)$-free. For these two cases, let us leave the list of embeddings in $\Phi^\mathrm{full}_{L_{(20|20)}}$ for clarification, and also rephrase these conditions on the $\Gal((K^{(1)}K^{(2)})^\mathrm{nc}/\Q)$ action in terms of $K^{(1)}$, $K^{(2)}$ and their actions.
\begin{itemize}
\item 
Firstly, let us choose (i-2) and (ii-2), i.e. $[L_{(20|20)} :K^{(1)}] = [L_{(20|20)}: K^{(2)}] = 2$ (so, $n_1 = n_2=:n$) and
$W_{(20|20)} = W_{(20|02)}$, to satisfy the conditions (i) and (ii). The contents of $\Phi^\mathrm{full}_{L_{(20|20)}}$ are
\begin{align}
 \Phi^\mathrm{full}_{L_{(20|20)}} & \; = \left\{ 
     \rho^{(1)}_{(20)} \otimes \rho^{(2)}_{(20)}, \; \rho^{(1)}_{(20)}\otimes \rho^{(2)}_{(02)}, \;
     \rho^{(1)}_{(02)} \otimes \rho^{(2)}_{(20)}, \; \rho^{(1)}_{(02)}\otimes \rho^{(2)}_{(02)}
       \right\} \nonumber \\ 
   & \; \qquad \cup  \left\{ \rho^{(1)}_a \otimes \rho^{(2)}_b \; | \; 
      3 \leq a \leq n, \; 3 \leq b \leq n, 
      \; a,b {\rm ~appearing~twice}\right\}.
\end{align}
There are $4 + 2\times (n-2) = 2n$ embeddings of $L_{(20|20)}=L_{(20|02)}$; 
$2n-2$ embeddings among them correspond to Hodge (2,2) components, and the other two are the (4,0) and (0,4) Hodge components; indeed there are no Hodge (3,1) or (1,3) components. This case turns out to happen if and only if 
\begin{equation}
\rho^{(1)}_{(20)}(K^{(1)}_0)=\rho^{(2)}_{(20)}(K^{(2)}_0)\subset\Qbar \quad \text{and}\quad \rho^{(1)}_{(20)}(K^{(1)})\neq\rho^{(2)}_{(20)}(K^{(2)}),
\end{equation}
where $K^{(1)}_0$ and $K^{(2)}_0$ are the maximal totally real subfields of $K^{(1)}$ and $K^{(2)}$, respectively.\footnote{
Let us formally state the claim and prove it here. The claim is that if and only if $[L_{(20|20)}:K^{(2)}]=2$ and $\Phi^\mathrm{full}_{L_{20|20)}}$ contains both $\rho^{(1)}_{(20)}\otimes\rho^{(2)}_{(20)}$ and  $\rho^{(1)}_{(02)}\otimes\rho^{(2)}_{(20)}$, then $\rho^{(1)}_{(20)}(K^{(1)}_0)=\rho^{(2)}_{(20)}(K^{(2)}_0)$ and $\rho^{(1)}_{(20)}(K^{(1)})\neq\rho^{(2)}_{(20)}(K^{(2)})$.

Recall that
$
    L_{(20,20)}=K^{(2)}[x]/g(x),
$
where $g\in K^{(2)}[x]$ and is a degree-2 polynomial. The two roots of $\rho^{(2)}_{(20)}(g(x))$, $\alpha_+$ and $\alpha_-$, must correspond to a simple generator $\alpha$ of $K^{(1)}$, i.e. $K^{(1)}=\Q(\alpha)$, such that $\alpha_+=\rho^{(1)}_{(20)}(\alpha)$ and $\alpha_-=\rho^{(1)}_{(02)}(\alpha)$.
This means that $\alpha_+$ and $\alpha_-$ are complex conjugate to each other. Let us explicitly define $g(x)=x^2+a_1 x+a_0$ with $a_1,a_0\in K^{(2)}$. Then from the explicit form of the roots, one can conclude that $a_0,a_1\in\R$.
This implies that $\Q(\alpha_+)=\rho^{(1)}_{(20)}(K^{(1)})$ is a degree-2 extension of a totally real field $\rho^{(2)}_{(20)}(K^{(2)}_0)$, which must equal to $\rho^{(1)}_{(20)}(K^{(1)}_0)$. Note that $\rho^{(2)}_{(20)}(K^{(2)})\neq \rho^{(2)}_{(20)}(K^{(1)})$ because $\alpha_+\not\in\rho^{(2)}_{(20)}(K^{(2)})$.

Conversely, let us assume $\rho^{(2)}_{(20)}(K^{(2)}_0)=\rho^{(1)}_{(20)}(K^{(1)}_0)$ and $\rho^{(2)}_{(20)}(K^{(2)})\neq \rho^{(2)}_{(20)}(K^{(1)})$. Denoting the totally real field $\rho^{(2)}_{(20)}(K^{(2)}_0)=\rho^{(1)}_{(20)}(K^{(1)}_0)$ by $K_0$, the composite field $\rho^{(1)}_{(20)}(K^{(1)})\rho^{(2)}_{(20)}(K^{(2)})$ can be rewritten as $\rho^{(1)}_{(20)}(K^{(1)})\rho^{(2)}_{(20)}(K^{(2)})=K_0(\eta^{(1)},\eta^{(2)})$ for some $\eta^{(1)}\in \rho^{(1)}_{(20)}(K^{(1)})$ and $\eta^{(2)}\in \rho^{(2)}_{(20)}(K^{(2)})$. The Galois action that maps $\eta^{(1)}$ to its complex conjugate and leaves everything else will map $\rho^{(1)}_{(20)}\otimes\rho^{(2)}_{(20)}$ to $\rho^{(1)}_{(02)}\otimes\rho^{(2)}_{(20)}$, so the latter is also contained in $\Phi^\mathrm{full}_{L_{(20|20)}}$. One can also see that $[\rho^{(1)}_{(20)}(K^{(1)})\rho^{(2)}_{(20)}(K^{(2)}):\rho^{(2)}_{(20)}(K^{(2)})]=2$, so $[L_{(20|20)}:K^{(2)}]=2$.
} % 
\item 
Alternatively, we can choose (i-1) and (ii-1) to satisfy the conditions (i) and (ii). In this case,
$[L_{(20|20)}: K^{(1)}] = [L_{(20|20)}: K^{(2)}] = 1$ and $W_{(20|20)} \neq W_{(20|02)}$. This happens if and only if
\begin{align}
  \rho^{(1)}_{(20)}(K^{(1)}) = \rho^{(2)}_{(20)}(K^{(2)}) \subset \overline{\Q}.
\end{align} 
The contents of $\Phi^\mathrm{full}_{L_{(20|20)}}$ are
\begin{align}
 \Phi^\mathrm{full}_{L_{(20|20)}} & \; = \left\{ 
     \rho^{(1)}_{(20)} \otimes \rho^{(2)}_{(20)}, \; \rho^{(1)}_{(02)}\otimes \rho^{(2)}_{(02)}
       \right\} \nonumber \\ 
   & \; \qquad \cup  \left\{ \rho^{(1)}_a \otimes \rho^{(2)}_b \; | \; 
      3 \leq a \leq n_1, \; 3 \leq b \leq n_2, 
      \; a,b {\rm ~appearing~once}\right\},
\end{align}
and there are $2+(n-2)=n$ embeddings of $L_{(20|20)}$; two of them correspond to the Hodge $(4,0)$ and $(0,4)$ components, and the rest correspond to $(2,2)$ components. There are no $(3,1)$ or $(1,3)$ components.
\end{itemize}

Let us move on to the $W_{(20|02)}$ component. This component is level-4 when $W_{(20|20)}=W_{(20|02)}$, and is level-0 or level-2 otherwise. We are interested in how $K^{(1)},K^{(2)}$ and their actions on $T_X^{(1)}$ and $T_X^{(2)}$ controls whether this component is (3,1)-free, especially whether it is level-0, or not. Almost the same analysis as above can be carried out, and there turn out to be only two cases where the component becomes $(3,1)$-free:
\begin{itemize}
\item The first case is where $[L_{(20|02)} :K^{(1)}] = [L_{(20|02)}: K^{(2)}] = 2$ and $W_{(20|20)} = W_{(20|02)}$ holds. The component is level-4, and this case has already been considered in the analysis of $W_{(20|20)}$ as the (i-2)-(ii-2) case.
\item The component is also $(3,1)$-free when $[L_{(20|02)}; K^{(1)}] = [L_{(20|02)}: K^{(2)}] = 1$ holds. This is equivalent to
\begin{equation}
  \rho^{(1)}_{(20)}(K^{(1)}) = \rho^{(2)}_{(02)}(K^{(2)}) \subset \overline{\Q}
\end{equation}
and in this case, $W_{(20|02)} \neq W_{(20|20)}$ holds, which means that the $W_{(20|02)}$ component is level-0. The contents of $ \Phi^\mathrm{full}_{L_{(20|02)}}$ are
\begin{align}
 \Phi^\mathrm{full}_{L_{(20|02)}} & \; = \left\{ 
     \rho^{(1)}_{(20)} \otimes \rho^{(2)}_{(02)}, \; \rho^{(1)}_{(02)}\otimes \rho^{(2)}_{(20)}
       \right\} \nonumber \\ 
   & \; \qquad \cup  \left\{ \rho^{(1)}_a \otimes \rho^{(2)}_b \; | \; 
      3 \leq a \leq n_1, \; 3 \leq b \leq n_2, 
      \; a,b {\rm ~appearing~once}\right\},
\end{align}
and all the embeddings are associated with Hodge $(2,2)$ components, and thus the component $W_{(20|02)}$ is indeed level-0.
Note that $\rho^{(1)}_{(20)}(K^{(1)}) = \rho^{(2)}_{(02)}(K^{(2)})$ is equivalent to $\rho^{(1)}_{(20)}(K^{(1)}) = \rho^{(2)}_{(20)}(K^{(2)})$, since $\rho^{(2)}_{(02)}(K^{(2)})$ is the complex conjugate of $\rho^{(2)}_{(20)}(K^{(2)})$, which is $\rho^{(2)}_{(20)}(K^{(2)})$ itself because it is a CM field. This means that, when $W_{(20|02)}\neq W_{(20|20)}$, $W_{(20|20)}$ is (3,1)-free if and only if $W_{(20|02)}$ is level-0.
\end{itemize}

The analysis of the Hodge structures of $W_{(20|20)}$ and $W_{(20|02)}$ in the last pages can be summarized in a very simple way:\label{pg:ABC-classifictn}
\renewcommand{\theenumi}{\Alph{enumi})}
\begin{enumerate}
    \item When\footnote{Let us remind ourselves that $K^{(i)}_0$ is defined to be the maximal totally real subfield of $K^{(i)}$ for $i=1,2$.}
    \begin{equation}
    \rho^{(1)}_{(20)}(K^{(1)}_0)=\rho^{(2)}_{(20)}(K^{(2)}_0)\subset\Qbar \quad \text{and}\quad \rho^{(1)}_{(20)}(K^{(1)})\neq\rho^{(2)}_{(20)}(K^{(2)}),\label{eq:flux-cond-math1}
    \end{equation}
    the component $W_{(20|20)}=W_{(20|02)}$ in \eref{eq:V1-V2-Chinese} is level-4 and $(3,1)$-free. All other $W_i$ components in \eref{eq:V1-V2-Chinese} are level-2.
    
    \item
    When 
\begin{equation}
  \rho^{(1)}_{(20)}(K^{(1)}) = \rho^{(2)}_{(20)}(K^{(2)}) \subset \overline{\Q},\label{eq:flux-cond-math2}
\end{equation}
then there are two components in \eref{eq:V1-V2-Chinese} that are not level-2: $W_{(20|20)}$ is level-4 and $(3,1)$-free, and $W_{(20|02)}$ is level-0. All other components are level-2.

\item
When neither \eref{eq:flux-cond-math1} nor \eref{eq:flux-cond-math2} is 
satisfied, none of the component $W_i$ in \eref{eq:V1-V2-Chinese} is (3,1)-free.
\end{enumerate}
\renewcommand{\theenumi}{\arabic{enumi}}

Before getting into the recap of the physical consequences, it is worth while to take a slightly different view on cases A) and B):
A non-trivial flux satisfying $DW=0$ condition is possible only when the condition \eref{eq:flux-cond-math1} or \eref{eq:flux-cond-math2} is satisfied. Noticing that $\rho^{(1)}_{(20)}(K^{(1)})=\rho^{(2)}_{(20)}(K^{(2)})$ implies $\rho^{(1)}_{(20)}(K^{(1)}_0)=\rho^{(2)}_{(20)}(K^{(2)}_0)$, it is clear that such flux configurations exist if and only if 
\begin{equation}
    \rho^{(1)}_{(20)}(K^{(1)}_0)=\rho^{(2)}_{(20)}(K^{(2)}_0)=:K_0
    \label{eq:flux-cond-p1}
\end{equation}
holds. Let us call the situation case A+B), since it combines the cases A) and B). Introducing some generators $\eta^{(1)},\eta^{(2)}\in\Qbar$ such that $K_0(\eta^{(1)})=\rho^{(1)}_{(20)}(K^{(1)})$ and $K_0(\eta^{(2)})=\rho^{(2)}_{(20)}(K^{(2)})$, one can see that the case B) is a non-generic situation where $K_0(\eta^{(1)})$ coincides with $K_0(\eta^{(2)})$, and the case A) is the generic situation complementary to it. 
The condition (\ref{eq:cond-DW=0-smpl-Hdge}) for a $DW=0$ flux has been 
translated into an arithmetic characterization (\ref{eq:flux-cond-p1}), and 
a stronger condition (\ref{eq:cond-DW=W=0-smpl-Hdge}) for a $DW=W=0$ 
flux into a stronger characterization (\ref{eq:flux-cond-math2}).

Step 3 (a physics recap): 
\label{pg:phys-recap}
Now let us discuss the physical consequences, although most of what follows is included in the discussion so far.
As a first physical consequence, one can see that there is no topological flux satisfying the $DW=0$ condition, if $n_1: = [K^{(1)}:\Q]$ is not equal to $n_2 := [K^{(2)}:\Q]$; as we have been assuming $T_X^{(1)} = T^{(1)}_0$ and $T_X^{(2)} = T_0^{(2)}$ in this section \ref{sssec:flux-tx-is-t0}, this condition is equivalent to ${\rm rank}(T^{(1)}_0) = {\rm rank}(T^{(2)}_0)$. Furthermore, a non-trivial supersymmetric flux exists only in either one of these: 
\begin{itemize}
    \item
    In case A), with the condition (\ref{eq:flux-cond-math1}), the component $W_{(20|20)}=W_{(20|02)}$ is level-4 and a $2\times(n=n_1=n_2)$-dimensional subspace of the $n^2$-dimensional vector space $V_1 \otimes_\Q V_2$. Any flux in this component satisfies the $DW=0$ condition but always violates the $W=0$ condition.
    
    \item
    In case B), with the condition (\ref{eq:flux-cond-math2}), $W_{(20|02)}$ is an $n$-dimensional subspace of $V_1 \otimes_\Q V_2$ and is level-0. One has $n$-dimensional degrees of freedom to turn on the flux in this component without violating $DW=0$ or $W=0$ conditions. Another $n$-dimensional subspace $W_{(20|20)}$ is also free of $(3,1)$-components, but since it contains the $(4,0)$ component by definition, turning on any flux in this component violates the $W=0$ condition.
    In summary, non-trivial flux vacua with $DW=0$ and $W=0$ is possible, if and only if \eref{eq:flux-cond-math2} is satisfied.
\end{itemize}
As a reminder, we did not study arithmetic characterization for the 
$H^1(Z_{(1)};\Q)\otimes H^1(Z_{(2)};\Q)$ component 
of (\ref{eq:H4-Hpart-inv-case}) to support a $DW=0$ flux ($W=0$ is automatic). 

The conclusion above is similar\footnote{
References \cite{AK, BHV, BKW-phys} considered compactification by $Y = X^{(1)} \times X^{(2)}$, but since they did not take an orbifold, their set-up is different from the one in this article. When it comes to the study of supersymmetric fluxes within $(T^{(1)}_X \otimes T^{(2)}_X) \otimes \Q$, however, cases in \cite{AK,BKW-phys} can be regarded as a special case of the study in this section. The scope of \cite{BHV} is not limited to ${\rm rank}(T_X^{(i)})=2$ or vacuum complex structure within ${\cal M}_{CM}^{X(T_X^{(1)})} \times {\cal M}_{CM}^{X(T_X^{(2)})}$, on the other hand. 
} %  
to, and also a generalization of the study by Aspinwall--Kallosh \cite{AK}. They chose the pair of K3 surfaces $X^{(1)}$ and $X^{(2)}$ to be attractive, that is, ${\rm rank}(T^{(1)}_X) = {\rm rank}(T^{(2)}_X) = 2$, and studied topological fluxes satisfying the $DW=0$ condition as well as ones satisfying both of the $DW=0$ and $W=0$ conditions. Note that attractive K3 surfaces are always of CM-type with endomorphism fields being imaginary quadratic fields. The condition (\ref{eq:flux-cond-p1}) follows immediately from their set-up because $K^{(1)}_0 = K^{(2)}_0 = \Q$ with $K^{(1)}=\Q(\sqrt{-d_1})$, $K^{(2)}=\Q(\sqrt{-d_2})$ in this case. The condition (\ref{eq:flux-cond-p1}) for non-trivial fluxes with $DW=0$ is regarded as a generalization of the ${\rm rank}(T^{(1)}_X) = {\rm rank}(T^{(2)}_X)=2$ setup in \cite{AK}. For fluxes with $DW=W=0$ to exist, \cite{AK} concluded that $K^{(1)}=\Q(\sqrt{-d_1})$ should be isomorphic to $K^{(2)}=\Q(\sqrt{-d_2})$; we have seen that this condition should be generalized to \eref{eq:flux-cond-math2}. See also footnote \ref{fn:compare-AK-BKW} in section \ref{ssec:Tx-neq-T0}. 

%%%%%%%%%%%%%%%%%%%%%%%%%%%%%%%%%%%%%%%%%%%%%%%%%%%%%%%
\subsubsection{Complex Structure Moduli Masses with $W=0$}
\label{sssec:mass-TX=T0}
%%%%%%%%%%%%%%%%%%%%%%%%%%%%%%%%%%%%%%%%%%%%%%%%%%%%%%%

Now that we have worked out the conditions for non-trivial supersymmetric 
flux to exist in terms of arithmetic of the endomorphisms fields $K^{(1)}$ 
and $K^{(2)}$, let us move on to see whether such fluxes generate mass
of complex structure moduli of M-theory compactification on 
$Y = (X^{(1)} \times X^{(2)})/\Z_2$. 
In this section \ref{sssec:mass-TX=T0}, we assume that the vacuum complex 
structure of the pair of K3 surfaces $X^{(1)}$ and $X^{(2)}$ 
are of CM-type, generic enough in $D(T_0^{(i)})$ so that $T_X^{(i)}=T_0^{(i)}$, 
and satisfy the condition (\ref{eq:flux-cond-math2}); low-energy 
effective field theory (including the the mass matrix)
of the fluctuation fields around the vacuum complex structure is 
studied in the following.\footnote{
We restrict our attention to the fields of complex structure  
deformation within ${\cal M}_{\rm cpx~str}^{[Y]BV}$, not to the 
full deformation in ${\cal M}_{\rm cpx~str}^{[Y]}$.
Nothing is lost when $g_{(1)}g_{(2)}=0$, because there is no complex 
structure moduli deforming away from the orbifold limit then. 
For cases $g_{(1)}g_{(2)} \neq 0$, we do not have something to add 
to what we have already written in section \ref{ssec:generic-noFlux}. 
} %

At the vacuum, a holomorphic (2,0) form $\Omega_{X^{(i)}}$ of the K3 surface
$X^{(i)}$ can be chosen to be $v^{(i)}_{(20)}$ (by choosing the normalization 
of $\Omega_{X^{(i)}}$). Over the moduli space ${\cal M}^{[X(T_0^{(i)})]}_{\rm cpx~str}$ 
around the vacuum $\vev{z_{(i)}}$, the 2-form $\Omega_{X^{(i)}}(z_{(i)})$ that 
is holomorphic and purely of Hodge (2,0)-type in the complex structure of 
$z_{(i)} \in {\cal M}^{[X(T_0^{(i)})]}_{\rm cpx~str}$ is parameterized by 
\begin{align}
  \Omega_{X^{(i)}} = v^{(i)}_{(20)} + t^{(i)}
     - \frac{(t^{(i)}, t^{(i)})_{T_X^{(i)} \otimes \C}}{2C^{(i)}}  \; v^{(i)}_{(02)}
  \label{eq:k3-cpx-str-parametrize-cm}
\end{align}
where $t^{(i)}$ collectively denotes\footnote{
See below (\ref{eq:moduli-eff-super-A}) for a component 
description of the moduli fluctuation fields. 
} %
 the local coordinates of the moduli 
space ${\cal M}^{[X(T_0^{(i)})]}_{\rm cpx~str}$ around the vacuum $\vev{z_{(i)}}$
(i.e., the moduli field fluctuations around the vacuum), and is regarded 
as an element of $[T_0^{(i)} \otimes \C]^{(1,1)}$---the (1,1) Hodge 
component with respect to $\vev{z_{(i)}}$; 
$v^{(i)}_{(20)}$ and $v^{(i)}_{(02)}$ are also fixed against $T_X^{(i)}\otimes \Q$
and provide a fixed frame\footnote{
We could use an integral basis of the lattices $T_0^{(i)}$ for a fixed frame, 
but the choice in the main text is obviously much more convenient for the 
discussion here.
} % 
with which we describe deformation of complex 
structure of $X^{(i)}$; finally, $C^{(i)} = (v^{(i)}_{(20)}, v^{(i)}_{(02)})$. 
The four-form $\Omega_Y = \Omega_{X^{(i)}} \wedge \Omega_{X^{(2)}}$ to be fed into 
the flux superpotential (\ref{eq:GVW}) is 
\begin{align}
  \Omega_Y & \; = v^{(1)}_{(20)}v^{(2)}_{(20)}
   + \left(v^{(1)}_{(20)}t^{(2)} + t^{(1)}v^{(2)}_{(20)} \right)  
    \label{eq:OmegaY-expansion-inv} \\
  &  - v^{(1)}_{(20)} v^{(2)}_{(02)} (2C^{(2)})^{-1} (t^{(2)},t^{(2)})_{T_X^{(2)}}
   - v^{(1)}_{(02)} v^{(2)}_{(20)} (2C^{(1)})^{-1} (t^{(1)},t^{(1)})_{T_X^{(1)}}
   + t^{(1)}t^{(2)}
   + {\cal O}(t^3).  \nonumber
\end{align}

Suppose that a non-trivial flux is in the $W_{(20|02)}$ component; 
the condition (\ref{eq:flux-cond-math2}) is implicit now.  
Then the contributions to $\Omega_Y$ in the first line of 
(\ref{eq:OmegaY-expansion-inv}) do not yield any terms in the effective 
superpotential, so both of the $DW=0$ and $W=0$ conditions are satisfied 
at the vacuum, as designed. 
A flux in $W_{(20|02)}$ gives rise to terms that are quadratic 
in the fluctuation of the $2(n_{1,2}-2) = 2(20-r_{(i=1,2)})$ moduli 
fields---$t^{(i=1,2)}$---that 
would deform the complex structure of $Y$ (within ${\cal M}_{\rm cpx~str}^{[Y]BV}$) 
from the CM-type vacuum complex structure $\vev{z}$.
Note also that a flux in $W_{(20|02)}$ does not generate a cubic 
or quartic terms of those moduli fields $t^{(i=1,2)}$, but just yields 
the mass terms.  

With a closer look, one finds that the mass matrix is Dirac type, and 
that the product of the mass eigenvalues is real. As a first step to 
see this, we write down the mass terms using the following notations
(so that we can keep track of Galois-conjugate relations among 
the coefficients in the effective superpotential). 
Under the condition (\ref{eq:flux-cond-math2}), we can fix one 
isomorphism from $K^{(2)}$ to $K^{(1)}$, 
\begin{align}
    (\rho_{(20)}^{(1)})^{-1} \circ \rho_{(02)}^{(2)}: K^{(2)} \rightarrow 
        \overline{\Q} \rightarrow K^{(1)}; 
\end{align}
this isomorphism can be used to set up a 1-to-1 correspondence between 
the embeddings of $K^{(1)}$ and those of $K^{(2)}$; 
\begin{align}
  \rho^{(2)}_{\beta(\alpha)} := \rho_\alpha^{(1)} \circ 
     \left( (\rho_{(20)}^{(1)})^{-1} \circ \rho_{(02)}^{(2)} \right), \qquad 
   \alpha \in \{(20), (02), 3,\cdots, n\}.
\end{align}
Then $\Phi^{\rm full}_{L_{(20|02)}} = \{ \rho^{(1)}_\alpha \otimes 
\rho^{(2)}_{\beta(\alpha)} \; | \; \alpha = (20), (02), 3,\cdots, n\}$ 
in this notation. The fact that a flux must be in the 
$\Q$-coefficient cohomology, rather than in the $\R$ or $\C$-coefficient
cohomology groups, is translated into the condition $n^{ij} \in \Q$ defined 
below, when we use $v^{(1)}_\alpha \otimes v^{(2)}_\beta$'s for a basis of 
the cohomology:  
\begin{align}
  \int_Y G \wedge (v^{(1)}_\alpha \otimes v^{(2)}_{\beta(\alpha)}) 
 =: \sum_{i,j} n^{ij} \rho_\alpha^{(1)}(y_i^{(1)}) \; \rho_{\beta(\alpha)}^{(2)}(y^{(2)}_j)
   =: G_{\alpha\beta(\alpha)}, \quad 
  \alpha \in \{ (20), (02), 3,\cdots, n\},  \nonumber 
\end{align}
where $\{ y^{(1)}_i \; | \; i=1,\cdots, n\}$ and 
$\{ y^{(2)}_j \; | \; j=1,\cdots, n\}$ are the basis of $K^{(1)}$ and 
$K^{(2)}$, respectively, over $\Q$, introduced in 
section \ref{sssec:FUP}. 
$G_{(20)(02)}$ must be an algebraic number within $\rho_{(20)}^{(1)}(K^{(1)}) 
= \rho^{(2)}_{(02)}(K^{(2)})$. Other $G_{\alpha\beta(\alpha)}$'s are Galois conjugate 
of $G_{(20)(02)}$: 
\begin{align}
   \sigma_\alpha ( G_{(20)(02)} ) = G_{\alpha\beta(\alpha)}, 
\end{align}
where $\sigma_a \in {\rm Gal}((\rho^{(1)}_{(20)}(K^{(1)}))^{\rm nc}/\Q)$ that 
brings $\rho^{(1)}_{(20)}$ to $\sigma_\alpha \cdot \rho^{(1)}_{(20)} = 
\rho^{(1)}_\alpha$ and $\rho^{(2)}_{(02)}$ to 
$\sigma_\alpha \cdot \rho^{(2)}_{(02)} = \rho^{(2)}_{\beta(\alpha)}$. 
So, the mass matrix is of the form 
\begin{align}
 W \propto  - \frac{G_{(20)(02)}}{2C^{(2)}} \; (t^{(2)},t^{(2)})_{T_X^{(2)}}
   - \frac{(G_{(20)(02)})^{\rm c.c.}}{2C^{(1)}} \; (t^{(1)},t^{(1)})_{T_X^{(1)}}
   + \sum_{a=3}^n \sigma_a(G_{(20)(02)}) t^{(1)}_a t^{(2)}_{a}, 
\label{eq:moduli-eff-super-A}
\end{align}
when we parametrize the moduli by $t^{(1)} = \sum_{a=3}^n t^{(1)}_a v^{(1)}_a$
and $t^{(2)} = \sum_{a=3}^n t^{(2)}_a v^{(2)}_{\beta(a)}$. 

The Dirac structure of the mass matrix becomes manifest only after examining 
the mass terms $\propto (t^{(2)}, t^{(2)})$ and $\propto (t^{(1)},t^{(1)})$  
that are apparently Majorana. A key observation is that 
$(v^{(i)}_{(20)},v^{(i)}_\gamma)=0$ for any $\gamma \in \{ (20), 3,\cdots, n\}$. 
Applying the Galois transformations,\footnote{
Note that the map $\Phi^{\rm full}_{K^{(i)}} \ni \rho^{(i)}_{\gamma} \mapsto 
\sigma_\alpha \cdot \rho^{(i)}_\gamma \in \Phi^{\rm full}_{K^{(i)}}$ is one-to-one map, 
and that the basis vectors $v_\gamma$ have a component 
description (\ref{eq:c-y-egvct-relation}) for a $\Q$-basis of $T_0^{(i)}$. 
} % 
 we see that 
\begin{align}
  (v_{\alpha}^{(i)}, v_{\overline{\alpha}}^{(i)}) = \sigma_{\alpha} (C^{(i)}),  \qquad 
  (v^{(i)}_\alpha, v^{(i)}_{\gamma}) = 0 \quad  {\rm for~} \gamma \neq
    \overline{\alpha}. 
  \label{eq:Gal-K3-intfrm}
\end{align}
Using this property, the moduli effective superpotential 
is written in the following form:\footnote{ 
Here is a little more set of notations.
The $n$ embeddings $\Phi_{K^{(i)}}^{\rm full}$ form $n/2$ pairs under the complex 
conjugations in $\overline{\Q}$ (and also in the CM fields $K^{(i)}$); 
$\rho^{(i)}_{\alpha'}$ is paired with 
${\rm cc} \cdot \rho^{(i)}_{\alpha'} = \rho^{(i)}_{\alpha'} \cdot {\rm conj.}$, 
which is denoted by $\rho^{(i)}_{\overline{\alpha'}}$;
the set $\Phi_{K^{(i)}}^{\rm full}$ can be grouped into two 
$\{ \rho^{(i)}_{\alpha'} \; | \; \alpha' \in \{ (20), 2,\cdots, n/2\} \}$ and 
$\{ \rho^{(i)}_{\overline{\alpha'}} \; | \; \alpha' \in \{ (20), 2,\cdots, n/2\} \}$;
a separation into two in this way is not unique.
Note also that $\overline{\beta(\alpha)} = \beta(\overline{\alpha})$. 
} % 
\begin{align}
 \sum_{a' \in }^{\{ 2,\cdots, n/2\}} (t^{(1)}_{\overline{a'}}, t^{(2)}_{a'})
   \left( \begin{array}{cc} 
     - (G_{(20)(02)})^{\rm c.c.} \frac{\sigma_{a'}(C^{(1)})}{C^{(1)}} & 
     \sigma_{\overline{a'}} (G_{(20)(02)}) \\
    \sigma_{a'}(G_{(20)(02)})  &
     - G_{(20)(02)} \frac{\sigma_{\overline{a'}}(C^{(2)})}{C^{(2)}} 
   \end{array} \right)
   \left( \begin{array}{c} t^{(1)}_{a'} \\ t^{(2)}_{\overline{a'}}
    \end{array} \right). 
\label{eq:moduli-eff-super-B}
\end{align}
This mass matrix is obviously Dirac type, and is furthermore split 
into $(n/2-1)$ blocks of $2 \times 2$ matrices. 

The product of all the mass eigenvalues is in $\R$. This is so 
even at the level of the individual $2 \times 2$ mass matrices; 
the product is the determinant of the mass matrix above, which is\footnote{
For a generic choice of a flux $G$ in $W_{(20|02)}$, this combination would 
not vanish, which means that all the $2(n-2)$ moduli fields have non-zero 
masses.
} % 
\begin{align}
 (C^{(1)}C^{(2)})^{-1} \; 
   \left( |G_{(20)(02)}|^2 \sigma_{a'}(C^{(1)}C^{(2)})
       - |\sigma_{a'}(G_{(20)(02)})|^2 C^{(1)}C^{(2)} \right) \in \R, 
\end{align}
because $C^{(1)}, C^{(2)} \in \R \cap \rho^{(1)}_{(20)}(K^{(1)})$.

To summarize, for a given vacuum complex structure 
in ${\cal M}_{\rm CM}^{[X(T_0^{(1)})]} \times {\cal M}_{\rm CM}^{[X(T_0^{(2)})]}$ 
satisfying the condition (\ref{eq:flux-cond-math2}), each choice of 
a flux from $W_{(20|02)} \simeq \Q^n$ is consistent with the 
$DW=0$ and $W=0$ conditions, and the $(n-2)$ Dirac mass eigenvalues 
(all the values in $\overline{\Q}$) can be computed systematically. 
As a reminder, $n = [K^{(1)}:\Q] = [K^{(2)}:\Q] = {\rm rank}(T_X^{(i)}) 
= 22-r_{(1,2)}$. 
The Dirac type mass matrix and the real nature of the product 
of the mass eigenvalues are a common (and unexpected!)\footnote{
The vacuum complex structure of the pair of K3 surfaces being CM-type 
does not imply at all such things as period integrals being real, 
or the field of moduli having embedding into $\R$. 
Here, we pay attention to the product of the mass eigenvalues 
as an exercise problem for potential applications to the strong 
CP problem. 
} %
 consequence the class of flux vacua under consideration. 

\vspace{5mm}

The moduli stabilization discussed above appears similar to the one 
in \cite{AK, BHV, BKW-phys}.\footnote{
The relation to \cite{DDFK} is discussed in section \ref{ssec:on-fibre}.
} %
 Direct comparison with \cite{AK, BHV, BKW-phys} 
is easier in the cases we discuss in section \ref{ssec:Tx-neq-T0} 
(see footnote \ref{fn:compare-AK-BKW}). 
When we take $T_X^{(i)}=T_0^{(i)}$ (as in this section \ref{ssec:Tx=T0})
and set ${\rm rank}(T_X^{(i)})=2$ (as in \cite{AK,BKW-phys}), 
all the complex structure moduli fields of $Y=(X^{(1)}\times X^{(2)})$ 
whose mass discussed in \cite{AK, BKW-phys} are now projected out 
in the orbifold $Y = (X^{(1)} \times X^{(2)})/\Z_2$ here. 
The mass and stabilization of the $(40-r_{(1)}-r_{(2)})$ moduli above 
is a special case of those in \cite{BHV} (in that the vacuum complex 
structure of $X^{(i)}$ is assumed to be CM-type in this article).

The moduli mass from fluxes (with $\vev{W}=0$) above is closer to 
the one in \cite{prev.paper.PRD}. Discussion there correspond to 
a special case of the above result in this article; 
Ref. \cite{prev.paper.PRD} was for $X^{(1)} = {\rm Km}(E_\phi \times E_\tau)$, 
the Type IIB orientifold set-up. 
Although \cite{prev.paper.PRD} only argued that the complex structure 
moduli of Type IIB Calabi--Yau threefold $M = (E_\tau \times X^{(2)})/\Z_2$ 
and the axi-dilaton chiral multiplets are stabilized along with $\vev{W}=0$, 
the discussion above shows that the moduli fields of D7-brane positions 
are also stabilized along with $\vev{W} = 0$. 

\vspace{1cm}

Having studied the mass terms in (\ref{eq:moduli-eff-super-A}, 
\ref{eq:moduli-eff-super-B}) let us now have a look at the whole 
low energy effective theory superpotential of the complex structure 
moduli fields $t^{(1,2)}$ from the perspective of symmetry. 
The superpotential have $\U(1)^{\frac{n}{2}-1} \times \U(1)_R$ symmetry. 
All the moduli fields $t^{(i)}_{a}$ have $+1$ charge under the $R$-symmetry; 
there is also one U(1) symmetry for each one of $a' \in \{ 2, \cdots, n/2\}$, 
where the chiral multiplets $t^{(1)}_{a'}$ and $t^{(2)}_{\overline{a'}}$ have charge 
$+1$, and the chiral multiplets $t^{(1)}_{\overline{a'}}$ and $t^{(2)}_{a'}$ 
charge $-1$.  
This symmetry is a part of the symmetry of the K\"{a}hler potential,
 which is 
\begin{align}
  K & \; = - \sum_{i=1}^2
     \ln \left( (\Omega_{X^{(i)}}, \overline{\Omega}_{X^{(i)}}) \right) 
    \label{eq:moduli-eff-Kahler} \\
 & \; =- \sum_{i=1}^2 \ln \left( C^{(i)} + \sum_{a}^{3\sim n} \sigma_{a}(C^{(i)}) 
    t^{(i)}_{a} (t^{(i)}_a)^\dagger  
     + \!\!\! \sum_{a',b'}^{2\sim n/2} \frac{
      \left( \sigma_{a'}(C^{(i)}) t^{(i)}_{a'}t^{(i)}_{\overline{a'}} \right)       
      \left( \sigma_{b'}(C^{(i)}) (t^{(i)}_{b'} t^{(i)}_{\overline{b'}})^\dagger \right)
    }{C^{(i)}}
     \right).    \nonumber 
\end{align}

To understand the origin and nature of the $\U(1)^{\frac{n}{2}-1}$ symmetry
in the moduli effective field theory, it helps to reflect more upon the 
symmetry of the K\"{a}hler potential, the non-linear sigma model 
metric (\ref{eq:moduli-eff-Kahler}). 
The target space ${\cal M}_{\rm cpx~str}^{[Y]BV}$ is a homogeneous space 
with the symmetry group
\begin{align}
  \mathbb{G}U{\rm Isom}(T_X^{(1)}; \C) \times \mathbb{G}U{\rm Isom}(T_X^{(2)} ; \C), 
\end{align}
where 
\begin{align}
  \mathbb{G}U{\rm Isom}(L ; \C) := \left\{ g \in {\rm Isom}(L\otimes \C) \; | \; 
     (g^\dagger \bar{x}, g x)_L = {}^\exists c_g(\Slash{x}) (\bar{x},x)_L 
   {\rm ~for~} {}^\forall x \in L\otimes \C \right\} 
\end{align}
for a lattice $L$; $c_g$ can be any constant independent of $x$.  
For any (not necessarily CM) point in 
${\cal M}_{\rm cpx~str}^{[Y]BV}$ chosen as a vacuum, 
the isotropy group---the symmetry group linearly realized on the 
fluctuations fields---is
\begin{align}
 \SO(n_1-2) \times \SO(n_2-2); 
\end{align}
we have seen that $n_1 = n_2 =: n$ under the 
condition (\ref{eq:flux-cond-math2}). 
The $\U(1)^{n/2 - 1}$ symmetry of the full moduli effective theory (including 
the moduli mass terms due to fluxes) is the Cartan part of the diagonal 
$\SO(n-2)$. They are global symmetries.\footnote{
\label{fn:gauged-discrt}
A non-trivial discrete subgroup in $\U(1)^{\frac{n}{2}-1}$ may be gauged, 
in the sense that a discrete subgroup of the symmetry of a vacuum complex 
structure may be regarded as a part of the isotropy group of the form 
${\rm Isom}(T_X^{(1)})^{\rm Hdg~Amp} \times {\rm Isom}(T_X^{(2)})^{\rm Hdg~Amp}$,  
which induces automorphisms (unphysical difference) of $X^{(1)}\times X^{(2)}$.
} %

It is worth noting that the presence of the symmetry $\U(1)^{n/2 - 1}$ 
in the non-linear sigma model in $\R^{2,1}$ in M-theory ($\R^{3,1}$ in F-theory) 
is essentially due to the nature of the period domain of a K3 
surface\footnote{ 
The complex structure moduli effective 
superpotential (\ref{eq:moduli-eff-super-A}) is of very specific---purely 
quadratic---form also essentially due to this.

There is an argument on the ground of genericity that $\vev{W}=0$ must be 
associated with some discrete R-symmetry (more than $\Z_2$), and then 
the moduli superpotential must be of the form $W \sim \sum_i X_i f_i(\phi)$, 
where $X_i$ are chiral fields that transform the same way as $W$ under the 
R-symmetry, and $\phi$'s other moduli fields that are neutral under the 
R-symmetry \cite{DS}. It then follows in this regime when all those fields 
have masses. The moduli superpotential on K3 x K3 orbifolds in this article
is not within this genericity regime.       
} %
rather than that of a Calabi--Yau manifold of higher dimensions. 
As already briefly referred to\footnote{
Interested readers are referred to \cite{Borcea, vMHS}.
} %
 in section \ref{ssec:CM-BV}, 
one could think of the space of rational Hodge structures on $H^3(M;\Q)$
for a family of Calabi--Yau threefolds $[M]$, which is a homogeneous 
space just like $D(T_0^{(i)})$'s are; the complex structure moduli space 
of $[M]$ is only a subspace of the homogeneous space, so the symmetry 
of the non-linear sigma model of the complex structure moduli of $[M]$ 
cannot be simply stated by just referring to the vector space $H^3(M;\Q)$
and the skew-symmetric intersection form on it. The same discussion 
applies also to Calabi--Yau fourfolds.

Furthermore, those continuous symmetries---either $\U(1)^{n/2-1}$ 
or $\SO(n-2) \times \SO(n-2)$---of the non-linear sigma model in $\R^{2,1}$ 
or $\R^{3,1}$ cannot be attributed to a symmetry of the geometry 
$X^{(1)} \times X^{(2)}$. 
A symmetry transformation on $X^{(i)}$ would manifest itself as a symmetry 
action on $H^2(X^{(i)};\Q)$; a transformation on $H^2(X^{(i)};\C)$ that cannot 
be derived from one on $H^2(X^{(i)};\Q)$ does not have an interpretation 
as an $X^{(i)} \rightarrow X^{(i)}$ map. Those continuous symmetries are 
not symmetries of the geometry $X^{(1)} \times X^{(2)}$, but are symmetries 
of their moduli spaces. They are accidental symmetry in the low-energy 
effective theory. 

The continuous $\U(1)^{\frac{n}{2}-1} \times \U(1)_R$ symmetry in the 
moduli effective theory are likely not to be an exact symmetry apart 
from its possible non-trivial discrete subgroup 
(cf footnote \ref{fn:gauged-discrt}). This expectation 
is from general arguments in quantum gravity; as for the $\U(1)_R$ 
part,\footnote{ 
There are tight constraints on how R-symmetry charge is assigned 
on the particles in supersymmetric Standard Models. On the other hand, 
we need to know how the moduli fields $t_a^{(i)}$ couple to the Standard 
Model particles to find out how (or whether) the $\U(1)^{n/2-1}$ symmetry 
can be extended to the whole low-energy effective theory. 
} %
one may also argue this by computing triangle anomalies against 
the Standard Model gauge groups (e.g., \cite{EIKY}).
The source of explicit breaking of the symmetry may be  
the anomalies with gauge fields, stringy non-perturbative effects, 
or just stringy perturbative corrections to the approximation 
$K= - \ln (\int_Y \Omega_Y \wedge \overline{\Omega}_Y)$ and 
$W \propto \int_Y G \wedge \Omega_Y$. Better understanding on the 
source of explicit breaking\footnote{
Alternatively, one may focus on the common subset of 
${\rm Isom}(T_X^{(i)}) \times {\rm Isom}(T_X^{(2)})$ and $\U(1)^{n/2-1}$, 
which will be a more mathematical study, to infer what the discrete 
gauged symmetry is. 
} % 
 will give us better hint on a discrete 
exact symmetry in the effective theory containing all of moduli, 
the supersymmetric Standard Model and anything else. In the case 
the discrete exact symmetry is larger than the symmetry acceptable 
at TeV scale (such as a subgroup of $\U(1)_R$ larger than $\Z_2$ 
R symmetry), the domain wall problem sets constraints on inflation 
and the thermal history after that.  If the explicit breaking leaves 
only the $\Z_2$ subgroup of the $\U(1)_R$ symmetry, then the source 
of the explicit breaking also determines the gravitino mass. 

%%%%%%%%%%%%%%%%%%%%%%%%%%%%%%%%%%%%%%%%%%%%%%%%%%%%%%%
\subsection{Cases with $T_X\subsetneq T_0$}
\label{ssec:Tx-neq-T0}
%%%%%%%%%%%%%%%%%%%%%%%%%%%%%%%%%%%%%%%%%%%%%%%%%%%%%%%

Think of a case where the vacuum complex structure of $X^{(i)}$ is still 
of CM-type, but not generic enough to have $T_X^{(i)}=T_0^{(i)}$ for 
at least one of $i=1,2$; $T_X^{(i)} \subsetneq T_0^{(i)}$ and 
$S_0^{(i)} \subsetneq S_X^{(i)}$ then. Put differently, the vacuum complex 
structure of $X^{(i)}$ is in a Noether--Lefschetz locus of 
$\overline{D(T_0^{(i)})}$, where there must be an element of 
$H^2({\rm K3};\Z)$ (Poincar\'e dual of a 2-cycle)\footnote{
memo:
The 2-cycle in question may or may not be norm (-2). If it is not norm (-2), 
then the limit vacuum complex structure is in a subvariety of $D(T_0^{(i)})$, 
and such a K3 surface $X^{(i)}$ is obtained by just tuning the complex 
structure. If it is norm (-2), then the limit vacuum complex structure 
is found only in the closure $\overline{D(T_0^{(i)})}$, not within 
$D(T_0^{(i)})$, if that matters. Such a K3 surface $X^{(i)}$ is obtained by 
taking a limit in the complex structure so an $A_1$-singularity emerges, 
and then by resolving it. } % 
that becomes algebraic. 
Now, the rank of $T_X^{(i)}$ is still even (because of its CM nature), 
but $T_0^{(i)}$ in Nikulin's list may be of odd rank.
We will see below that much the same story unfolds for a $DW=0$ flux, and 
also for a $DW=W=0$ flux; one difference, though, is that 
there is one more way (without the relation (\ref{eq:flux-cond-math2})) 
to stabilize moduli in ${\cal M}_{\rm cpx~str}^{[Y]BV}$ by a $DW=W=0$ flux, 
when $T_X^{(i)} \subsetneq T_0^{(i)}$ for both $i=1,2$. 

Let $\overline{T}_0^{(i)}$ be the negative definite lattice 
$[(T_X^{(i)})^\perp \subset T_0^{(i)}]$, so that 
\begin{align}
  T_0^{(i)} \otimes \Q \cong (T_X^{(i)} \otimes \Q) \oplus 
(\overline{T}_0^{(i)} \otimes \Q).
\end{align}
The $(T_0^{(1)} \otimes T_0^{(2)}) \otimes \Q$ component of 
$H^4_H(Y;\Q)$ is then expanded as follows:
\begin{equation}
    (T_0^{(1)} \otimes T_0^{(2)})=(T_X^{(1)} \otimes T_X^{(2)})
 \oplus (T_X^{(1)}\otimes \overline{T}_0^{(2)}) 
 \oplus (\overline{T}_0^{(1)} \otimes T_X^{(2)}) 
 \oplus (\overline{T}_0^{(1)} \otimes \overline{T}_0^{(2)}).
\end{equation}
The two components 
$(T_X^{(1)} \otimes \overline{T}_0^{(2)})$ and 
$(\overline{T}_0^{(1)} \otimes T_X^{(2)})$ with rational Hodge substructure
are always level-2, and the 
$(\overline{T}_0^{(1)} \otimes \overline{T}_0^{(2)})$ component 
always level-0, if it is non-empty. 
The component $T_X^{(1)}\otimes T_X^{(2)}$ 
contains a level-4 component; whether the rational Hodge structure 
on this component is simple or not depends. 

Suppose that the condition (\ref{eq:flux-cond-math2}) is satisfied. 
Then any rational flux in $\overline{T}_0^{(1)}\otimes \overline{T}_0^{(2)} 
\otimes \Q$ and the $i=(20|02)$ component of 
$T_X^{(1)}\otimes T_X^{(2)}\otimes\Q$ satisfies 
the $DW=0$ and $W=0$ conditions.\footnote{
\label{fn:compare-AK-BKW}
This mechanism is quite close to the one in \cite{AK, BHV, BKW-phys}; 
in the terminology of \cite{AK,BKW-phys}, the flux in 
$(\overline{T}_0^{(1)} \otimes \overline{T}_0^{(2)}) \otimes \Q$ 
corresponds to a part of $G_0$-flux, and the one in
$(T_X^{(1)} \otimes T_X^{(2)}) \otimes \Q$ to $G_1$-flux.
}  % 
When a flux is non-zero only in the $W_{(20|02)}$ component 
within $T_X^{(1)}\otimes T_X^{(2)}$, the moduli effective theory 
superpotential (\ref{eq:moduli-eff-super-A}) remains as it is 
if it is interpreted as follows; 
the third term of (\ref{eq:moduli-eff-super-A}) only involves the 
moduli fluctuation fields within $D(T_X^{(1)}) \times D(T_X^{(2)})$, 
while $(t^{(2)},t^{(2)})$ and $(t^{(1)},t^{(1)})$ in the first two terms 
of (\ref{eq:moduli-eff-super-A}) are meant to include all the fluctuations 
fields in $D(T_0^{(1)})\times D(T_0^{(2)})$. 
When a non-zero flux is in the 
$\overline{T}_0^{(1)}\otimes \overline{T}_0^{(2)}\otimes \Q$, there is one 
more term in the effective superpotential, which is the Dirac-type 
mass term of the moduli fluctuation fields in $N_{D(T_X^{(1)})|D(T_0^{(1)})}$
(normal directions) and $N_{D(T_X^{(2)})|D(T_0^{(2)})}$. 
Because the (stabilizing) mass terms for the fluctuations within 
$D(T_X^{(1)})\times D(T_X^{(2)})$ rely on the flux in $W_{(20|02)}$, 
the condition (\ref{eq:flux-cond-math2}) is necessary (apart from the 
caveat mentioned below). This moduli effective theory has an 
U(1) R-symmetry, where all the moduli fluctuation fields have $+1$ R-charge; 
to see this, we almost have to repeat the argument in 
section \ref{sssec:mass-TX=T0}, and the fact that a flux in 
$\overline{T}_0^{(1)}\otimes \overline{T}_0^{(2)}$ also generates only the 
mass term. There is no additional non-R U(1) symmetry where the moduli 
fluctuation fields in $N_{D(T_X^{(1)})|D(T_0^{(1)})}$ and $N_{D(T_X^{(2)})|D(T_0^{(2)})}$ 
are charged, however. This is because there is no Dirac-like structure 
for those moduli fields in the $(t^{(2)},t^{(2)})$ and $(t^{(1)},t^{(1)})$ 
in the first two terms in the superpotential (\ref{eq:moduli-eff-super-A}).

One caveat in the argument above is the case there is no moduli fluctuation 
fields within $D(T_X^{(1)}) \times D(T_X^{(2)})$, which is when 
both $X^{(1)}$ and $X^{(2)}$ are attractive (${\rm rank}(T_X^{(i)})=2$, 
${\rm rank}(S_X^{(i)})=20$) K3 surfaces. Even when the condition 
(\ref{eq:flux-cond-math2}) is not satisfied, 
a flux in $\overline{T}_0^{(1)} \otimes \overline{T}_0^{(2)}$ provide 
a mass term for all the moduli fluctuation fields in 
$D(T_0^{(1)})\times D(T_0^{(2)})$ if the condition 
\begin{align}
  {\rm rank}(\overline{T}_0^{(1)}) = {\rm rank}(\overline{T}_0^{(2)})
   \label{eq:flux-cond-math3}
\end{align}
is satisfied. The mass matrix is Dirac type then. 
This is because the mass matrix from the flux here is always Dirac type. 

%%%%%%%%%%%%%%%%%%%%%%%%%%%%%%%%%%%%%%%%%%%%%%%%%%
\section{General K3 x K3 Orbifolds}
\label{sec:non-Z2-orbifold}
%%%%%%%%%%%%%%%%%%%%%%%%%%%%%%%%%%%%%%%%%%%%%%%%%%

The Borcea--Voisin orbifold $(X^{(1)}\times X^{(2)})/\Z_2$ in the previous 
section is regarded as one way to construct a Calabi--Yau variety of 
higher dimensions by using K3 surfaces (and/or elliptic curves). 
So, there is an obvious generalization; think of any supersymmetry-preserving
orbifold of a product of K3 surfaces (and/or elliptic curves); the 
orbifold group $\Gamma$ is not necessarily $\Z_2$ \cite{Generalized-BV}.  
In this section \ref{sec:non-Z2-orbifold}, we bring known materials 
together from the literatures, to have a broad brush picture 
of possible variety in the construction, to identify open math problems 
for a complete classification, and to repeat the same study as in 
sections \ref{ssec:Tx=T0} and \ref{ssec:Tx-neq-T0} for the cases with 
$\Gamma \neq \Z_2$. 

%=====================================%
\subsection{K3 x K3 Orbifold}
\label{ssec:gen-BV}
%======================================%

Consider an orbifold $Y_0 = (X^{(1)} \times X^{(2)})/\Gamma$, where $X^{(1)}$ 
and $X^{(2)}$ is a pair of K3 surfaces. The orbifold group $\Gamma$ should 
be a subgroup of ${\rm Aut}(X^{(1)}) \times {\rm Aut}(X^{(2)})$, first of all. 
For the action of the orbifold group $\Gamma$ to preserve supersymmetry, 
one more condition needs to be imposed. To state the condition, 
we prepare some notations.  

Under the projection $p_i: {\rm Aut}(X^{(1)}) \times {\rm Aut}(X^{(2)}) 
\rightarrow {\rm Aut}(X^{(i)})$, let $G_i := p_i(\Gamma)$. 
Let $\alpha'_i: {\rm Aut}(X^{(i)}) \rightarrow {\rm Isom}(T^{(i)}_X)^{\rm Hdg~Amp}$ 
be the projection that fits into the exact sequence\footnote{
In this section, we use the same notation as in \cite{BKW-math, BKW-phys} 
without spelling out their definitions. 
Reviews in \cite{Asp-K3, huybrechts2016lectures, BKW-math, BKW-phys} 
will also be useful.
} % 
\begin{align}
  1 \rightarrow {\rm Aut}_N(X^{(i)}) \rightarrow {\rm Aut}(X^{(i)}) \rightarrow 
  {\rm Isom}(T_X^{(i)})^{\rm Hodge~Amp} \rightarrow 1.
\end{align}
Because the elements of ${\rm Isom}(T_X^{(i)})^{\rm Hdg~Amp}$ acts on 
the holomorphic (2,0) form $\Omega_{X^{(i)}}$ faithfully, 
$\alpha'_i(\sigma_{(i)}) \in {\rm Isom}(T^{(i)}_X)^{\rm Hdg~Amp}$ for 
$\sigma_{(i)} \in {\rm Aut}(X^{(i)})$ 
may well be identified with the complex phase $\alpha_i(\sigma_{(i)})$ in 
$\sigma_{(i)}^*\Omega_{X^{(i)}} = \alpha_i(\sigma_{(i)}) \Omega_{X^{(i)}}$. 
With those preparations, the supersymmetry condition is written as 
\begin{align}
  {}^\forall \sigma \in \Gamma, \qquad 
    \alpha_1(p_1(\sigma)) \; \alpha_2(p_2(\sigma)) = 1 \in \C. 
\end{align}
We will discuss only the cases that the group $\Gamma$ has 
a finite number of elements.\footnote{
It sounds like an orbifold $(X^{(1)}\times X^{(2)})/\Gamma$ with 
$|\Gamma| = \infty$ would yield a pathological ``Calabi--Yau fourfold'', 
although we are not absolutely sure if such possibilities should be completely ruled out.
} %

An equivalent way to state the supersymmetry condition is that 
there is a group $\Delta$, so that\footnote{
The isomorphism $\alpha'_1(G_1) \cong \alpha'_2(G_2)$ should be such that 
their representations $\alpha_1$ and $\alpha_2$ are complex conjugate. 
}\raisebox{4pt}{,}\footnote{
Complete classification of $(S^{(i)},T^{(i)},G_i; G_{s,i}, \Delta)$ 
for $i=1,2$ and $\Gamma \subset G_1 \times_\Delta G_2$ will be redundant 
for classification of variety in the generalized Borcea--Voisin fourfolds
for compactification. For example, in a case $\Gamma$ has a structure 
of $\Gamma \cong \Gamma_0 \times G'_1 \times G'_2$ with 
$\Gamma_0 \subset {\rm Aut}(X^{(1)}) \times_\Delta {\rm Aut}(X^{(2)})$ 
and $G'_i \subset {\rm Aut}_N(X^{(i)})$ for $i=1,2$, one may replace 
a compactification by $(X^{(1)}\times X^{(2)})/\Gamma$ with 
a compactification by $(X^{(1)}_{\rm cr} \times X^{(2)}_{\rm cr})/\Gamma_0$, 
where $X^{(i)}_{\rm cr}$ is a crepant resolution of $X^{(i)}/G'_i$. 
} % 
\begin{align}
  \alpha'_i(G_i) \cong \Delta, \qquad \Gamma \subset G_1 \times_\Delta G_2.
\end{align}
When we impose the Calabi--Yau condition ($h^{p,0}(Y_0)=0$ for $p=1,2,3$ 
in addition to $h^{4,0}(Y_0) = 1$), the group $\Delta$ needs to be something 
other than\footnote{
It can be shown (\cite{nikulin1980finite} Thm. 0.1 (a), Thm. 3.1 (a) and Cor 3.2) that a K3 surface
with $\Delta \neq \{1\}$ is always algebraic.
} %
 the trivial group $\{ 1\}$. 

The two K3 surfaces $X^{(1)}$ and $X^{(2)}$ for an M-theory/F-theory 
compactification come with one K\"{a}hler form for each one of them. 
The orbifold group action by $G_i$ for $i=1,2$ should preserve the 
K\"{a}hler form on $X^{(i)}$ (so the orbifold defines a consistent 
theory).\footnote{
Here is a version of this statement that reflects the underlying 
theoretical principles more directly:  
The metric on $X^{(i=1,2)}$ is expressed as a positive definite 3-plane
within $H^2(X^{(i)};\R)$, and all the elements $g$ in $G_i$ of the 
orbifold group should preserve this 3-plane as a whole; $g$ induces 
an SO(3) rotation on the 3-plane, which is regarded as a rotation 
around one axis. 
We require further that this rotation axis is common for all 
$g \in G_i$, in which case the orbifold becomes Calabi--Yau.  

The common axis of rotation within the 3-plane is identified with 
the K\"{a}hler form modulo normalization, and the other two directions 
orthogonal to the axis are identified with $\Omega_{X^{(i)}}$. 
} %

%================================================%
\subsection{K3 Surfaces with Non-symplectic Automorphisms}
\label{ssec:K3-nonsymp-Aut}
%================================================%

%%%%%%%%%%%%%%%%%%%%%%%%%%%%%%%%%%%%%%%%%%%%%%%%%%%%%%%%%%%%%%%%%%
\subsubsection{Discrete Classification}
\label{sssec:discr-classify}
%%%%%%%%%%%%%%%%%%%%%%%%%%%%%%%%%%%%%%%%%%%%%%%%%%%%%%%%%%%%%%%%%%

Just like we used Nikulin's classification in the previous section, 
one can think of a similar classification problem whose answer can be used 
for this general form of the Borcea--Voisin orbifolds. Here is how we 
formulate the problem: how many different choices of $(S\oplus T, G)$ there 
are modulo ${\rm Isom}({\rm II}_{3,19})$, subject to the conditions 
\begin{itemize}
\item $G$ is a finite subgroup of ${\rm Isom}({\rm II}_{3,19})$, and 
   $S$ and $T$ are mutually orthogonal primitive sublattices of 
   ${\rm II}_{3,19} \cong H^2(K3;\Z)$ such that 
   $(S\oplus T)\otimes \Q \cong {\rm II}_{3,19}\otimes \Q$, 
\item $g(T) = T$ (and also $g(S) = S$) for any $g \in G$, 
\item for any $g \in G$ whose $g|_T$ is non-trivial, $g|_T$ is not identity 
on any vector subspace of $T \otimes \Q$.
\item $S$ and $T$ have signature $(1,r-1)$ and $(2,20-r)$ 
  (with $1\leq r \leq 20$), 
\item for any $g \in G$, the sublattice $S^g := \{ x \in S \; | \; 
     g \cdot x = x\}$ contains 1 signature-positive 
     direction.
\end{itemize}
The first three conditions characterize $G$, $T$ and $S$ as 
a set of automorphism group, transcendental lattice and N\'eron--Severi 
lattice of a K3 surface.\footnote{
See \cite[Thm. 3.1 (b)]{nikulin1980finite} for the third condition.
} %
The last two conditions reflect\footnote{
See \cite[Thm. 0.1 (a) and 3.1 (a)]{nikulin1980finite} for the 
fourth condition, and 
\cite[Lemma 4.2 (a)]{nikulin1980finite} for the fifth condition. 
} %
 the Calabi--Yau condition of $Y_0$ 
($\Delta \neq \{1\}$) and the $\Gamma$-invariance of the K\"{a}hler 
parameter discussed in section \ref{ssec:gen-BV}. 

For one choice $(S\oplus T, G)$, we can determine two groups 
\begin{align}
  G_{\rm s} & \; := {\rm Ker} \left(G \rightarrow {\rm Isom}(T)\right), \\
  \Delta & \; := {\rm Im}\left( G \rightarrow {\rm Isom}(T) \right).
\end{align}
So, the classification of $(S\oplus T, G)$ may well be regarded as 
classification of the data $(S \oplus T, G; G_s, \Delta)$. 
Furthermore, one may state the result of the classification 
by listing up all possible choices 
$1 \rightarrow G_s \rightarrow G \rightarrow \Delta \rightarrow 1$ first, 
and then by listing up of all possible lattice pairs $S\oplus T$ for 
$(G; G_s,\Delta)$.
Nikulin's classification implies that the case $G_s \cong \{1\}$ and 
$G \cong \Delta \cong \Z_2$ contains\footnote{
\label{fn:def-purely-nonsymp}
Nikulin's list contain $S \oplus T$ where $G \cong \Delta \cong \Z_2$ 
acts trivially on $S$. When the last condition is relaxed, then there may 
be more choices of $S \oplus T$ for the case of $G_s \cong \{1\}$ and 
$G\cong \Delta \cong \Z_2$. When the latter condition is satisfied, 
we say that the action of $G \cong \Delta$ is {\bf purely non-symplectic}.
}\raisebox{4pt}{,}\footnote{
\label{fn:more-purely-nonsymp}
In the case of $G_s \cong \{1\}$ and $G \cong \Delta \cong \Z_p$ for 
a prime number $p$, it is enough to list up all the $(S, T)$'s 
where $G \cong \Delta$ acts trivially on $S$. Even when one finds 
$(S', T')$ where $G \cong \Delta$ acts non-trivially on $S'$, one may find 
$S\subsetneq S'$ where $G \cong \Delta$ acts trivially; a K3 surface 
with its complex structure in $D(T')$ may be regarded as a special case 
of a K3 surface with its complex structure in $D([S^\perp])$.  
}  %
 75 different choices of $S\oplus T$, as we have referred to 
in section \ref{sec:flux-analysis}.  

There are at most 41 different choices of the group $\Delta$ including 
$\Delta \cong \{1\}$ % (which would lead to a foufold with $h^{2,0} = 1$)
(\cite[Cor 3.2]{nikulin1980finite} and \cite{MO}); 
all the possible groups $\Delta$ are cyclic groups $\Z_m$ for some 
$m \in \N_{>0}$ 
(\cite[Thm. 3.1 (b,c)]{nikulin1980finite}, \cite[Lemma 2.1]{Sterk}, 
\cite[Cor 3.3.4]{huybrechts2016lectures}), 
 and the list of 41 $m$'s (Table 1 of \cite{MO}) are reprinted explicitly 
in Table \ref{tab:list-transc-indx} here for convenience of the readers. 
%%%%%%%%%%%%%%%%%%%%%%%%%%%%%%%%%%%%%%%%%%%%%%%%%%%%%%%%%%%%%%%%%%%%%%%%% 
\begin{table}[tbp]
\begin{center}
\begin{tabular}{cc}
\begin{tabular}{cccccc}
  $1_1$ & $2_1$ & $\! 4_2$ & $8_4$ & $16_8$ & $32_{16}$ \\
  $3_2$ & $6_2$ & $\! 12_4$ & $24_{8}$ & $48_{16}$ \\
  $9_6$ & $18_6$ & $\! 36_{12}$ \\
  $27_{18}$ & $54_{18}$
\end{tabular}
&
\begin{tabular}{cccccccc}
  $1_1$ & $5_4$ & $7_6$ & $11_{10}$ & $13_{12}$ & $17_{16}$ & $19_{18}$ & $25_{20}$ \\
  $2_1$ & $10_{4}$ & $14_6$ & $22_{10}$ & $26_{12}$ & $34_{16}$ & $38_{16}$ & $50_{20}$ \\
  $3_2$ & $15_{8}$ & $21_{12}$ & $33_{20}$ \\
  $4_2$ & $20_{8}$ & $28_{12}$ & $44_{20}$ \\
  $6_2$ & $30_{8}$ & $42_{12}$ & $66_{20}$ \\
  $8_4$ & $40_{16}$ \\
  $12_4$ & $60_{16}$
\end{tabular}
\end{tabular}
 \caption{\label{tab:list-transc-indx}The list of $m:= |\Delta|$ 
and the corresponding $\varphi (m)$. 
The $m_{\varphi(m)}$'s with an $m$ of the form of $m=2^p 3^q$ are shown 
in the table on the left, while the table on the right lists up $m$'s 
that are not divisible by $2^4$, $3^2$, or $24$.
$m=1,2,3,4,6,8,12$ overlap in the two Tables.  }
\end{center}
\end{table}
%%%%%%%%%%%%%%%%%%%%%%%%%%%%%%%%%%%%%%%%%%%%%%%%%%%%%%%%%%%%%%%%%%%%%%%%%%
The rank $(22-r)$ of the lattice $T$ should be divisible by $\varphi(m)$. 

There are at most 82 different choices of $G_s$ including $G_s \cong \{1\}$.
They should be a subgroup of 11 different finite groups listed 
in \cite{Mukai88} (two of them are $\mathfrak{A}_6$ and $\mathfrak{S}_5$; 
see \cite{Mukai88} for nine others). See \cite{XiaoGal} for the list 
of those all those 82 finite groups.\footnote{
\label{fn:finite-Aut-symp}
Among the 82 of them are 15 abelian groups (including $\{1\}$) worked out 
by \cite[Thm 4.5]{nikulin1980finite}. If we are to demand that $|{\rm Aut}(X)|<\infty$ 
(not just $|G|<\infty$), then just the three in the list, 
$G_s = \{1\}$, $\Z_2$ and $S_3$, are possible \cite{Kondo89-1,Kondo89-2}.
} %
It is also known that if $G_s$ has an element $g$ of order $n>1$, then 
it must be\footnote{
An immediate consequence of this fact is that, if we are to choose 
a finite subgroup $G \subset {\rm Aut}(X)$, 
then any element in $G$ has an order not larger than $8 \cdot 66$; 
in fact there is no such an element of order $8 \times 66$
because the unique K3 surface admitting $\Delta  \cong \Z_{66}$ does not 
have symplectic automorphisms. The true upper bound is known 
to be 66 \cite{Keumorder}. 
} %
 that $n \leq 8$ first of all, and secondly, 
${\rm rank}(S)\geq 9$ if $n=2$, 
${\rm rank}(S)\geq 13$ if $n=3$, 
${\rm rank}(S)\geq 15$ if $n=4$, 
${\rm rank}(S)\geq 17$ if $n=5,6$, and 
${\rm rank}(S)\geq 19$ if $n=7,8$ \cite[Cor 15.1.8]{huybrechts2016lectures}. 

Therefore, there can be at most [82 x 41] different choices of the 
finite groups $G_s$ and $\Delta$; the choice $G_s \cong \{1\}$ and 
$\Delta \cong \Z_2$ in section \ref{sec:flux-analysis} is one of 
this [82 x 41] choices. 
In fact, not all the 82 x 41 choices can be realized. 
A group $\Delta \cong \Z_m$ with larger $m$ requires that the lattice $T$ 
has a larger rank because $\varphi(m)|{\rm rk}(T)$, 
whereas a larger group $G_s$ requires $S$ with a larger rank (and $T$ 
with a smaller rank). Using the data available in \cite{XiaoGal},
%%%%%%%%%%%%%%%%%%%%%%%%%%%%%%%%%%%%%%%%%%%%%%%%%%%%%%%%%
\begin{table}[tbp]
\begin{center}
\begin{tabular}{rc|l}
 $G_s$ in \cite{XiaoGal} & $c$ & $\Delta \cong \Z_m$ \\
 \hline 
 $\{ 1\}$ & & any 41 $m$'s \\
 \#1 & ($c=8$)  & 26 $m_{\varphi(m)}$'s with $\varphi(m) \leq 12$ \\
 \#2, 3 & ($c=12$) & 18 $m_{\varphi(m)}$'s with $\varphi(m) \leq 8$ \\
 \# 4,6,9,10,21 & $c=14,15$ & 13 $m_{\varphi(m)}$'s with $\varphi(m) \leq 6$ \\
 17 $G_s$'s & $c = 16,17$ & 9 $m_{\varphi(m)}$'s with $\varphi(m) \leq 4$ \\
 56 $G_s$'s & $c = 18,19$ & $m_{\varphi(m)} \in \{ 1_1,2_1,3_2,4_2,6_2\}$ \\
  \hline 
\end{tabular}
\caption{\label{tab:Gs-m-triangle}The range of $m_{\varphi(m)}$ such that 
$\Delta \cong \Z_m$ can be combined with a given $G_s$ to form $G$.
The 82 choices of $G_s$ 
are grouped into six by their value of $c$ listed in Table 2 
of \cite{XiaoGal}. For those six groups of $G_s$'s, the possible range 
of $m_{\varphi(m)}$ is determined by the condition $\varphi(m) \leq 21-c$, 
shown on the right. }
\end{center}
\end{table}
%%%%%%%%%%%%%%%%%%%%%%%%%%%%%%%%%%%%%%%%%%%%%%%%%%%%%%%%
the range of $m_{\varphi(m)}$ can be narrowed down for each choice of $G_s$;
Table \ref{tab:Gs-m-triangle} is the summary 
(cf. also \cite{KeumMax, Keumorder}).

For $\Delta \cong \Z_m$ with $m=66,44,33,50,25,40$, and 60, 
for which $G_s = \{1\}$ (and $G \cong \Delta$) is the only option, 
all the possible $S \oplus T$'s have been worked out by using lattice 
theory and a bit of geometry  \cite[Lemma (1.2)]{MO}. It turns out that 
there is just one choice of $S\oplus T$ for each one of $m=66,44,33,50,25,40$,
and that $G \cong \Delta$ happens to act on $S$ trivially. There is no 
choice of $S \oplus T$ where $m=60$ \cite{Zhang}. 
For $\Delta \cong \Z_m$ with $m=17$ and 19, see \cite{OZ99}.

For general $(G; G_s, \Delta)$, complete classification of the choices 
of $S\oplus T$ is not available yet. For cases with $G_s = \{1\}$ 
(so $G \cong \Delta$), all the possibilities of $(S \oplus T)$ with 
$G\cong \Delta \cong \Z_m$ acting trivially on $S$ have been classified, 
however. For cases with an $m$ that is divisible by two (or more) prime 
numbers (such as $m = 6, 10, 15, \cdots$), it turns out that both $S$ and 
$T$ have to be unimodular; see \cite{Kondo92} for the list of $S \oplus T$ 
for the $m$'s that are not in the form of $m=p^k$ for a single prime 
number $p$. For cases with $m = p^k$,  
this is an immediate generalization of the classification of Nikulin
\cite{Nikulin-factor-long}. 
See \cite{Taki12, Schuett10, TST14} 
for the $m=2^2$ case (there are 12 $S\oplus T$), 
the $m=2^3$ case (there are 3 $S\oplus T$), and 
the $m=2^4$ case ($S=UD_4$ unique), while there is no choice of 
$S\oplus T$ for the case $m=2^5$ \cite{Vorontsov83, Kondo92}.
For the cases with 
$m=3^k$, and $m = 5, 7, 11, 13, 17, 19$, see \cite{AS08, Taki11, Taki10} 
and \cite{Artebani2011}, respectively. 

For the cases with $G_s = \{1\}$ (so $G \cong \Delta \cong \Z_m$), one 
may think of the classification of $(S\oplus T)$'s where $G$ may 
act on $S$ non-trivially. Only partial results are known. 
Results for $m=17,19,40,25,50,33,44,66,60$ have been quoted earlier already. 
The $m=2^5$ case has just one choice of $S\oplus T$, where $G \cong \Z_{32}$ 
acts on the rank-6 $S = UD_4$ (the same as the unique choice for the $m=2^4$
case) non-trivially on a 2-dimensional subspace through a quotient 
$\Z_{32}\rightarrow \Z_{4}$ (and trivially on a 4-dimensional subspace) 
\cite{Oguiso93, TakionO}. For a similar study in the case of 
$G_s = \{1\}$ and $G \cong \Delta \cong \Z_m$ with $m=2^4$, $m=2^3$, and 
$m=2^2$, see \cite{TST14, Tabbaa-thesis, TS, AS15}. 

Just like there is only small number of choices of $(S\oplus T)$ is 
available for a large $\Delta$, it is also known that there are tight 
constraints on the possible choices of $(S\oplus T)$ when $G_s$ is large. 
See such references as \cite[Thm. 4.7]{nikulin1980finite}, \cite{Hashimoto},
\cite{GS07}, \cite{GS09}, \cite{Kondomax}, and \cite{OZ168}. 

This section \ref{sssec:discr-classify} is a literature survey, relying 
mostly on \cite{huybrechts2016lectures} as a guide. 
We wished to learn what is known as well as what has not been known 
about how much the $\Z_2$ orbifold in section \ref{sec:flux-analysis} 
can be generalized.  

%%%%%%%%%%%%%%%%%%%%%%%%%%%%%%%%%%%%%%%%%%%%%%%%%%%%%%%%%%%%%%%
\subsubsection{Period Domains for K3 Surfaces with Automorphisms}
%%%%%%%%%%%%%%%%%%%%%%%%%%%%%%%%%%%%%%%%%%%%%%%%%%%%%%%%%%%%%%%

The period integrals (complex structure) of a K3 surface $X$ should be 
in the period domain $D(T)$ for one of the choice $(S\oplus T)$, when 
$X$ has an automorphism $(G; G_s,\Delta)$, but the converse 
is not true. For a complex structure to be consistent with the 
automorphism group $(G; G_s,\Delta)$ with $\Delta \neq \{1\}$, 
$G|_T =\Delta \cong \Z_m$ needs to be a Hodge isometry.  

The subspace in $D(T)$ consistent with such a non-symplectic automorphism 
group is specified as follows. 
Note that the action of $\Delta \cong \Z_m$ on $T\otimes \Q$ is always 
of the form of 
\begin{align}
 T\otimes \Q \cong (N_m)^{ \oplus \ell} \qquad {\rm with} \quad 
  \ell := {\rm rk}(T)/\varphi(m)
\end{align}
where $\Z_m$ acts as $\Q$-valued matrices on a $\varphi(m)$-dimensional 
vector space $N_m$ over $\Q$; the generator $[[\sigma]]$ of $\Z_m$ has  
the set of eigenvalues $\{ \zeta_m^a \; | \; a \in [\Z_m]^\times\}$. 
So, $T\otimes \C$ is divided into $\varphi(m)$ distinct eigenspaces of 
$\Z_m$, $\oplus_{a \in [\Z_m]^\times} V_a$, where $[[\sigma]]|_{V_a} = \zeta_m^a$.
Individual $V_a$'s are of $\ell$-dimensions over $\C$. So, the complex 
structure should be in
% 
% \footnote{
% %
% The K\"{a}hler moduli should also be in the subspace of $S \otimes \R$
% under the action of $(G;G_s,\Delta)$. So, this must be a non-trivial 
% subspace when $G_s$ is non-trivial, and/or non-symplectic automorphisms 
% (lifts of $\Delta$ in $G$) act non-trivially on $S$. 
% cf \cite{nikulin1980finite}.
% } % 
%
\begin{align}
 D(V_a) := \P[V_a] \cap D(T)
\end{align}
for some $a \in [\Z_m]^\times$. The subvariety $D(V_a)$ of $D(T)$ is 
determined only by $\Delta \cong \Z_m$, independent of the symplectic 
subgroup of the automorphism $G_s$.

This extra condition on the complex structure moduli space was 
absent in the case of $\Z_2$ orbifold in the previous section, 
because $\varphi(m=2)=1$, and $T\otimes \Q \cong V_{a=1}$. 
For the cases with $m > 2$, however, $V_a \subsetneq T\otimes \C$, and 
$D(V_a) \subsetneq D(T)$. In fact, there is just one pair of 
$D(V_a)$ and $D(V_{a'})$ with $a,a' \in [\Z_m]^\times$ and 
$a' = -a \in \Z_m$; that is because the intersection matrix remains 
non-zero only between $V_b$--$V_{b'}$ pairs with $b ' = -b \in \Z_m$
(remember that $[[\sigma]]$ is an isometry of $T$), and the 2-dimensional 
positive signature directions must be contained only in one of those pairs. 
In that non-empty pair $D(V_{a0})$ and $D(V_{-a0})$, the $\Omega^2 =0$ condition 
is automatically satisfied in $\P(V_{a0})$ and $\P(V_{-a0})$, so $D(V_{a0})$
and $D(V_{-a0})$ are open subspace of $\P^{\ell-1}$ specified by the 
$(\Omega, \overline{\Omega})>0$ condition \cite{Dolgachev2007, Artebani2011}. 

In the cases with $\varphi(m) = {\rm rk}(T)$, so $\ell = 1$, the 
subvarieties $D(V_a)$ are of 0-dimensions, so they are isolated 
points.\footnote{
Those isolated points are further subject to identification by 
a certain finite index subgroup of ${\rm Isom}(T)$. More is known 
in the literature about the identification of those isolated points 
(e.g., \cite[\S5]{Kondo92}).
} %
This is consistent with the fact that a CM-type K3 surface corresponds 
to an isolated point on the moduli space, as discussed later.

%%%%%%%%%%%%%%%%%%%%%%%%%%%
\subsubsection{K3 Surfaces of CM-type and with Non-symplectic Automorphisms}
\label{sssec:CM-pts-K3}
%%%%%%%%%%%%%%%%%%%%%%%%%%%

Not all the points in the moduli space $D(V_{a0})$ of K3 surfaces with 
non-symplectic automorphisms correspond to K3 surfaces of CM-type. 
The subspace of $D(V_{a0})$ corresponding to CM-type K3 surfaces is 
characterized as in the discussion in the following. 
We focus on the cases with $\Delta \cong \Z_m$ for $m>2$, 
but some parts of the discussion applies to the cases of involution, $m=2$.

In the cases with $\ell =1$ and $\varphi(m)={\rm rk}(T)$, the one point 
$D(V_{a0})$ corresponds to a CM-type K3 surface (cf \cite{0904.1922}). 
This is because 
the algebra ${\rm Span}_\Q\{ [[\sigma]] \in \Delta \}$ is a part of 
the endomorphism algebra ${\rm End}(T)^{\rm Hdg}$, and already 
$\dim_\Q({\rm Span}_\Q \{ [[\sigma]] \}) = {\rm rk}(T)$. 
The endomorphism field is isomorphic to $\Q(\zeta_m)$. 

In the cases with $\ell := {\rm rk}(T)/\varphi(m) > 1$, if a CM point 
is contained $D(V_{a0})$ outside of Noether--Lefschetz loci, then 
the CM field $K$ must be an extension of $\Q([[\sigma]]) \cong \Q(\zeta_m)$. 
Beyond that, however, the authors have not been able to find 
a comprehensive and concise statement\footnote{
\label{fn:PSSarth-Taelmn}
For a given CM field $K$, one can construct an even lattice $T$ of signature 
$(2,[K:\Q]-2)$ with a simple rational Hodge structure of CM type by $K$. 
This is done \cite{PSS-arth} by choosing $\lambda \in K_0^\times$ of certain 
kinds, introducing a $\Q$-bilinear form $q_\lambda$ on $K$ by using $\lambda$, 
and identifying a free rank-$[K:\Q]$ abelian subgroup of $K$ as $T$. 

Conversely, for a given even lattice $T$ of signature $(2,[K:\Q]-2)$ with 
a simple rational Hodge structure of CM-type by a CM field $K$, one can 
always find an appropriate $\lambda \in K_0^\times$ and an embedding 
$T \hookrightarrow K$ so that $q_\lambda|_T=(-,-)_T$; this can be seen 
by exploiting the property (\ref{eq:Gal-K3-intfrm}). 
So, all the CM points in $D(T)$ and their CM field $K$ should satisfy 
the $K$-and-$T$ relation in the construction of \cite{PSS-arth}. 

It therefore follows, in particular, that $D(T)$ admits a CM point with 
the CM field $K$ only when ${\rm discr}(T) \sim (-1)^{[K:\Q]/2} D_{K/\Q}$ mod 
$(\Q^\times)^2$; here, $D_{K/\Q}$ is the discriminant of the field extension 
$K/\Q$ (see \cite[\S3]{Taelman} and references therein).
} %
about how to find out all possible 
$K$'s for a given lattice $T$. For a given $T$ and $K$, CM points with the 
CM field $K$ form orbits under $\mathbb{G}{\rm Isom}(T \otimes \Q;\Q)^{[[\sigma]]}$; we do not know whether this action is transitive, or whether there is 
an action of a larger group. 

%%%%%%%%%%%%%%%%%%%%%%%%%%%%%%%%%%%%%%%%%%%%%%
\subsubsection{Bonus Symmetry}
\label{sssec:bonus}
%%%%%%%%%%%%%%%%%%%%%%%%%%%%%%%%%%%%%%%%%%%%%%

By construction, the K3 surfaces $X^{(1)}$ and $X^{(2)}$ to be used 
in the orbifold construction have certain amount of automorphisms, 
$G_1$ and $G_2$, respectively. It happens to be the case for some 
$(S, T, G; G_s, \Delta)$, though, that a K3 surface $X$ 
with a generic complex structure in $D(T)$ has ${\rm Aut}(X)$
larger than $G$. 

For example, think of $(G;G_s, \Delta) = (\Z_m; \{1\}, \Z_m)$ with an 
$m_{\varphi(m)}$ in Table \ref{tab:list-transc-indx} such that 
$\varphi(m)$ divides either one of $4, 12, 20$. 
Then $(S,T)$ can be unimodular lattices of rank $(18,4)$, $(10,12)$ 
and $(2,20)$. For a unimodular $T$, a K3 surface with a generic complex 
structure in $D(T)$ has a $\Z_2$ purely non-symplectic 
automorphism (generated by the combination of $(-1)$ multiplication on $T$ 
and id. on $S$) \cite{Kondo92}. 
This automorphism is a part of the symmetry $\Delta \cong \Z_m$
that we imposed, if $m$ is even. For an odd $m$, namely  
$m=5, 7, 11, 13, 25, 3, 21, 33, 9$, however, we have more automorphisms 
($\Z_{2m} \subset {\rm Aut}(X)$) than we imposed for orbifold construction
($G = \Delta = \Z_m$).   
See also \cite{1006.1604}, where similar enhancement of automorphism groups 
are discussed in the case $(S, T)$ are not necessarily unimodular. 
 
As another class of examples, we may think of a case of 
$(G; G_s, \Delta)$ with $G_s \neq \{0\}, \Z_2, S_3$. 
It is then known that $|{\rm Aut}_s(X^{(i)})| = \infty$ 
(see footnote \ref{fn:finite-Aut-symp} and also \cite{1010.3904}). 
So, there are more automorphisms than we impose in this class of examples. 

Those bonus automorphisms are available for any point in the period domain 
$D(T)$; this means that they act trivially on $D(T)$. The automorphisms
can be realized linearly in the effective theory (not a broken symmetry), 
because a choice of a complex structure in $D(T)$ is not shifted by the 
automorphisms. This observation also indicates that the fluctuation 
fields of complex structure within $D(T)$ are neutral under these symmetry 
transformation.\footnote{
Discussion here focuses on automorphisms available for a generic point 
in $D(T)$, once $(G;G_s, \Delta)$ and $(S, T)$ are given. 
For special loci in $D(T)$, there can be larger group of 
automorphisms. 
 
We will see in section \ref{ssec:mass-genOrbfld} 
in the case of $\Delta \cong \Z_m$ with $m>2$ that $D(T)$ moduli 
are stabilized by a $DW=W=0$ flux only in the case the vacuum 
complex structure minimizes the rank of the transcendental lattice $T_X$ 
to $\varphi(m)$. So, the question of real interest is not necessarily 
about a generic point in $D(T)$.
} %
 Because those bonus automorphisms\footnote{
Besides the bonus automorphisms $({\rm Aut}(X^{(1)}) \times {\rm Aut}(X^{(2)}))/
(G_1 \times G_2)$, there are also automorphisms $(G_1 \times G_2)/\Gamma$
acting non-trivially on the orbifold $Y_0=(X^{(1)}\times X^{(2)})/\Gamma$ by 
construction.  
Note that $\Gamma \subset G_1 \times_\Delta G_2 \subsetneq G_1 \times G_2$
(because $\Delta \neq \{ 1\}$). 
Both of the bonus automorphisms and the by-construction automorphisms 
present themselves as symmetries of the low-energy effective theory. 
} %
 act on $X^{(i)}$, and are 
non-trivial transformation on the orbifold geometry 
$Y_0 = (X^{(1)}\times X^{(2)})/\Gamma$, they are still non-trivial information 
on the effective field theory.\footnote{
Some of those bonus automorphisms (symmetries) may be broken 
by a non-trivial flux in $H^4(Y;\Q)$. 
It is the symmetry respected by the flux that matters in the low-energy 
effective theory and cosmology after inflation.   
} %
 When one considers F-theory applications 
(where the orbifold $Y_0$ and its crepant resolutions are  
replaced by a birationally equivalent fourfold $\widetilde{Y}$ and some 
K\"{a}hler parameters are brought to zero (see section \ref{sec:particle})), 
one will be interested in working out how the symmetry acts\footnote{
Choices of configuration of metric and other fields that become 
equivalent under automorphisms are regarded as one and the same point 
in the space of path integral. So, an automorphism may be regarded as 
a gauge symmetry. 

It makes sense to study non-trivial representations 
of those automorphisms (gauge symmetries) on field fluctuations instead 
of throwing away all the modes in non-trivial representations, because 
two particle excitation state can be gauge-symmetry neutral, while 
each particle is not. 
} %
 on fluctuation 
fields other than the complex structure moduli in $D(T)$. That is beyond the 
scope of this article, however.

%%%%%%%%%%%%%%%%%%%%%%%%%%%%%%%%%%%%%%%%%%%%%%%%%%%%%%%%
\subsection{Complex Structure Moduli Masses with $W=0$}
\label{ssec:mass-genOrbfld}
%%%%%%%%%%%%%%%%%%%%%%%%%%%%%%%%%%%%%%%%%%%%%%%%%%%%%%%%

For a general choice of the orbifold group 
$\Gamma \subset G_1 \times_\Delta G_2$, we do not try to say 
what the cohomology $H^4(Y;\Q)$ is like (a statement analogous 
to (\ref{eq:H4-full-inv-case}) and/or (\ref{eq:H4-Hpart-inv-case})) 
for $Y$, a minimal crepant resolution the orbifold 
$(X^{(1)}\times X^{(2)})/\Gamma$. In the cases 
$\Gamma \cong G_1 \cong G_2 \cong \Delta = \Z_m$, 
the cohomology group $H^4(Y;\Q)$ contains \cite{CR}
% 
% \footnote{
% %
% This statement is meant to be a guess (than something reliable). 
% Details on the last two terms do not matter very much in the discussion 
% in this section \ref{ssec:mass-genOrbfld}, however.
% } % 
%
\begin{align}
 [(T_0^{(1)} \otimes T_0^{(2)})\otimes \Q]^{[[\sigma]]} \oplus 
  \left( H^2(Z_{(1)} \times Z_{(2)};\Q) \right)^{\oplus (m-1)} \oplus 
  \left( \cdots \right), 
\end{align}
where $Z_{(1)}$ and $Z_{(2)}$ are curve loci of points in $X^{(1)}$ and $X^{(2)}$, 
respectively, fixed under the group $\Delta$. The last term stands for 
possible contributions from fixed loci $Z_{(1)} \times ({\rm isolated~pts})$,  
$({\rm isolated~pts}) \times Z_{(2)}$, and $({\rm isolated~pts}) \times 
({\rm isolated~pts})$ in $X^{(1)} \times X^{(2)}$. The second term has 
a Hodge structure of level 2 (for a vacuum complex structure within 
${\cal M}_{\rm cpx~str}^{[Y]BV}$), and hosts the fluctuation fields of complex 
structure deforming the $\C^2/\Z_m$ orbifold singularity.
The $H^2(Z_{(1)}\times Z_{(2)};\Q)$ component may contain 
a level-0 rational Hodge structure in some cases (see 
footnote \ref{fn:Z1xZ2-correspondence} 
in section \ref{ssec:generic-noFlux}).
Possible contributions $(\cdots )$ are also level 0. 
So, $W=DW=0$ flux are available in those level-0 components. 
Those fluxes may (or may not) give rise to the  
complex structure moduli fluctuation fields that move away 
from ${\cal M}_{\rm cpx~str}^{[Y]BV}$ into ${\cal M}_{\rm cpx~str}^{[Y]}$, 
but they do not generate a mass term or interaction term of moduli 
fluctuation fields within ${\cal M}_{\rm cpx~str}^{[Y]BV}$.  

Let us now focus on supersymmetric fluxes available within the 
$[(T_0^{(1)}\otimes T_0^{(2)})\otimes \Q]^{[[\sigma]]}$ component; only such 
fluxes can stabilize (generate mass terms) of the 
$(\ell^{(1)}-1) + ( \ell^{(2)}-1)$ moduli fluctuation fields\footnote{
$(\ell^{(i)}-2)$ instead of $\ell^{(i)}-1$ in the case of $m=2$.
} % 
 in ${\cal M}_{\rm cpx~str}^{[Y]BV}$.
In the case of $\vev{[[\sigma]]} = \Delta \cong \Z_2$, we have nothing 
to modify\footnote{
The cohomology group $H^4(Y;\Q)$ outside of the 
$[(T_0^{(1)}\otimes T_0^{(2)})\otimes \Q]^{[[\sigma]]}$ should be modified, 
when $\Gamma$ acts non-trivially on $S_0^{(i)} := 
[(T_0^{(i)})^\perp \subset H^2(X^{(i)};\Z)]$, for $i=1$ or 2.  
} %
 in the discussions in sections \ref{ssec:Tx=T0} 
and \ref{ssec:Tx-neq-T0}. In a case of $\Delta \cong \Z_m$ with $m>2$, 
let us start our discussion with an assumption that $T_X^{(i)} = T_0^{(i)}$, 
i.e., a generic CM complex structure available within 
$D(V_{a0}) \times D(V_{-a0})$. A few observations to be added to the 
discussions in sections \ref{sssec:math-preparation-tensor} 
and \ref{sssec:flux-tx-is-t0} are the following. 

First, only a proper subspace of $(T_0^{(1)}\otimes T_0^{(2)}) \otimes \Q$ 
survives the orbifold projection, as we consider a case with $m>2$ now. 
Second, the decomposition (\ref{eq:V1-V2-Chinese}) of the vector space 
$(T_0^{(1)}\otimes T_0^{(2)}) \otimes \Q$ is still useful; keeping in mind 
the fact that individual components $W_i$ in (\ref{eq:V1-V2-Chinese}) 
are in one-to-one with the orbits $\Phi_{L_i}^{\rm full}$ under the 
action of ${\rm Gal}((K^{(1)}K^{(2)})^{\rm nc}/\Q)$, and also the fact that both 
$K^{(1)}$ and $K^{(2)}$ contain a subfield $\Q([[\sigma]])$, one concludes 
that $[[\sigma]]$ acts on each one of $W_i$'s either trivially entirely or 
non-trivially entirely on that $W_i$. So, we have 
\begin{align}
  [(T_0^{(1)} \otimes T_0^{(2)})\otimes \Q]^{[[\sigma]]} \cong 
   \oplus_{i \in}^{(\Delta{\rm neut}) \subset \{1,\cdots, r\}} W_i,
  \label{eq:V1-V2-Chinese-orbfldProjtn}
\end{align}
where only the subset of $\{1,\cdots, r\}$ where $W_i$ is neutral under 
$\Delta$---denoted by $\Delta{\rm neut}$---is retained on the right 
hand side. Finally, the component with $i=(20|20)$ is in the subset 
$(\Delta{\rm neut})$, but the component $i=(20|02)$ is not.\footnote{
Note that $\zeta_m^{2a_0}=1$ and $a_0 \in [\Z_m]^\times$, only when 
$m=2$ and $a_0 = 1$. 
} %

We can review the conclusions in section \ref{sssec:flux-tx-is-t0} with 
those three observations in mind.  Now (for $m>2$), the case-A in 
page \pageref{pg:ABC-classifictn} is not logically possible. Besides 
the case-C, where there is no flux with the 
$DW=0$ condition available, the only possibility for a supersymmetric flux 
is the case-B in page \pageref{pg:ABC-classifictn}, where 
all but one components $W_i$ in (\ref{eq:V1-V2-Chinese-orbfldProjtn}) 
are level-2, and the remaining $W_{(20|20)}$ is simple and level-4. 
So, to conclude (when $m>2$ and $T_X^{(i)}=T_0^{(i)}$), a $DW=0$ flux is 
possible if and only if the condition (\ref{eq:flux-cond-math2}) is satisfied;
such a flux is in the level-4 $W_{(20|20)}$ component, so $\vev{W} \neq 0$.
There is no chance for a $DW=W=0$ flux in 
$[(T_0^{(1)}\otimes T_0^{(2)})\otimes \Q]^{[[\sigma]]}$ when $m>2$ and 
$T_X^{(i)}=T_0^{(i)}$, because the level-0 $W_{(20|02)}$ component does not 
survive the orbifold projection when $m>2$.
 
A $DW=W=0$ flux is possible within 
$[(T_0^{(1)} \otimes T_0^{(2)})]^{[[\sigma]]}$ if and only if 
$T_x^{(i)}\subsetneq T_0^{(i)}$ for both $i=1,2$; it is not enough to have 
$T_X^{(i)}\subsetneq T_0^{(i)}$ for just one of $i=1,2$. To see this, 
remember that 
\begin{align}
& [(T_0^{(1)} \otimes T_0^{(2)})]^{[[\sigma]]} \\
\cong &\;  
 [(T_X^{(1)} \otimes T_X^{(2)})]^{[[\sigma]]} \oplus
 [(T_X^{(1)} \otimes \overline{T}_0^{(2)})]^{[[\sigma]]} \oplus 
 [(\overline{T}_0^{(1)} \otimes T_X^{(2)})]^{[[\sigma]]} \oplus 
  [(\overline{T}_0^{(1)} \otimes \overline{T}_0^{(2)})]^{[[\sigma]]};  \nonumber 
\end{align}
the middle two components on the right-hand side are purely level-2, and 
the first component consists only of level-2 and level-4 Hodge components; 
the latter statement is obtained by repeating the argument above 
(for $T_0^{(1)} \otimes T_0^{(2)}$ with $T_X^{(i)}=T_0^{(i)}$ there). 
So, a $DW=W=0$ flux can only be in the last component.
Such a flux cannot generate a mass term for the moduli field 
fluctuations in $D(T_X^{(1)}) \cap D(V_{a0})$ and $D(T_X^{(2)})\cap D(V_{-a_0})$, 
however. 

Therefore, the only possibility for a $DW=W=0$ flux stabilizing all the 
complex structure moduli, if $m > 2$, is when ${\rm rank}(T_X^{(1)}) 
= {\rm rank}(T_X^{(2)}) = \varphi(m)$ so that there is no moduli within 
$[(T_X^{(1)}\otimes T_X^{(2)})\otimes \C]^{[[\sigma]]}$. 
The condition (\ref{eq:flux-cond-math2}) is satisfied automatically then  
($K^{(1)} \cong K^{(2)} \cong \Q(\zeta_m)$), but there is no $W_{(20|02)}$ 
component to support a $DW=W=0$ flux when $m>2$.  
The $(\ell^{(1)}-1) + (\ell^{(2)}-1)$ moduli field fluctuations have Dirac 
type mass terms from a flux in the $(\overline{T}_0^{(1)} \otimes 
\overline{T}_0^{(2)})\otimes \Q$ component.\footnote{
One may also notice (when $m  > 2$) that the third term in the expansion 
in (\ref{eq:k3-cpx-str-parametrize-cm}) vanishes. So, in the expression 
for the K\"{a}hler potential (\ref{eq:moduli-eff-Kahler}), 
the third term in the second line vanishes.  
} % 
So, for all those moduli fields to have masses, $\ell^{(1)} = \ell^{(2)}$ 
is also necessary, just like the condition (\ref{eq:flux-cond-math3}) 
in section \ref{ssec:Tx-neq-T0}. 
Just like in sections \ref{sssec:mass-TX=T0} and \ref{ssec:Tx-neq-T0}, 
this moduli effective field theory has an approximate 
$\U(1) \times \U(1)_R$ symmetry. 

%%%%%%%%%%%%%%%%%%%%%%%%%%%%%%%%%%%%%%%%%%%%%%%%%%%%%%%%%%
\section{F-theory Applications and Particle Physics Aspects}
\label{sec:particle}
%%%%%%%%%%%%%%%%%%%%%%%%%%%%%%%%%%%%%%%%%%%%%%%%%%%%%%%%%%

In the earlier sections, we have discussed the supersymmetry 
conditions (\ref{eq:cond-DW=0}, \ref{eq:cond-W=0}) of fluxes on 
CM-type Borcea--Voisin Calabi--Yau fourfolds $Y=(X^{(1)}\times X^{(2)})/\Z_2$, 
and also stabilization of the complex structure moduli. 
The analysis in sections \ref{sec:flux-analysis} 
and \ref{sec:non-Z2-orbifold} 
can be read in the context of M-theory compactification on such fourfolds 
down to $(2+1)$-dimensions; the orbifold geometry 
$Y_0=(X^{(1)}\times X^{(2)})/\Z_2$ is singular, but the study in 
sections \ref{sec:flux-analysis} and \ref{sec:non-Z2-orbifold} in that 
M-theory context should be read\footnote{
For example, eq. (\ref{eq:H4-full-inv-case}) is justified for smooth 
manifolds $Y=Y^{BV}$. 
} %
 as that for a fourfold  
$Y^{BV}$ which is the minimal and crepant resolution of $Y_0$, with 
positive values of K\"{a}hler parameters for the exceptional cycles. 
 
To think of an F-theory compactification down to $(3+1)$-dimensions, 
however, we need a Calabi--Yau fourfold $\widetilde{Y}$ that has 
a flat\footnote{
This condition is for absence of an exotic particle spectrum on $\R^{3,1}$.
So, this is a phenomenological constraint.
} % 
elliptic fibration.\footnote{
We also need to take the limit in the K\"{a}hler moduli so that the 
volume of the elliptic fibre vanishes, and to keep some part of the 
purely vertical part of $H^4(Y;\Q)$ free from fluxes \cite{DRS}, 
in order to restore the $\SO(3,1)$ symmetry.  
} %
When F-theory is compactified on $\widetilde{Y}$ such that 
${\cal M}^{[\widetilde{Y}]}_{\rm cpx~str}$ is contained in 
${\cal M}^{[Y]BV}_{\rm cpx~str}$, the analysis for presence of a non-trivial 
supersymmetric flux and stabilization 
of moduli in ${\cal M}^{[\widetilde{Y}]}_{\rm cpx~str}$ is still valid. 

In a large fraction of this section, we will be concerned about 
when and how one can find $\widetilde{Y}$ birational to $Y^{BV}$. 
When a $\widetilde{Y}$ is available, its geometry should determine 
gauge groups and possible matter representations in the effective theory 
on $(3+1)$-dimensions, motivated by $\vev{W} \simeq 0$. 
We will take steps to read out those implications. 

%%%%%%%%%%%%%%%%%%%%%%%%%%%%%%%%%%%%%%%%%%%%%%%%%%%%%%%%%
\subsection{Elliptic Fibred K3 Surface with a Non-symplectic Involution}
\label{ssec:K3-Weierstrass}
%%%%%%%%%%%%%%%%%%%%%%%%%%%%%%%%%%%%%%%%%%%%%%%%%%%%%%%%%

One of the technical problems that we face in the context of F-theory 
compactification is to find, for a given Calabi--Yau variety $Y$ for 
an M-theory compactification, a set of $(Y, B, \pi)$, where 
$\pi:Y \rightarrow B$ is a flat elliptic fibration and $B$ a base manifold. 
This is much easier in a lower dimensional set-up; the classification 
problem has a long history in the case of $\dim_\C Y=2$.
Nearly a complete classification of $(Y=X^{(1)}, B=\P^1_{(1)}, \pi_{X^{(1)}}^f)$
is available (\cite{Cmp-Garbag, 1806.03097} and references therein) for 
K3 surfaces $X^{(1)}$ associated with the 75 choices of 
$(S_0^{(1)}, T_0^{(1)}, \sigma_{(1)})$ of Nikulin, as we will review 
briefly in this section \ref{ssec:K3-Weierstrass}.
Such an elliptic fibration 
$(X^{(1)}, \P^1_{(1)}, \pi^{[f]}_{X^{(1)}})$ is used in 
sections \ref{ssec:on-fibre} 
and \ref{ssec:on-base} to construct a fourfold $\widetilde{Y}$ birational to 
a Borcea--Voisin orbifold $Y_0 = (X^{(1)}\times X^{(2)})/\Z_2$ where there 
is a flat elliptic fibration morphism $\pi: \widetilde{Y} \rightarrow B_3$.
%
% one approach is to use lattice theory 
% (e.g., \cite{P-SS, Nik-lattice, Nishiyama} and also in this article) 
% and another is to construct toric fibration morphism from a toric ambient 
% space of $Y$ \cite{Gray}. 
%

We begin with recalling known facts about how we find elliptic fibration 
morphisms from an algebraic K3 surface to $\P^1$. 
There exists a genus-one curve fibration\footnote{
In a {\bf genus-one curve fibration}, the fibre of a generic 
point in the base is a curve of genus one; a section $s: \P^1 
\rightarrow X^{(1)}$ that covers the base just once does not necessarily 
exist. In an {\bf elliptic fibration}, we require that such a section 
exists; the image of a section is often denoted by $s$ (by abusing 
notation). 
This is to follow the terminology in F-theory community. 
Genus-one curve fibrations here correspond to {\bf elliptic fibrations} 
in math literatures, and elliptic fibrations here to {\bf Jacobian elliptic 
fibrations} there. In this article, we stick to the terminology of 
F-theory community.
} % 
morphism from a K3 surface $X^{(1)}$ 
to $\P^1$ if and only if there exists a divisor class $[f] \in S_{X}^{(1)}$
with $[f]^2 = 0$ \cite{P-SS}. The corresponding fibration morphism is 
denoted by $\pi_{X^{(1)}}^{[f]}: X^{(1)} \rightarrow \P^1_{(1)}$. For a genus-one 
fibration morphism $\pi_{X^{(1)}}^{[f]}: X^{(1)} \rightarrow \P^1_{(1)}$ to be an 
elliptic fibration morphism,\footnote{
In this article, we consider only F-theory compactifications down to 
(3+1)-dimensional space-time, by taking the limit of the K\"{a}hler moduli 
of the fibre elliptic curve. In this context, a fourfold $Y_1$ with a 
genus-one fibration over a threefold $B$ yields the same effective theory 
on 3+1-dimensions as a fourfold $\widetilde{Y}$ with an elliptic fibration 
over $B$, when $\widetilde{Y}$ is the Jacobian fibration of $Y_1$. 
For this reason, it is fine to restrict our attention only to fourfolds 
with elliptic fibration morphisms; it should be remembered however that 
$\widetilde{Y}$ cannot necessarily be made non-singular and Calabi--Yau 
even when $Y_1$ is \cite{BM, MT}. So, one should be careful about 
what kind of singularity still leads to sensible physics when one deals 
exclusively with fourfolds with elliptic fibrations \cite{singlrty}. 

In this article, however, we do not try to explore that borderline, 
and restrict our attention only to F-theory compactifications that can 
be associated with non-singular Calabi--Yaus $\widetilde{Y}$ with elliptic 
fibration morphisms. 
} %
 there must exist another divisor class 
$[s] \in S_X^{(1)}$ satisfying $(s, f) = +1$ and 
$(s, s)= -2$. The primitive sublattice generated by $[f]$ and 
$[s]$ within $S_X^{(1)}$ is isomorphic to $U$ then. 
To repeat, existence of an elliptic fibration is equivalent to existence 
of a factor $U$ in $S_X^{(1)}$.  

In the context of F-theory applications, when we write 
$S_X^{(1)} = U \oplus W$, the lattice $W$ contains the information of 
non-Abelian 7-brane gauge groups, the number of U(1) gauge fields, 
and also the spectrum of charges under those gauge groups in 7+1-dimensions. 
So, a well-motivated classification of elliptic fibration morphisms of 
$X^{(1)}$ is equivalent to classifying\footnote{
A review addressed to string theorists is found in \cite{BKW-math}. 
} %
 primitive embeddings of $U$ 
into $S_X^{(1)}$ modulo isometry of the lattice $S_X^{(1)}$.   
One and the same K3 surface $X^{(1)}$ (with a common $S_X^{(1)}$ 
and $T_X^{(1)}$) may have multiple different types of elliptic fibration 
morphisms in this classification; one of the most famous examples is 
the case $S_X = U \oplus E_8^{\oplus 2} \cong 
{\rm II}_{1,17} \cong U \oplus (D_{16};\Z_2)$. 
An F-theory limit takes the volume of a fibre elliptic curve class 
$[f] \in U$ to zero, so different choices of $U \subset S_X^{(1)}$ 
correspond to different F-theory limits.

In general, $S_0^{(1)} \subset S_X^{(1)}$; when the complex structure 
of the K3 surface $X^{(1)}$ is not fully generic in the period domain 
of the lattice $T_0^{(1)}$, $S_X^{(1)}$ is strictly larger than $S_0^{(1)}$
(and $T_X^{(1)}$ is strictly smaller than $T_0^{(1)}$).
Although it is enough to find a factor $U$ within $S_X^{(1)}$ in constructing 
an elliptic fibration $\pi_{X^{(1)}}: X^{(1)} \rightarrow \P^1_{(1)}$, we wish 
to use the elliptic fibration morphism to construct an elliptic fibration 
$\pi_Y: Y \rightarrow B_3$ with some threefold $B_3$ (which is to be 
constructed in the following). 
So, we need to be concerned how the elliptic fibration morphism 
$\pi_{X^{(1)}}: X^{(1)} \rightarrow \P^1_{(1)}$ behaves under the generator 
$\sigma$ of the $\Z_2$ orbifold. We stick to the simplest case 
where the $U$ sublattice is within $S_0^{(1)} \subset S_X^{(1)}$, 
which means that 
\begin{align}
  \sigma_{(1)}^* : [f] \mapsto [f], \qquad
  \sigma_{(1)}^*: [s] \mapsto [s].
\end{align}

There are two types in the way the involution $\sigma_{(1)}$ acts 
on a K3 surface with elliptic fibration $(X^{(1)}, \P^1_{(1)}, \pi_{X^{(1)}}^f)$
\cite[Prop. 2.3]{1806.03097}. It always maps the zero section $s$ to itself, 
but it may be either an identity $\sigma_{(1)}|_s = {\rm id}_{s}$ 
(Type 1 (referred to 
as type b in \cite{Cmp-Garbag})), or a non-trivial holomorphic involution 
(Type 2 (referred to as type a in \cite{Cmp-Garbag})). 
An involution of Type 1 acts on individual fibre elliptic curves, 
while an involution of Type 2 exchanges two fibre curves (except 
the fibres over the two $\sigma_{(1)}|_s$-fixed points in the base $\P^1_{(1)}$).
%
% Regardless of which one is the case for a given choice of 
% $U \subset S_0^{(1)}$,
% %
% \begin{align}
%    \pi_{X^{(1)}}^{[f]} \circ \sigma_{(1)} = \sigma_{(1)}|_s \circ \pi_{X^{(1)}}^{[f]}:
%     X^{(1)} \rightarrow \P^1_{(1)}.
%  \label{eq:ellK3-invl-comm}
% \end{align}
%

Let us take a few examples from the 75 choices of 
$(S_0^{(1)}, T_0^{(1)}, \sigma_{(1)})$ of \cite{Nikulin-factor-long}. 
For $S_0^{(1)} = U[2]E_8[2]$, there is no primitive embedding of $U$ 
into $S_0^{(1)}$, so there is no elliptic 
fibration \cite[Thm. 2.6.(i)]{1806.03097}. 
For $S_0^{(1)}=U$, there is unique elliptic fibration, which is Type 1. 
Think of the case $S_0^{(1)} = UE_8[2]$, next. An obvious primitive embedding 
of $U$ into $S_0^{(1)}$ corresponds to an elliptic fibration of Type 2; 
this embedding is actually the only one available for this $S_0^{(1)}$  
\cite[Thm. 2.6.(ii)]{1806.03097}.
For the choice $(S_0^{(1)},T_0^{(1)})$ for $T_0 = U[2]^{\oplus 2}$, which is for 
$X^{(1)} = {\rm Km}(E \times E')$ for mutually non-isogenous elliptic curves 
$E$ and $E'$, there are 11 different elliptic fibrations (modulo 
${\rm Isom}(S_0^{(1)})$) \cite{Og-Km}; three out of the 11 elliptic fibrations 
(${\cal J}_{1,2,3}$ in \cite{Og-Km}) are Type 2, and the remaining eight 
(${\cal J}_{4,\cdots,11}$) are Type 1.
In the study of \cite{Cmp-Garbag, 1806.03097}, it turns out 
that more than 60 choices out of the 75 in \cite{Nikulin-factor-long}
admit at least one elliptic fibration; choices with 
larger [resp. smaller] $g_{(1)} = (22-r_{(1)}-a_{(1)})/2$ tend to have 
less [resp. more] inequivalent primitive embeddings 
$U \hookrightarrow S_0^{(1)}$ and inequivalent elliptic fibrations 
consequently. A pair $(\pi_{X^{(1)}}^f, \sigma_{(1)})$ of Type 2 is rare relatively 
to one of Type 1, and is possible only for the choices with $g_{(1)} \leq 1$. 
For more information, see \cite{Cmp-Garbag, 1806.03097} and references therein. 

%%%%%%%%%%%%%%%%%%%%%%%%%%%%%%%%%%%%%%%%%%%%%%%%%%%%%%
\subsection{Borcea--Voisin Manifold and Weierstrass Model}
%%%%%%%%%%%%%%%%%%%%%%%%%%%%%%%%%%%%%%%%%%%%%%%%%%%%%%

For an F-theory compactification, we need a Calabi--Yau fourfold $Y$ 
that has an elliptic fibration $\pi: Y \rightarrow B_3$ and its section
$\sigma: B_3 \rightarrow Y$. It is not obvious in F-theory (due to the lack 
of its theoretical formulation) which one of $Y$ and $Y'$ should be regarded 
as input data of compactification, when there is a birational 
pair of Calabi--Yau varieties $Y$ and $Y'$ with no difference in 
cycles of finite volume or the number of complex structure deformation 
parameters.
This article deals with F-theory compactification on such an equivalence 
class\footnote{
Although we attempted to write down the equivalence relation explicitly 
above, the choice of the relations may have to be refined or corrected 
from the version written there. 
} %
 of Calabi--Yau fourfolds that is represented by a non-singular model 
$\widetilde{Y}$ with a flat elliptic fibration, $\widetilde{Y}\rightarrow 
B_3$.
%
% , if the effective field theory on $\R^{3,1}$ is to be free from 
% exotics (such as quantized tensionless strings).  
%
Although the Borcea--Voisin manifold $Y^{BV}$---the minimal resolution of 
the Borcea--Voisin orbifold $Y_0 = (X^{(1)}\times X^{(2)})/\Z_2$---is 
non-singular, it is hard, or even seems to be impossible for some 
choices\footnote{
\label{fn:F-geometry-1}
Suppose that a singular fibre of $\pi_{X^{(1)}}^f: X^{(1)}\rightarrow \P_{(1)}^1$ 
contains both an irreducible component in $Z_{(1)}$ and also $\P^1$ 
not in $Z_{(1)}$------(**). Then the fibration 
${\rm Bl}_{\sigma{\rm -fixed}}(X^{(1)}\times X^{(2)}) \rightarrow \P^1_{(1)}$ 
is not flat. 
Apart from the choice of $S_0^{(1)}=UE_8[2]$, which has an elliptic fibration 
of Type 2, all other Type 2 elliptic fibrations available in K3 surfaces 
with a non-symplectic involutions fall into the category (**). 
Elliptic fibrations of Type 1 that stay out of the category (**) 
is when $S_0^{(1)}=U \oplus W_0$, with $W_0$ containing only $A_1$'s and 
the Mordell--Weil group, but no other $ADE$-type lattices. Such 
$S_0^{(1)} = U \oplus W_0$ constitutes a small fraction of the tables 
in \cite{1806.03097}.    
} %
 of $(S_0^{(1)}, T_0^{(1)}, [f])$, to find a 
flat elliptic fibration on $Y^{\text{BV}}$. So, for F-theory applications, 
let us find $\widetilde{Y}$ that is birational
% 
% \footnote{
% \label{fn:F-geometry-1}
% %
% The fourfold $\widetilde{Y}$ being birational to $Y^{BV}$ is a sufficient 
% condition for the phycisal moduli space of the F-theory compactification 
% on $\widetilde{Y}$ to contain $D(T_X^{(1)}) \times D(T_X^{(2)})$, but 
% we do not have a proof that it is necessary.
% } %
% 
 to $Y^{BV}$, along with
a threefold $B_3$ so that there is a flat elliptic fibration 
$\widetilde{Y} \rightarrow B_3$.

We will find such $\widetilde{Y}$ and $B_3$ in 
sections \ref{ssec:on-fibre} and \ref{ssec:on-base} as a resolution of 
a Weierstrass 
model fourfold $Y^W$; see (\ref{eq:birat-net-invF}) 
and (\ref{eq:birat-net-invB}). 
As a first step for that purpose, consider an orbifold\footnote{
\label{fn:F-geometry-2}
It is likely that the constructions of $(\widetilde{Y}, B_3, \pi)$ starting 
from here are not the most general ones with a moduli space containing 
$D(T_X^{(1)}) \times D(T_X^{(2)})$. The authors are not yet ready to 
write down a broader class of constructions, however. 
} % 
$Y_0^W = (X^{(1)W} \times X^{(2)})/\Z_2$. $X^{(1)W}$ is the Weierstrass model 
of a non-singular K3 surface $X^{(1)}$, which is obtained from 
$(X^{(1)}, \P^1_{(1)}, \pi_{X^{(1)}}^{[f]})$ discussed in 
section \ref{ssec:K3-Weierstrass} by collapsing $(-2)$-curves in the 
singular fibres of $\pi_{X^{(1)}}^{[f]}$ except those that intersect the 
section $s$ of $\pi_{X^{(1)}}^{[f]}$. The $\Z_2$ quotient is by 
$(\sigma_{(1)W}, \sigma_{(2)})$, where $\sigma_{(1)W}$ is described below. 

A K3 surface $X^{(1)}$ of interest in this article is in a family 
parameterized by the (CM points in the) period domain $D(T_0^{(1)})$ 
characterized by the pair $(S_0^{(1)}, T_0^{(1)})$, where $\sigma_{(1)}$ acts 
identically on $S_0^{(1)}$ and by $[(-1)\times]$ on $T_0^{(1)}$, 
as a reminder. Its Weierstrass model $X^{(1)W}$, however, is regarded 
as $X^{(1)}$ with $S_0^{(1)W}=U$ in the Type 1 case, and the period domain 
$D(T_0^{(1)})$ as a special subspace in $\overline{D(T_0^{(1)W})}$; 
$T_0^{(1)W}=U^{\oplus 2}E_8^{\oplus 2}$ now.     
The involution $\sigma_{(1)W}$ on $X^{(1)W}$ is that of $X^{(1)}$ with\footnote{
Complex structure can be tuned continuously from the bulk of $D(T_0^{(1)W})$
to $D(T_0^{(1)})$, but the process of blowing-up singularity of $X^{(1)W}$
(or collapsing $(-2)$-curves of $X^{(1)}$ in the other way around) 
is not a continuous process; 
the involution on the cohomology group $H^2(K3;\Z)$ changes in this 
discontinuous process. 
} % 
$S_0^{(1)}=U$, which multiplies $(-1)$ to the $y$ coordinate of 
the Weierstrass equation $y^2 = x^3 + f(t)x + g(t)$. 

In the Type 2 case, its Weierstrass model $X^{(1)W}$ is regarded as 
$X^{(1)}$ with $S_0^{(1)W}=U\oplus E_8[2]$, and the period domain 
$D(T_0^{(1)})$ as a special subspace in $\overline{D(T_0^{(1)W})}$;
$T_0^{(1)W}=U^{\oplus 2}\oplus E_8[2]$ now. The involution $\sigma_{(1)W}$ on 
$X^{(1)W}$ is that of $X^{(1)}$ with $S_0^{(1)}=U\oplus E_8[2]$, which multiplies 
$(-1)$ to the inhomogeneous coordinate $t$ of the base $\P^1_{(1)}$, where 
the Weierstrass equation is 
$y^2 = x^3 + f(t^2)x + g(t^2)$  \cite{Cmp-Garbag, 1806.03097}. 

The orbifold $Y_0^W$ is now well-defined; we claim now that 
there is a regular map $Y_0 \rightarrow Y_0^W$, and that this map 
is birational. 
To see that they are birational, note first that 
there is a field isomorphism $\C(X^{(1)W}) \C(X^{(2)}) \cong 
\C(X^{(1)})\C(X^{(2)})$ because of the birationality between $X^{(1)}$ and 
$X^{(1)W}$. The action of $(\sigma_{(1)W},\sigma_{(2)})$ on the left and that of 
$(\sigma_{(1)},\sigma_{(2)})$ on the right are compatible with this field 
isomorphism, so we have 
\begin{align}
  \C(Y_0^W) \cong \left[ \C(X^{(1)W}) \C(X^{(2)}) \right]^{\Z_2} \cong 
   \left[ \C(X^{(1)}) \C(X^{(2)}) \right]^{\Z_2} \cong \C(Y_0).
\end{align}
So, they are birational indeed. 
The regularity of the map $Y_0 \rightarrow Y_0^W$ follows from 
\begin{align}
   \left( \C[U_i] \C[V] \right)^{\Z_2} \hookrightarrow 
   \left( \C[X^{(1)}_i] \C[V] \right)^{\Z_2}
\end{align}
for open patches $V$ of $X^{(2)}$; here, $U_i$'s are open patches of 
$X^{(1)W}$ and $X_i^{(1)}$'s those of $X^{(1)}$ so that $X_i^{(1)}$'s are 
mapped to $U_i$'s under the regular map $X^{(1)} \rightarrow X^{(1)W}$.  

Construction of $Y_0^W$ from $Y^{BV}$ or from $Y_0$ is essentially the same 
for both Type 1 and Type 2. 
From this point on, however, we need to deal with the Type 1 and 2 cases
separately 
in the construction of a Weierstrass-model fourfold $Y^W$ and 
a non-singular model $\widetilde{Y}$ with a flat fibration.

%%%%%%%%%%%%%%%%%%%%%%%%%%%%%%%%%%%%%%%%%%%%%%%%%%%%%%
\subsection{Fibration and Involution of Type 1}
\label{ssec:on-fibre}
%%%%%%%%%%%%%%%%%%%%%%%%%%%%%%%%%%%%%%%%%%%%%%%%%%%%%%

\subsubsection{Construction of $\widetilde{Y}$, and Gauge Group and 
Matter Representations}

In the case of a pair of fibration and involution of Type 1, 
a Weierstrass model $Y^W$ is obtained by once blowing up 
$Y_0^W$ ($Y^{W'} \rightarrow Y_0^W$), and then blowing it down  
($Y^{W'} \rightarrow Y^W$), as we elaborate a bit more in the following. 

\begin{figure}[tbp]
 \begin{center}
  \begin{tabular}{ccccccc}
   \includegraphics[width=.1\linewidth]{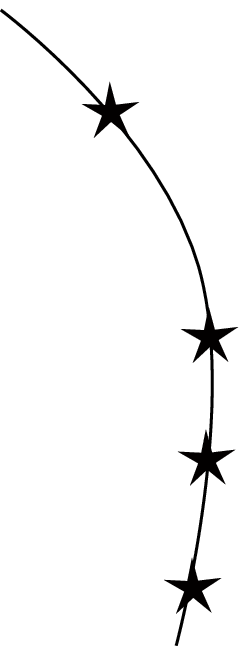}  & $\qquad$ & 
   \includegraphics[width=.12\linewidth]{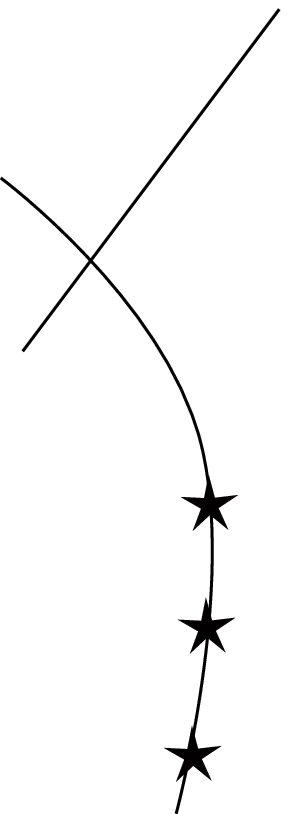} & $\qquad$ & 
   \includegraphics[width=.14\linewidth]{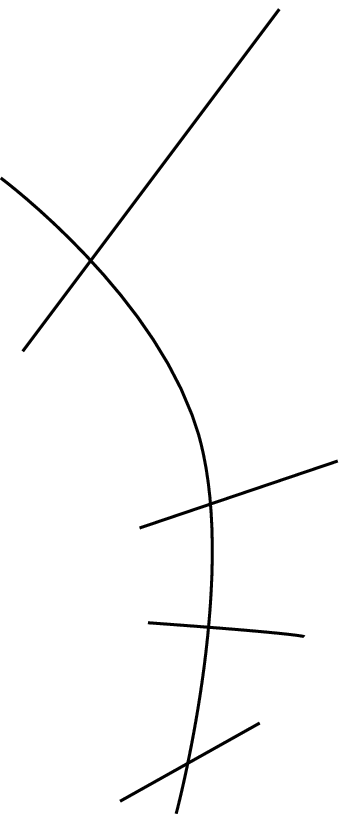} & $\qquad$ &
   \includegraphics[width=.1\linewidth]{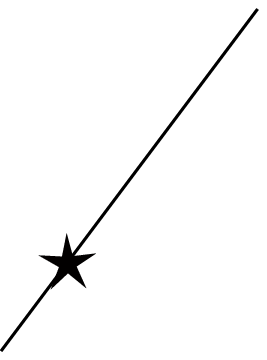} \\
   (a) & & (b) & & (c) & & (d)  
  \end{tabular}
  \caption{\label{fig:invF-singl-fib}
Schematic picture of the singular fibre geometry for a generic point 
in $[Z_{(2)}]$ in (a) $Y_0^W$, (b) $Y^{W'}$, (c) $Y^{W''}$, and (d) $Y^W$. }
\end{center}
\end{figure}

Construction of $Y^{W'}$ from $Y_0^W$ is as follows. 
The $\Z_2$-orbifold $Y_0^W$ has a two-dimensional locus of singularity
that is $A_1$-type for each isolated component of $[Z_{(2)}] \subset B^{(2)}$. 
The two transverse 
directions are the transverse direction of $[Z_{(2)}]$ in $B^{(2)}$ and 
also the elliptic fibre direction.  For a generic point in 
$[Z_{(2)}] \subset B^{(2)}$, the locus of $A_1$-singularity consists 
of two pieces of curves, one of which is a three-fold cover over $\P^1_{(1)}$
and the other a one-fold cover. The proper transform of $Y_0^W$ in a blow-up 
centred along the latter singular locus (the one-fold covering one) 
is denoted by $Y^{W'}$; one may also think of the blow-up along both of the 
singular loci, where the proper transform is denoted by $Y^{W''}$. 
See (\ref{eq:birat-net-invF}) and Fig. \ref{fig:invF-singl-fib}. 

The Weierstrass model $Y^W$ is obtained from $Y^{W'}$ by collapsing  
the divisors over $[Z_{(2)}] \times \P^1_{(1)}$ that are non-exceptional 
in the blow-up $Y^{W'} \rightarrow Y_0^W$ (see Fig. \ref{fig:invF-singl-fib}). 
This variety $Y^W$ has a projection 
$\pi: Y^W \rightarrow B_w := (\P^1_{(1)} \times B^{(2)})$, and is given by 
\begin{align}
  \tilde{y}^2 = \tilde{x}^3 + V^2 f(t) \tilde{x} + V^3 g(t)
      \label{eq:weierstrass_eq_fiber}
\end{align}
in one of its Affine patch. 
%%%%%%%%%%%%%%%%%%%%%%%%%%%%%%%%%%%%%%%%%%%%%%%%%%%%%%%%%
%%%%%%%%%%%%%%%%%%%%%%%%%%%%%%%%%%%%%%%%%%%%%%%%%%%%%%%%%
The Affine coordinates 
$(\tilde{x},\tilde{y},t,V,u)$ of $Y^W$ are related with the coordinates 
$(x,y,t,v,u)$ of $X^{(1)W}$ and $X^{(2)}$ through 
\begin{align}
  t = t, \quad u = u, \quad V = v^2, \quad
    \tilde{x} = x v^2, \quad \tilde{y} = y v^3.
\end{align}
Remember that the involution $\sigma_{(1)W}$ acts trivially on $t$, $u$, 
and $x$, and by $[(-1)\times]$ on $y$ and $v$. 
Here is a summary (all the arrows between $Y$'s are regular and birational):
\begin{align}
 \xymatrix{
    Y^{BV} \ar[d] & & Y^{W''} \ar[d] & & & & \widetilde{Y} \ar[d] \\
    Y_0 \ar[r] & Y_0^W & Y^{W'} \ar[l] \ar[r] & Y^W \ar[d]
         & \nu^*(Y^W) \ar[l] \ar[d] & \overline{\nu^*(Y^{W})} \ar[l] \ar[r] & 
       Y \ar[dll] \\
       & & & B_w & B_3 \ar[l]
  }
  \label{eq:birat-net-invF}
\end{align}
See the following discussions for 
$B_3 = {\rm Bl}_{{\rm pt}_*\times [Z_{(2)}]}(B_w)$, 
$\nu^*(Y^W)$, $Y$, and $\widetilde{Y}$.

So long as complex structure of $X^{(1)}$ is that of a generic one in 
$D(T_0^{(1)W}) = D({\rm II}_{2,18})$, which means that $S_X^{(1)} = S_0^{(1)} = U$,
there is no difference between $Y_0$ and $Y_0^W$; $Y^{W''}$ is nothing 
but $Y^{BV}$; the projection $\pi: Y^{W''} = Y^{BV} \rightarrow 
(\P^1_{(1)}\times B^{(2)})$ yields a flat elliptic fibration, so 
$\widetilde{Y} =Y^{W''}$ and $B_3 = B_w$.   
The discriminant locus $\Delta_{\rm discr}$ of the elliptic fibration 
$Y^W \rightarrow B_w$ is of the form 
\begin{align}
\Delta_{\rm discr} = \Delta_f + \Delta_b, \qquad 
\Delta_f = (24{\rm pts}) \times B^{(2)}, \quad {\rm and} \quad
\Delta_b = 6(\P^1_{(1)} \times [Z_{(2)}]). 
\end{align}
On a generic point in $\Delta_b$, the singular fibre in 
$\widetilde{Y}=Y^{W''}$ is the $I_0^*$-type in the Kodaira classification
\cite{CY4BV-invF}, and the equation (\ref{eq:weierstrass_eq_fiber}) is 
completely in the non-split type over $\P^1_{(1)}$ (and also 
over $\P^1_{(1)}\times [Z_{(2)}]$) \cite{6auth}. 
So, the ${\cal N}=1$ supersymmetric effective theory on $\R^{3,1}$ has 
one vector multiplet with the gauge group $G_2$ for each one of the 
isolated components\footnote{
There are $k_{2}+1$ isolated components for 
all the (75-2) choices of $(S_0^{(2)}, T_0^{(2)})$ from Nikulin's list.
} % 
of $[Z_{(2)}]$. The 7-branes $\Delta_f$ do not yield a massless vector 
multiplet on the effective theory on $\R^{3,1}$. 
 
There may be massless ${\cal N}=1$ chiral multiplets (matter fields) charged 
under those $G_2$ gauge groups, possibly in the adjoint representation, 
and also possibly in the ${\bf 7}$-representation, because matter 
hypermultiplets in those two representations can be present in 
F-theory compactifications to 5+1-dimensions \cite{6auth, hep-th/0002012}.
None of them must be charged under multiple $G_2$'s, because all the 
irreducible components of $[Z_{(2)}] \subset B^{(2)}$ are disjoint from 
each other \cite{Nikulin-factor-long}. 
All those matter fields are in self-real 
representations, so there is no such things as a formula for the net 
chirality. Although the $(20-r_{(1)})+(20-r_{(2)}) = 38-r_{(2)}$ complex 
structure moduli of $Y^{W''}=Y^{BV}$ remain to be gauge-group neutral 
moduli chiral multiplets in the effective theory on $\R^{3,1}$, 
the $g_{(1)}g_{(2)}=10g_{(2)}$ complex structure moduli of $Y^{BV}$ are 
likely to be part of $G_2$-charged matter chiral multiplets; solid evidence 
for this statement can be provided by studying F-theory compactification 
on a threefold $M^{BV}$ as the crepant resolution of 
$(X^{(1)} \times E_\tau)/\Z_2$. 

Now, let us turn to cases where $X^{(1)}$ and $X^{(1)W}$ are not isomorphic.   
In terms of the lattice, let $S_X^{(1)} =: U \oplus W$, and $R$ denote the 
sublattice of $W$ generated by the norm-$(-2)$ elements of $W$; 
$X^{(1)}$ and $X^{(1)W}$ are not mutually isomorphic if and only if 
$R$ is non-empty. $Y^{BV}$ and $Y^{W''}$ are not mutually isomorphic either 
in such cases. A flat elliptic 
fibration
%
% \footnote{
% %
% A projection morphism $Y^{BV} \rightarrow B_w$ can be well-defined, 
% and is an elliptic fibration in an invF case, but it is not flat when 
% $S_0^{(1)}$ is something other than $U$. 
% } % 
%
$(\widetilde{Y}, B_3, \pi)$ is constructed as reviewed below 
by starting from $Y^W \rightarrow B_w$, or from $Y^{W''}\rightarrow B_w$.  

Suppose first that the lattice $R$ contains only $A_n$'s, not $D_n$'s or 
$E_n$'s. It is then known that we can take $B_w$ as $B_3$; $Y^{W''}$ has 
singularity of $A_n$ type 
over the 7-brane $(n+1)({\rm pt}_* \times B^{(2)}) \subset \Delta_f$, 
so those codimension-2 singularity is resolved canonically; 
after a small resolution, a non-singular Calabi--Yau fourfold 
$\widetilde{Y}$ is obtained in this case \cite{Miranda, collision}.\footnote{
\label{fn:F-geometry-3}
In the construction of $\widetilde{Y}$ in the main text, we consider 
choosing a complex structure of $X^{(1)}$ from $D(T_0^{(1)}) = D({\rm II}_{2,18})$
in such a way that $S_X^{(1)}$ is enhanced from $S_0^{(1)}=U$ to 
$U\oplus W$ with $W$ containing $A_n$'s. 
When we consider a complex structure so that $S_0^{(1)}=U\oplus W_0$, $W_0 = W$,
and $R \subset W_0$ contains only $A_1$'s (cf footnote \ref{fn:F-geometry-1}), 
however, one may think of another construction of $(Y,B_3,\pi)$. 
That is to choose $Y^{BV}$ as the fourfold, 
and $B_3 = \P^1_{(1)}\times B^{(2)}$; this is a flat elliptic fibration
\cite[Prop. 3.1]{CY4BV-invF} in such a case. It is a question of interest 
whether $Y^{BV}$ is isomorphic to $\widetilde{Y}$ in the main text 
and whether the matter spectra are the same or not.
} %
The gauge group on the 7-brane ${\rm pt}_* \times B^{(2)}$ becomes 
${\rm Sp}$-type in the effective theory on $\R^{3,1}$ (and a product of $G_2$'s 
is from $\P^1_{(1)} \times [Z_{(2)}]$ as before).\footnote{
This gauge group $G_2$ in the $\R^{3,1}$ effective theory is enhanced to 
$\SO(7)$ or $\SO(8)$, for example, when the cubic polynomial $x^3+f(t)x + g(t)$ 
is factorized into a product of a pair of linear and quadratic polynomials, 
or of three linear polynomials. 
} % 
The matter fields must be in the bifundamental representation
of $G_2$ and ${\rm Sp}$ \cite{KV}, besides those in the adjoint 
representations, $G_2$-${\bf 7}$, and the ${\rm Sp}$ rank-2 antisymmetric 
representation \cite{collision, hep-th/0002012} (consistent with the 
Type IIB brane constructions). All the fourfolds $Y^{BV}$, $Y_0$, $Y^{W'}$, 
$Y^{W''}$ and $\widetilde{Y}$ are Calabi-Yau and are birational, and 
no cycles of finite volume are added or removed. 
The $(20-r_{(1)}-{\rm rk}(R))+(20-r_{(2)})$ complex structure moduli 
remain neutral chiral multiplets on $\R^{3,1}$; other complex structure moduli 
of $Y^{BV}$ will remain massless chiral multiplets on $\R^{3,1}$, but 
as a part of gauge-charged matter fields (they are the $g_{(1)}g_{(2)}=10g_{(2)}$
moduli deforming the $\C^2/\Z_2$ singularity of $Y_0$ and the ${\rm rk}(R)$ 
moduli of $X^{(1)}$ that reduces\footnote{
\label{fn:F-geometry-5}
It is desirable to carry out the Higgs cascade analysis \cite{MV-2, 6auth}
of all those kinds of constructions in sections \ref{ssec:on-fibre} 
and \ref{ssec:on-base}, where F-theory prediction including matter 
multiplicity information is compared against symmetry breaking
processes in the effective field theory on (3+1)-dimensions (or 
on (5+1)-dimensions). 
The authors consider that such a study will uncover much more aspects 
of F-theory compactification on K3 x K3 orbifolds (or on Borcea--Voisin 
orbifolds) than those presented in sections \ref{ssec:on-fibre} 
and \ref{ssec:on-base}. 
} %
 $S_X^{(1)}$ back to $S_0^{(1)}$).  

Suppose next that the lattice $R$ contains a factor $D_n$ or $E_6$, 
corresponding to a singular fibre of $I_{n-4}^*$ type or ${\rm IV}^*$ in 
$X^{(1)}$ over ${\rm pt}_* \in \P^1_{(1)}$. The known prescription\footnote{
The prescription of Ref. \cite{collision} is to replace $\nu^*(Y^W)$ by 
the Affine part of $Y$ and then to add the zero section by hand, 
without discussing birational map between them. 
In that prescription, the Calabi--Yau condition of $Y$ had to be tested 
independently from the Calabi--Yau nature of $Y^W$.
} %
 is to set\footnote{
We have in mind that the K\"{a}hler parameter is such that the 
exceptional divisor in the blow-up $B_3 \rightarrow B_w$ has a non-zero 
positive volume.
 } %
$B_3 ={\rm Bl}_{{\rm pt}_* \times [Z_{(2)}]}B_w$, and think of $\nu^*(Y^W)$ with 
a Weierstrass fibration over $B_3$ for a moment; $\nu: B_3 \rightarrow B_w$
is the blow-up map. The fourfold $\nu^*(Y^W)$ has a parabolic singularity at 
$\{ \tilde{y}=\tilde{x}=0 \}$ in the fibre of the exceptional locus $E$ of 
$B_3 = {\rm Bl}_{{\rm pt}_* \times [Z_{(2)}]}B_w$. The ambient space of $\nu^*(Y^W)$ 
is blown-up three times with the centre in the fibre of $E$, and now the proper
transform $\overline{\nu^*(Y^W)}$ has only $A_{n-5}$ singularity 
(assuming an even $n > 4$; none for $I_0^*$ or ${\rm IV}^*$). 
The fourfold $\overline{\nu^*(Y^W)}$ is not Calabi--Yau due to the 
non-trivial morphisms $\overline{\nu^*(Y^W)} \rightarrow \nu^*(Y^W) 
\rightarrow Y^W$, but there is a morphism 
$\overline{\nu^*(Y^W)} \rightarrow Y$ to a fourfold ramified along 
the canonical divisor of $\overline{\nu^*(Y^W)}$, so $Y$ is a Calabi--Yau 
fourfold. There is also a projection morphism $Y\rightarrow B_3$  
(see (\ref{eq:birat-net-invF})). 
The fourfold $Y$ has $D_4$ singularity in the fibre of  
$\overline{\Delta_b}$ (the proper transform of $\Delta_b$ under 
$\nu: B_3 \rightarrow B_w$).

In the case of $I_{n-4}^*$, the fourfold $Y$ also has $D_{4+n}$ singularity 
in the fibre of 
$\overline{{\rm pt}_* \times B^{(2)}}$, there is also $A_{n-5}$ singularity 
(if $n>4$) in the fibre of the exceptional divisor $E$ (statements 
in the rest of this paragraph is for an even $n$).   
Those singularities in $Y$ should be resolved canonically; further small 
resolution in the fibre of codimension-2 loci in $B_3$ yields $\widetilde{Y}$ 
that has a flat elliptic fibration over $B_3$ \cite{Miranda}. The 7-brane 
$\overline{{\rm pt}_*\times B^{(2)}}$ yields $\SO(2n)$ gauge group 
in the effective theory on $\R^{3,1}$; the effective theory also has 
an $({\rm Sp}((n-4)/2))^{k_2+1}$ gauge group (for an even $n>4$).\footnote{
${\rm Sp}(n) = {\rm USp}(2n)$ in this notation.
} %
A $I_0^*$--$I_{n-4}^*$ collision may yield chiral multiplets of 
4D ${\cal N}=1$ supersymmetry in the $G_2$--${\rm Sp}((n-4)/2)$ bifundamental, 
and in the ${\rm Sp}((n-4)/2)$--$\SO(2n)$ bifundamental 
representations (in the case $n=4$ there is no matter fields associated 
particularly with the $I_0^*$--$I_0^*$ collision) \cite{collision}. 
Cases with an odd $n>4$ are less trivial, but remain 
similar \cite{6auth, hep-th/0002012}. 

In the case of ${\rm IV}^*$, we have an $F_4$ vector multiplet on $\R^{3,1}$ 
from the brane $\overline{{\rm pt}_* \times B^{(2)}}$ \cite{6auth,collision}.
Chiral multiplets may arise from the $I_0^*$--${\rm IV}^*$ collision, which 
are in the fundamental representations of $G_2$ and $F_4$, 
but there is no matter in a mixed representation \cite{collision, hep-th/0002012}. 
The types of matter representations available are the same 
for all $(k_2+1)$ singularity collisions along the $(k_2+1)$ disjoint 
components of ${\rm pt}_* \times [Z_{(2)}]$.
Details of the massless spectrum may be different due to a choice of 
a four-form flux in the non-horizontal part of $H^4(\widetilde{Y};\Q)$.  

In the case the lattice $R \subset W$ contains a factor
 $E_7$, $B_3$ is obtained by blowing-up $B_w$ twice; 
$\widetilde{Y}$ is also obtained in a similar procedure.  
The gauge group on $\R^{3,1}$ becomes $(G_2 \times \SU(2))^{k_2+1} \times E_7$. 
Matter chiral multiplets charged under $E_7$ can only be in the 
${\bf 56}$ representation of $E_7$ and singlet under 
$(G_2 \times \SU(2))^{k_2+1}$ \cite{collision}. Such matter $E_7$-${\bf 56}$
fields, however, are not associated with $I_0^*$--${\rm III}^*$ collision;  
the base $B_3$ has divisors $\overline{\P^1 \times [Z_{(2)}]}$, two exceptional 
divisors in $B_3 \rightarrow B_w$, and $\overline{{\rm pt}_* \times B^{(2)}}$, 
over which a generic fibre is of $I_0^*$-type, type-${\rm III}$, 
non-singular, and of type-${\rm III}^*$, respectively; there is no 
enhancement of singularity over the non-singular--${\rm III}^*$ collisions, 
so there is no $E_7$-${\bf 56}$ matter arising from there. 
The $E_7$-${\bf 56}$ matter field may arise from the intersection of 
$\overline{{\rm pt}_* \times B^{(2)}}$ with other irreducible components 
of the discriminant locus, but there is no such intersection in the K3 x K3 
orbifolds we consider in this article. So, there is no $E_7$-charged 
matter field. 

\subsubsection{More Consequences in Physics} 

In all those cases\footnote{
\label{fn:F-geometry-4}
One may think of a case a complex structure is tuned in 
$D(T_0^{(1)} = {\rm II}_{2,18})$ so that $(S_X^{(1)}, T_X^{(1)})$ just happens 
to be identical to one of $(S_0, T_0)$ in Nikulin's list. It is a question 
of interest whether there is an isomorphism between $Y^{BV}$
and $\widetilde{Y}$ constructed as in the main text. 
} %
 where $R$ involves $D_n$, $E_6$ or $E_7$, 
birational morphisms between the two Calabi--Yau's $Y^{BV}$ and $Y$ 
in (\ref{eq:birat-net-invF}) can be constructed for any choice of moduli 
in $D([(S_X^{(1)})^\perp]) \times D(T_0^{(2)})$. So, those deformation 
degrees of freedom and their corresponding cohomology groups 
(i.e., $T_X^{(1)} \otimes T_0^{(2)}$) will remain 
to be there for $Y$ and $\widetilde{Y}$. 
The $g_{(1)}g_{(2)}$ complex structure moduli of $Y^{BV}$ (and the 
$H^2(Z_{(1)}\times Z_{(2)};\Q)$ component in $H^4(Y^{BV};\Q)$) may or may not 
be present in $\widetilde{Y}$, but even when they are present, they will 
be part of $G_2$-charged matter fields. The ${\rm rk}(R)$ moduli fields 
necessary in enhancing the $D_n$, $E_6$ or $E_7$ singularity in $Y^W$ 
may either be part of gauge-charged matter fields, or be absent as massless 
degrees of freedom in the F-theory compactification. 

If there were any chance of accommodating grand unification of the Standard 
Model in this Type 1 framework, a GUT gauge group such as 
$\SU(5)$, $\SO(10)$, $E_6$, and $E_7$ at the level of (7+1)-dimensions 
could only be from $\Delta_f$, because those gauge groups do not fit within 
$\SO(8)$. We have seen above that implementing $A_4=\SU(5)$ or $E_6$ 
in $R \subset W$ of $X^{(1)}$ does not result in an SU(5) or $E_6$ 
gauge group on (3+1)-dimensions due to the monodromy at the $I_0^*$--$R$ 
collision.\footnote{
An exception is when $S_0^{(2)}=U[2]E_8[2]$ in the list of Nikulin, because 
$Z_{(2)}$ is empty and there is no $I_0^*$--$R$ collision; this scenario 
is still not suitable for GUT, however, because there is no massless 
matter chiral multiplets charged under $R$.
} % 
Even when we require $D_5$ or $E_7$ within $R \subset W$, there is no 
chance having a matter field in the spinor representation of $\SO(10)$, 
or in the ${\bf 56}$ representation of $E_7$, as stated earlier. 
Furthermore, there is no massless adjoint chiral multiplets of $E_7$ 
because $h^{0,1}(B^{(2)})=h^{0,2}(B^{(2)})=0$.

The conditions for a $DW=W=0$ flux (or a $DW=0$ flux) and study of 
complex structure moduli stabilization in  
sections \ref{ssec:Tx=T0} and \ref{ssec:Tx-neq-T0} can be recycled 
without modifications for F-theory, as we see below. 
We stick to the Type 1 case available for $S_0^{(1)} = U$ and 
$T_0^{(1)} = {\rm II}_{2,18}$. 
For a CM-type vacuum complex structure such that $T_X^{(1)}=T_0^{(1)}$, 
then $T_X^{(2)}$ should\footnote{
This ${\rm rank}(T_X^{(2)}) = [K^{(1)}:\Q]$ condition is only a necessary 
condition for an existence of such a CM point 
(cf footnote \ref{fn:PSSarth-Taelmn}). 
At least in the case of $T_0^{(2)}=T_X^{(1)}$, we are sure that 
$D(T_0^{(2)})$ contains a CM point whose CM field is isomorphic to the 
CM field $K^{(1)}$ of a CM point in $D(T_X^{(1)})$.
} %
 also be of rank 20,
 so that $T_X^{(2)}$ has a CM point in $D(T_X^{(2)})$ with 
the CM field $K^{(2)}$ satisfying the condition (\ref{eq:flux-cond-math2}) 
(or (\ref{eq:flux-cond-math1})). This means that ${\rm rank}(T_0^{(2)})$ 
is either 20 (when ${\rm rank}(S_0^{(2)})=2$) or 21 
(when ${\rm rank}(S_0^{(2)})=1$); there are three such pairs 
$(S_0^{(2)},T_0^{(2)})$ in Nikulin's list ($S_0^{(2)}=\vev{+2}, U, U[2]$). 
For any one of the three choices of $(S_0^{(2)}, T_0^{(2)})$, 
all the 18+(19 or 18) complex structure fluctuation fields 
in $D(T_0^{(1)}) \times D(T_0^{(2)})$ are valid Calabi--Yau deformations 
of $\widetilde{Y}=Y^{W''}$, not just of $Y^{BV}$.  
A $DW=W=0$ flux provides large supersymmetric masses to all those 
complex structure moduli fluctuations. 

For a CM-type vacuum complex structure with $T_X^{(1)} \subsetneq T_0^{(1)}$, 
for example, when $S_X^{(1)} \supset U \oplus R$,
the CM field $K^{(1)}$ has a degree $[K^{(1)}:\Q] = {\rm rank}(T_X^{(1)}) < 20$. 
So, the necessary condition ${\rm rank}(T_X^{(1)}) = {\rm rank}(T_X^{(2)})$ for 
(\ref{eq:flux-cond-math2}), which is also for a non-trivial $DW = W=0$ 
flux, allows a choice of $(S_0^{(2)}, T_0^{(2)})$ from a broader subset 
of Nikulin's list. 
% 
% It should be remembered however, that 
% there is a chain of discrepant birational transformations between $Y^W$ 
% and $Y$ in (\ref{eq:birat-net-invF}) when $R$ in $S_X^{(1)}=U \oplus R$ 
% contains either $D_n$ or $E_n$; the fourfolds in between them are not 
% Calabi--Yau. 
% Some of Calabi--Yau complex structure deformations of $Y^W$ (including 
% the tangent space of $D(T_0^{(1)})\times D(T_0^{(2)})$) are not available 
% as Calabi--Yau complex deformations of $Y$ (cf \cite{collision}).  
% 
The complex structure deformation fields in $D(T_X^{(1)}) \times D(T_0^{(2)})$ 
obtain large supersymmetric masses by a $DW=W=0$ flux, which one can see 
by repeating the same discussion as in section \ref{ssec:Tx-neq-T0}. 

The complex structure moduli stabilization in \cite{DDFK} can be regarded 
as a special case of the general discussion above. Our interpretation is 
that the fourfolds for F-theory in \cite{DDFK} correspond to 
$(S_0^{(1)}, T_0^{(1)})=(U, {\rm II}_{2,18})$ as stated above, 
$(S_0^{(2)}, T_0^{(2)})$ that of a Kummer surface ($r_{(2)}=18$,  
$a_{(2)}=4$, $k_2=7$ and $g_{(2)}=0$), 
$T_X^{(1)} = U[2]^{\oplus 2} \subsetneq T_0^{(1)}$ and  
$T_X^{(2)}=T_0^{(2)}=U[2]^{\oplus 2}$. The discussion above further indicates 
that there should be a flux with the vev $\vev{W}=0$,   
when we choose the vacuum complex structure of all the tori in 
$X^{(1)} \sim (T^2 \times T^2)/\Z_2$ and 
$X^{(2)} \sim (T^2 \times T^2)/\Z_2$ so that they all have complex 
multiplication, and the condition (\ref{eq:flux-cond-math2}) is satisfied.

%
% It should be reminded, though, that we did not study whether the moduli 
% fields corresponding to 
% $N_{{\cal M}_{\rm cpx~str}^{[Y]BV}|{\cal M}_{\rm cpx~str}^{[Y]}}$, 
% or arithmetic conditions for $DW=0$ flux in the 
% non-$(T_0^{(1)}\otimes T_0^{(2)})$ components in $H^4(\widetilde{Y};\Q)$. 

%%%%%%%%%%%%%%%%%%%%%%%%%%%%%%%%%%%%%%%%%%%%%%%%%%%%%%
\subsection{Fibration and Involution of Type 2}
\label{ssec:on-base}
%%%%%%%%%%%%%%%%%%%%%%%%%%%%%%%%%%%%%%%%%%%%%%%%%%%%%%

In the Type 2 case, we start from the projection map 
$Y_0^W \rightarrow B_{w0}$, which is in between singular varieties;
$B_{w0} := \left( \P^1_{(1)} \times X^{(2)}\right)/\Z_2$. 
Consider the canonical resolution of the $A_1$-singularity of $B_{w0}$,   
$\nu: B_w := \widetilde{B}_{w0} \rightarrow B_{w0}$, and set 
$Y^W := \nu^*(Y_0^W)$. Now the projection $Y^W \rightarrow B_w$ 
is a Weierstrass model over a non-singular threefold $B_w$.  
The fourfold $Y^W$ satisfies the Calabi--Yau condition because 
$\nu: B_w \rightarrow B_{w0}$ is crepant. 
\begin{align}
  \xymatrix{
    Y^{BV} \ar[d] & & \widetilde{Y} \ar[d] & & & \widetilde{Y} \ar[d] \\
    Y_0 \ar[r] & Y_0^W \ar[d] & Y^W \ar[l] \ar[d] &
      \nu^{'*}(Y^W) \ar[l] \ar[d] & \overline{\nu^{'*}(Y^W)} \ar[l]\ar[r] & Y \ar[lld] \\
       & B_{w0}=(\P^1_{(1)} \times X^{(2)})/\Z_2 & B_w = \widetilde{B}_{w0} \ar[l] & B_3 \ar[l]
}
  \label{eq:birat-net-invB}
\end{align}
See the following discussions for $\nu'$, $Y$, and $\widetilde{Y}$.

So long as complex structure of $X^{(1)}$ corresponds to a generic point in 
$D(T_0^{(1)W}) = D(U^{\oplus 2}E_8[2])$, which means that $S_X^{(1)} = 
S_0^{(1)} = U \oplus E_8[2]$, the Weierstrass model $Y^W$ 
is already non-singular; the projection $Y^W\rightarrow B_w$ is a 
flat elliptic fibration, so we can set $\widetilde{Y} = Y^W$ 
and $B_3 = B_w$. The base threefold $B_3$ is a $\P^1$-fibration 
over $B^{(2)}$; the $\P^1$-fibre degenerates into three irreducible 
pieces ($\P^1 + 2\P^1 + \P^1$) over $[Z_{(2)}] \subset B^{(2)}$.
Note that there is no difference between $Y_0$ and $Y_0^W$, 
and that $Y^{BV}$ and $Y^W$ are identical in this generic complex 
structure. The discriminant locus $\Delta_{\rm discr}$ of the elliptic fibration 
$Y^W \rightarrow B_w$ consists of 12 isolated components, 
each one of which is a double cover over $B^{(2)}$ ramified 
over $[Z_{(2)}]$; each piece is isomorphic to the K3 surface $X^{(2)}$.  
Here, we assume on the ground of genericity that the 12 pairs of 
$I_1$ fibres of $X^{(1)}$ stay away from the two fixed points of 
$\P^1_{(1)}$ under the action of $\sigma_{(1)}$. 
There is no non-abelian gauge group in the effective theory on 
$\R^{3,1}$ then. 

When the vacuum complex structure of $X^{(1)}$ is tuned so that 
some of the 12 pairs of $I_1$ fibre come on top of each other 
(but remain distant from the $\sigma_{(1)}$-fixed locus), 
$S_X^{(1)}$ may be different from $S_0^{(1)} = U \oplus E_8[2]$, 
and in particular, the sublattice $R$ of $W$ in 
$S_X^{(1)} =: U \oplus W$ may contain a pair of copies of an ADE-type 
root lattice. Because the discriminant locus of the ADE-type fibre 
forms a single irreducible component, the effective theory on $\R^{3,1}$ 
will have a gauge group of that ADE type, with one chiral multiplet in the 
adjoint representation (because $h^{0,2}(X^{(2)})=1$).  
A non-trivial gauge flux on these 7-branes may reduce the ADE symmetry 
further down to a smaller non-abelian gauge group, but we cannot 
obtain a chiral spectrum on $\R^{3,1}$ in this way (note that $c_1(X^{(2)})=0$).  

Consider instead an $X^{(1)}$ that has a singular fibre at a fixed 
point of $\sigma_{(1)}$ in $\P^1_{(1)}$. 
Suppose that the singular fibre is $I_{2n}$ [resp. $\IV^*$  
or $\I_0^*$],\footnote{ 
There is a rule on the Kodaira type of a singular fibre that can 
appear over the base point of $\P^1_{(1)}$ fixed by the Type 2 involution
\cite{Kl, Cmp-Garbag, 1806.03097}. Those Kodaira types are consistent 
with the rule.  
} %   
and all the other singular fibres of $X^{(1)}$ are of $I_1$ type 
and are away from the $\sigma_{(1)}$-fixed points. 
The discriminant $\Delta_{\rm discr}$ consists of three distinct groups of 
components. One of them consists of $(12-n)$ [resp. 8 or 9] 
copies of $X^{(2)}$ that do not yield a non-abelian gauge group 
on $\R^{3,1}$. Another is a section of the $\P^1$-fibration 
over $B_{(2)}$, which yields the $\SU(2n)$ [resp. $E_6$, or $\SO(7)$ 
(due to monodromy)] gauge group on $\R^{3,1}$. 
The last group of 7-branes is 
the $(k_2+1)$ isolated pieces of the exceptional divisors associated 
with the $\sigma_{(1)}$-fixed point in $\P^1_{(1)}$ where $X^{(1)}$ has 
the singular fibre. Each one of those 7-branes yields a gauge group  
$\SU(n)$ [resp. $\SU(3)$ or $\SU(2)$ (monodromy is absent)] on $\R^{3,1}$. 

In the case of $I_{2n}$ [resp. $I_0^*$], we can set $B_3 = B_w$, and 
$\widetilde{Y}$ as the canonical resolution of $Y^W$ for its 
codimension-2 singularities followed by a small resolution 
in the fibre of $I_{2n}$--$I_n$ collision 
[resp. ${\rm I}_0^*$--${\rm III}$ collision]. The projection 
$\widetilde{Y} \rightarrow Y^W \rightarrow B_3$ is flat \cite{Miranda}. 
In the case of ${\rm IV}^*$, we can use as $B_3$ the blow-up of $B_w$ 
centred at the intersection of the $E_6$ (Kodaira type ${\rm IV}^*$) 
7-brane and the SU(3) (Kodaira type ${\rm IV}$) 7-branes. $Y^W$ is pulled 
backed to be $\nu^{'*}(Y^W)$ fibred over $B_3$; it will be possible 
to construct birational and regular maps $\overline{\nu^{'*}(Y^W)} \rightarrow 
\nu^{'*}(Y^W)$ and $\overline{\nu^{'*}(Y^W)} \rightarrow Y$ (as in 
section \ref{ssec:on-fibre}), where $Y$ is Calabi--Yau \cite{collision}, 
and $\widetilde{Y}$ is obtained as a canonical resolution of 
the codimension-2 singularities of $Y$. 

If there is any chance of accommodating a GUT gauge group, 
one might first consider SU(5) as a part of SU(6). 
In this case, there may be 4D ${\cal N}$=1 chiral multiplets 
in the SU(6)--SU(3) bifundamental representation localized 
at the $I_6$--$I_3$ collision matter curves. But, there 
is no matter in the rank-2 anti-symmetric representation. 
The other possibility is $E_6$. But, there is no 
matter fields in the $E_6$-{\bf 27} representation; to see this, 
note that there is no singularity enhancement at the 
intersection of the $E_6$ 7-brane with the exceptional divisor 
of $\nu': B_3 \rightarrow B_w$, and that there can be no singularity 
enhancement away from the orbifold loci. There may be $E_6$-adjoint chiral 
multiplets from the $E_6$ 7-brane, but its irreducible decomposition 
to SU(5) subgroup cannot yield a reasonably successful phenomenology 
\cite{TW-06}.   
To summarize, it is not possible to implement GUT phenomenology   
in any one of the constructions considered in this 
section \ref{ssec:on-base}.

There is not much to add particularly for the Type 2 case on the 
flux-induced supersymmetric mass terms of the complex structure 
moduli fields. The discussion at the end of section \ref{ssec:on-fibre} 
can be repeated with minimal changes;\footnote{
The $\C^2/\Z_2$-deforming moduli of $Y^{BV}$ and corresponding 
deformations in $\widetilde{Y}$ are localized in the fibre of non-abelian 
7-branes in the case of Type 1, but that is not the case generically for 
a Type 2 fibration and involution. So, there are gauge-neutral moduli fields 
whose stabilization / mass term is not discussed in this article
 (cf. discussion at the end of section \ref{ssec:generic-noFlux} 
and footnote \ref{fn:heuristic}). 
} %
 the only difference from the Type 1 
case is that $(S_0^{(1)}, T_0^{(1)}) = (UE_8[2], U^{\oplus 2}E_8[2])$ rather than 
$(U, U^{\oplus 2}E_8^{\oplus 2})$.  

For a K3 surface $X^{(1)}$ that corresponds to $S_0 = U \oplus E_8[2]$, there 
automatically exists two non-symplectic involutions. One acts on the 
base, and the other on the fibre. So, their combination also yields 
a non-trivial symplectic subgroup of the automorphisms. 
This means that all the compactifications in a Type 2 case  
has an extra $\Z_2$ symplectic (=non-R) symmetry in the effective theory 
(unless the flux breaks it).

%%%%%%%%%%%%%%%%%%%%%%%%%%%%%%%%%%%%%%%%%%%%%%%%%%%%%%%%%%%%%%%%
\section*{Acknowledgements}  % BrE with e
%%%%%%%%%%%%%%%%%%%%%%%%%%%%%%%%%%%%%%%%%%%%%%%%%%%%%%%%%%%%%%%%

We thank T. T. Yanagida who has reminded us for more than a decade 
of the importance of how to obtain $\vev{W}=0$ in string compactification, 
and also thank Zheng Sun for useful communications.  
This work is supported in part by the World Premier International Research Center Initiative (WPI), Grant-in-Aid New Area no. 6003
 (K.K. and T.W. so far),  
Advanced Leading Graduate Course for Photon Science grant and 
JSPS Research Fellowship for Young Scientists (K.K.),
all from MEXT of Japan. 

\appendix

%%%%%%%%%%%%%%%%%%%%%%%%%%%%%%%%%%%%%%%%%%%%%%%%%%%%%%%%%%%%
\section{Type IIB Orientifold Case}
%%%%%%%%%%%%%%%%%%%%%%%%%%%%%%%%%%%%%%%%%%%%%%%%%%%%%%%%%%%

As a special case of the analysis of supersymmetric flux configurations 
for M/F-theory in section \ref{sec:flux-analysis}, the case of 
Type IIB orientifold compactification on a Borcea--Voisin threefold 
$M=(E_\tau \times X^{(2)})/\Z_2$ is covered (see (\ref{eq:BV-IIB-as-Kummer-K3})); 
$X^{(1)}={\rm Km}(E_\phi \times E_\tau)$ corresponds to 
a choice of $(S_0^{(1)}, T_0^{(1)},\sigma_{(1)})$ 
from \cite{Nikulin-factor-long}
where $T_0^{(1)}=U[2]U[2]$, $r_{(1)}=18$, $a_{(1)}=4$ and $g_{(1)}=0$. 
The conditions (\ref{eq:flux-cond-math1}, \ref{eq:flux-cond-math2}) 
for the case of $X^{(1)} = {\rm Km}(E_\phi \times E_\tau)$ should therefore 
be equivalent\footnote{
\label{fn:IIB-DDFK}
In section 3.3 of \cite{prev.paper.PRD}, we worked out orientifold 
projection on the moduli of the threefold $M$, and found that the 
twisted sector moduli of complex structure of $M$ are projected out. 
In this article, the absence of such moduli is understood as absence 
of the $H^1(Z_{(1)};\Q)\otimes H^1(Z_{(2)};\Q)$ component; it is of 
$g_{(1)}g_{(2)}=0$ dimension. 
} %
 to the conditions worked out in \cite{prev.paper.PRD}.
The two sets of conditions do not look similar at first sight
(as reviewed below), but we confirm in the following that they are 
equivalent indeed. 
So, this appendix is regarded as a supplementary note 
to \cite{prev.paper.PRD}; 
consistency check in this appendix also gives 
confidence in the study in section \ref{sec:flux-analysis} in this article. 

Let us start off by recalling the Type IIB conditions in \cite{prev.paper.PRD}
for a non-trivial supersymmetric flux. $K^{(2)}$ and $K_E$ are the endomorphism 
fields of the CM-type Hodge structure on $T_X^{(2)}$ and $H^1(E_\tau;\Q)$, 
respectively. $n:= {\rm rank}(T_X^{(2)})$. 

When the untwisted sector $T_X^{(2)} \otimes_\Q H^1(E_\tau;\Q)$ 
is itself a simple component of the rational Hodge structure,\footnote{
Then there is no chance for a non-trivial flux with $W=0$.
} %  
it is level-3  
and $K^{(2)} \otimes_\Q K_E$ is the endomorphism field. A non-trivial $DW=0$ 
flux exists if and only if 
\begin{align}
 \left( K^{(2)}  \otimes_\Q K_E \right)^r \cong \Q(\phi),
     \qquad [\Q(\phi):\Q] = 2.
   \label{eq:flux-cond-IIB-1}
\end{align}
The half set\footnote{
Recall that we always consider the reflex field in the sense of 
Weil intermediate Jacobian, i.e. the Jacobian $J_W(M)$ associated with 
$H^{0,3}(M)\oplus H^{2,1}(M)$.
} % 
\begin{align}
  \Phi = \left\{ \rho^{(2)}_{(20)} \otimes \rho^\tau_{(10)}, \; 
      \rho^{(2)}_{a=1,\cdots,n-2} \otimes \rho^\tau_{(01)}, \;
      \rho^{(2)}_{(02)}\otimes \rho^\tau_{(10)} \right\}
\end{align}
of all the $2n$ embeddings $K^{(2)} \otimes K_E \rightarrow \overline{\Q}$
is used in determining the reflex field.\footnote{
Note that we started out in 
F/M-theory analysis in section \ref{sec:flux-analysis} in this article 
by assuming that $X^{(1)}={\rm Km}(E_\phi \times E_\tau)$ is of CM type 
(that both $E_\tau$ and $E_\phi$ are CM elliptic curves). In the analysis of 
\cite{prev.paper.PRD}, however, the CM nature of $E_\phi$, namely 
$[\Q(\phi):\Q]=2$, follows from the CM nature of $X^{(2)}$ and $E_\tau$
and the supersymmetry conditions on a non-trivial flux.
} % 

When $T_X^{(2)} \otimes_\Q H^1(E_\tau;\Q)$ is not a simple component, instead, 
$K^{(2)}$ has a structure of $K_0 \Q(\xi_S)$ for its totally real subfield $K_0$ 
and an imaginary quadratic field $\Q(\xi_S)$ isomorphic to $K_E$, and 
$T_X^{(2)} \otimes_\Q H^1(E_\tau;\Q)$ has a structure $K^{(2)} \oplus K^{(2)}$ 
under the action of the algebra $K^{(2)} \otimes_\Q K_E$ ($K_E$ acts through 
an isomorphisms $\Q(\xi_S) \cong K_E$). For a non-trivial $DW=0$ flux 
to exist, it is necessary and sufficient that
% 
% \footnote{
% %
% A non-trivial $DW=0$ made possible by this condition automatically 
% leads to $W=0$.  {\bf true?}
% } % 
%
\begin{align}
  ( K^{(2)} )^r \cong \Q(\phi) \cong \Q(\tau). 
    \label{eq:flux-cond-IIB-2}
\end{align}
A few more words are necessary for this condition to have a clear 
meaning. Let $\theta_{a=1,\cdots, n/2}$ be the embeddings 
$K_0 \rightarrow \overline{\Q}$, and $\theta_{a}^\pm$ those of $K^{(2)}$ 
so that their restriction on $K_0$ are $\theta_a$, and $\theta_a^+(\xi_S)$
[resp. $\theta^-_a(\xi_S)$] is in the upper [resp. lower] complex half plane. 
The reflex field $(K^{(2)})^r$ in the condition (\ref{eq:flux-cond-IIB-2}) 
should be for the half set\footnote{
It was not clearly stated 
in \cite{prev.paper.PRD} which half set of the $n$ embeddings of $K^{(2)}$ 
should be used in determining the reflex field in (\ref{eq:flux-cond-IIB-2}).
} % 
\begin{align}
  \left\{ \theta_{a=1}^+, \; \theta_{a =2,\cdots, n/2}^- \right\}.
  \label{eq:flux-cond-IIB-2prm}
\end{align}

{\bf The case $T_X^{(2)}\otimes H^1(E_\tau;\Q)$ is simple:}  
Now, we begin with making the condition (\ref{eq:flux-cond-IIB-1}) 
more explicit. To this end, a set of notations is introduced in order 
to capture 
the structure of the fields $K^{(2)}$ and $K_E$. As a general property 
of CM fields, $K^{(2)}$ has a structure of $K_0(\underline{x})$ where 
$K_0$ is the totally real subfield of $K^{(2)}$, and $\underline{x}$ 
an element of $K^{(2)}$ with the following properties: 
$\underline{x}^2 \in K_0$, and the element $Q := - \underline{x}^2$ in $K_0$
is mapped onto the real positive axis by all the $[K_0:\Q]=n/2$ embeddings 
$K_0 \rightarrow \overline{\Q}$.
Similarly, $K_E = \Q(\underline{\tau})$ for some $\underline{\tau} \in K_E$
such that $p:= - \underline{\tau}^2 \in \Q_{>0}$. The vector space 
$K^{(2)}\otimes_\Q K_E$ is regarded 4-dimensional over $K_0$ generated by 
$\{1, \underline{x}, \underline{\tau}, \underline{x}\underline{\tau}\}$; 
the totally real subfield of $K^{(2)}\otimes K_E$---denoted by 
$K_0^{\rm tot}$---is 2-dimensional over $K_0$ generated by 
$\{1, \underline{x}\underline{\tau}\}$. 

The condition that the reflex field in (\ref{eq:flux-cond-IIB-1}) 
is an imaginary quadratic extension of $\Q$ is equivalent to existence 
of $\underline{\eta} \in K^{(2)}\otimes K_E$ such that its images by the 
$n$ embeddings in $\Phi$ are all identical $\eta \in \overline{\Q}$ 
which generates an imaginary quadratic field $\Q(\eta)$.
For 
\begin{align}
  \underline{\eta} = A + B \underline{x} + C \underline{\tau}
 + D \underline{x}\underline{\tau} \in K^{(2)} \otimes K_E, 
    \qquad A,B,C,D \in K_0, 
\end{align}
the condition $\underline{\eta}^2 \in \Q$ is equivalent to 
\begin{align}
 AC = QBD, \quad AB = pCD, \quad AD + BC = 0, \quad 
 (A^2-QB^2-pC^2+pQD^2) \in \Q. 
  \label{eq:IIB-limit-temp1}
\end{align}
This leaves five distinct possibilities: i) none of $A$, $B$, $C$, $D$ is 
zero, ii-A) $A \neq 0$, and $B=C=D=0$, ii-B) $B \neq 0$ and three others 
are zero, ii-C) $C \neq 0$ and three others are zero, and ii-D) $D\neq 0$ 
and three others are zero. 

In fact, only the possibility ii-C) is viable. 
The possibility i) runs into a contradiction: $(B/D)$ is 
a well-defined element of the totally real field $K_0$ in this possibility, 
and yet one can derive that $(B/D)^2 = -p \in \Q_{<0}$. 
In the possibilities ii-A) and ii-D), the element $\underline{\eta} = A$ 
or $\underline{\eta}=D\underline{x}\underline{\tau}$ would not generate 
a totally imaginary extension over $K_0^{\rm tot}$. The possibility ii-B) 
cannot be consistent with the condition that the images of 
$\underline{\eta} = B \underline{x}$ under the $n$ embeddings in $\Phi$ 
should be all identical; 
$\rho^{(2)}_{(20)}(B\underline{x}) = - \rho^{(2)}_{(02)}(B\underline{x}) \neq 0$.

Let us focus on the remaining ii-C) possibility.
The condition (\ref{eq:IIB-limit-temp1}) implies that $C^2 \in \Q$. 
There are two cases, (*1) $C_{\neq 0} \in \Q$, and (*2) $C \Slash{\in} \Q$ 
whose square 
is a positive rational number $r \in \Q_{>0}$ that is not a square.

In the case (*1), the condition that all the $n$ images of 
$\underline{\eta}=C\underline{\tau}$ are identical is satisfied if and 
only if $n=2$; if $n>2$, then $\rho^{(2)}_{a > 2}(\underline{\eta})
 = C \rho^{\tau}_{(01)}(\underline{\tau})$ cannot be the same as the images
$\rho^{(2)}_{(20)}(C)\rho^\tau_{(10)}(\underline{\tau})$ and 
$\rho^{(2)}_{(02)}(C)\rho^\tau_{(10)}(\underline{\tau)}$. 
So, $K^{(2)}=\Q(\underline{x})$ must be some imaginary quadratic field, 
and the reflex field $(K^{(2)}\otimes K_E)^r$ must be $\Q(\sqrt{-p}) \cong K_E$.
It follows that $E_\phi$ also has the endomorphism field $\Q(\sqrt{-p})$, 
$E_\tau$ and $E_\phi$ are isogenous (and are both CM), and 
$X^{(1)} = {\rm Km}(E_\phi \times E_\tau)$ has a rank-20 N\'eron--Severi lattice.
So, to conclude, the case (*1) solution to the 
condition (\ref{eq:IIB-limit-temp1}) implies that 
$T_X^{(1)} \subsetneq T_0^{(1)}$, $K^{(1)} \cong \Q(\phi) \cong K_E \cong 
\Q(\sqrt{-p})$, $K^{(2)}$ is an imaginary quadratic field (and is not 
isomorphic to $\Q(\sqrt{-p})$ as assumed before (\ref{eq:flux-cond-IIB-1})), 
and the condition (\ref{eq:flux-cond-math1}) is satisfied; 
both $\rho^{(1)}_{(20)}(K^{(1)}_0)=\rho^{(2)}_{(20)}(K^{(2)}_0)=\Q$. 

In the case (*2), the totally real field $K_0$ must be a real quadratic field. 
To see this, note that $K_0$ contains $\Q(C) \cong \Q(\sqrt{r})$, 
which means that $n/2 \geq 2$. 
The condition that all the $n$ images of $\underline{\eta}=C\underline{\tau}$ 
should be the same now implies that $\rho^{(2)}_{a > 2}(C) = - \rho^{(2)}_{(20)}(C)$.
Because the $n/2$ embeddings of $K_0$ should yield the same 
number of two different embeddings of the subfield $\Q(C)$, 
$(n-2)/2$ must be equal to $2/2$; $n=4$. Therefore, 
$K^{(2)} \cong \Q(\underline{x},C)$, its totally real subfield 
must be $K^{(2)}_0 \cong \Q(C)$, and $(K^{(2)}\otimes K_E)^r \cong \Q(\sqrt{-pr})$.
Now, the remaining condition in (\ref{eq:flux-cond-IIB-1}) is 
$\Q(\phi) \cong \Q(\sqrt{-pr})$. So, it turns out that 
$K^{(1)} \cong \Q(\sqrt{-p}, \sqrt{-pr})$, and $K^{(1)}_0 \cong \Q(\sqrt{r})$. 
Thus, to summarize, the case (*2) solution to the condition 
(\ref{eq:flux-cond-IIB-1}) implies that $T_X^{(1)}=T_0^{(1)}$, 
$K^{(1)}_0 \cong K^{(2)}_0 \cong \Q(\sqrt{r})$, and hence the 
condition (\ref{eq:flux-cond-math1}), in particular.  

{\bf The case $T_X^{(2)}\otimes_\Q H^1(E_\tau;\Q)$ is not simple:}
Let us now turn to the case $T_X^{(2)} \otimes_\Q H^1(E_\tau;\Q)$ is not 
itself a simple component of the rational Hodge structure. 
The condition that the reflex field $(K^{(2)})^r$ {\it with respect to 
the half set} (\ref{eq:flux-cond-IIB-2prm}) should be imaginary and 
quadratic implies in fact that $n/2=1$. So, $K^{(2)}$ also needs to be 
an imaginary quadratic field.
To conclude, the condition (\ref{eq:flux-cond-IIB-2prm}) 
implies $K^{(1)} \cong \Q(\phi) \cong K_E$, $T_X^{(1)}\subsetneq T_0^{(1)}$, 
and $K^{(2)}$ is also isomorphic to $\Q(\phi)$; the 
condition (\ref{eq:flux-cond-math2}) is satisfied. 

{\bf To wrap up}, here is what we learned in this appendix, stated 
in a colloquial language. 
Although it is not apparent from the Type IIB conditions 
(\ref{eq:flux-cond-IIB-1}, \ref{eq:flux-cond-IIB-2prm}) 
in \cite{prev.paper.PRD}, only small classes of CM fields $K^{(2)}$ can 
satisfy either one of those conditions; the analysis in this appendix 
left $[K^{(2)}:\Q] = 2,4$ as the only possibilities, in particular. 
The M/F-theory condition (\ref{eq:flux-cond-math1}, \ref{eq:flux-cond-math2}) 
in the main text of this article also imply $[K^{(2)}:\Q] = 2,4$, because 
the CM field $K^{(1)}$ for $X^{(1)} = {\rm Km}(E_\tau \times E_\phi)$ can only 
be degree-4 or degree-2 extension over $\Q$.  So, both perspectives 
led us to the same result. 

%%%%%%%%%%%%%%%%%%%%%%%%%%%%%%%%%%%%%%%%%%%%%%%%%%%%%%%%%%%%%%%%%%%%%%%%%


\begin{thebibliography}{99}
%%%%%%%%%%%%%%%%%%%%%%%%%%%%%%%%%%%%%%%%%%%%%%%%%%%%%%%%%%%%%%%%%%%%%%%%%

\bibitem{flux-qntztn}
%
E.~Witten,
``On flux quantization in M theory and the effective action,''
J. Geom. Phys. \textbf{22} (1997), 1-13
% doi:10.1016/S0393-0440(96)00042-3
[arXiv:hep-th/9609122 [hep-th]].
%
A.~Collinucci and R.~Savelli,
``On Flux Quantization in F-Theory,''
JHEP \textbf{02} (2012), 015
% doi:10.1007/JHEP02(2012)015
[arXiv:1011.6388 [hep-th]].
%
A.~Collinucci and R.~Savelli,
``On Flux Quantization in F-Theory II: Unitary and Symplectic Gauge Groups,''
JHEP \textbf{08} (2012), 094
% doi:10.1007/JHEP08(2012)094
[arXiv:1203.4542 [hep-th]].
%
\bibitem{DDFK}
%
F.~Denef, M.~R.~Douglas, B.~Florea, A.~Grassi and S.~Kachru,
``Fixing all moduli in a simple f-theory compactification,''
Adv. Theor. Math. Phys. \textbf{9} (2005) no.6, 861-929
% doi:10.4310/ATMP.2005.v9.n6.a1
[arXiv:hep-th/0503124 [hep-th]].
%
\bibitem{MooreTASI}
%
G.~W.~Moore,
``Strings and Arithmetic,''
% doi:10.1007/978-3-540-30308-4\_8
[arXiv:hep-th/0401049 [hep-th]].
%
\bibitem{enumerate}
%
O.~DeWolfe, A.~Giryavets, S.~Kachru and W.~Taylor,
``Enumerating flux vacua with enhanced symmetries,''
JHEP \textbf{02} (2005), 037
% doi:10.1088/1126-6708/2005/02/037
[arXiv:hep-th/0411061 [hep-th]].
%
\bibitem{prev.paper.PRD}
%
K.~Kanno and T.~Watari, ``{Revisiting arithmetic solutions to the $W=0$
  condition},''{\emph{Phys. Rev.} {\bfseries D96} (2017) 106001},
  [arXiv:1705.05110 [hep-th]].

\bibitem{prev.paper.preprint}
%
the arXiv version of \cite{prev.paper.PRD}.
This version contains more systematic review on math background, while 
the version \cite{prev.paper.PRD} contains discussion on orientifold 
projection.
%
%
\bibitem{GV}
%
S.~Gukov and C.~Vafa,
``Rational conformal field theories and complex multiplication,''
Commun. Math. Phys. \textbf{246} (2004), 181-210
% doi:10.1007/s00220-003-1032-0
[arXiv:hep-th/0203213 [hep-th]].
%
\bibitem{PSS-arth}
%
I. Piatetski-Shapiro and I. R. Shafarevich, ``The arithmetic of K3 surfaces,'' 
in {\it Proc. Steklov Inst. Math.}, vol. 132, 1973; a copy is also available 
from J. Cogdell, ed., {\it Selected works of Ilya Piatetski-Shapiro}. 
Amer. Math. Soc., 2000. \\
%
Another record Proc. of the Int'l Conference on Number Theory {\bf 132} (1975) p.45 [Russian original is Trudy Mat. Inst. Steklov. {\bf 132} (1973) 45].
%
%
\bibitem{Rizov}
%
J. Rizov, ``Complex multiplication for K3 surfaces,'' 
arXiv:math/0508018 [math.AG].
%
\bibitem{Candelas}
%
P.~Candelas, X.~de la Ossa, M.~Elmi and D.~Van Straten,
``A One Parameter Family of Calabi-Yau Manifolds with Attractor Points of Rank Two,''
JHEP \textbf{10} (2020), 202
% doi:10.1007/JHEP10(2020)202
[arXiv:1912.06146 [hep-th]].
%
\bibitem{Kach20}
%
S.~Kachru, R.~Nally and W.~Yang,
``Supersymmetric Flux Compactifications and Calabi-Yau Modularity,''
[arXiv:2001.06022 [hep-th]].
%
\bibitem{Schimm}
%
R.~Schimmrigk,
``On flux vacua and modularity,''
JHEP \textbf{09} (2020), 061
% doi:10.1007/JHEP09(2020)061
[arXiv:2003.01056 [hep-th]].
%
\bibitem{Kach20-2}
%
S.~Kachru, R.~Nally and W.~Yang,
``Flux Modularity, F-Theory, and Rational Models,''
[arXiv:2010.07285 [hep-th]].
%
\bibitem{BV}
%
A.~P.~Braun and R.~Valandro,
``$G_4$ Flux, Algebraic Cycles and Complex Structure Moduli Stabilization,''
[arXiv:2009.11873 [hep-th]].
%
\bibitem{YL}
%
N. Yui, ``Update on the Modularity of Calabi--Yau Varieties'' 
with Appendix by H. Verrill, in N. Yui and J. Lewis (eds.) 
{\it Calabi--Yau Varieties and Mirror Symmetry}, Fields Inst. Comm., 2003.
%
\bibitem{Borcea}
%
C. Borcea, ``Calabi--Yau Threefolds and Complex Multiplication,''
  in {\it Essays on Mirror Manifolds}, (S.-T. Yau ed.) 
International Press, 1992.
%
\bibitem{voisinhodge}
C.~Voisin, \emph{Th\'eorie de Hodge et g\'eom\'etrie alg\'ebrique complexe},
  vol.~10 of \emph{Cours Sp\'ecialis\'es}.
\newblock Soci\'et\'e Math\'ematique de France, 2002.
%
\bibitem{Nikulin-factor-long}
V.~V. Nikulin, ``Quotient-groups of groups of automorphisms of hyperbolic forms
  of subgroups generated by {$2$}-reflections: Algebro-geometric applications,''
J. Soviet Math. {\bf 22} (1983) 1401--1475 [Russian original: 
Itogi Nauki i Tekhniki. Ser. Sovrem. Probl. Mat. {\bf 18} VINITI Moscow, (1981) 3--114]. 
The English title may sometimes be spelled as ``Factor groups of groups 
of .... ,'' instead of starting with ``Quotient-grups of ....'' 
% but this article is not the same as \cite{Nikulin-factor-short}.
% %
% \bibitem[Nik-inv-short]{Nikulin-factor-short}
% V.~V. Nikulin, 
% ``On factor groups of the automorphism groups of hyperbolic
%   forms modulo subgroups generated by 2-reflections,'' {\emph{Soviet Math. Dokl.} {\bfseries 20} (1979) 1156--1158}, 
% [Russian original: {\emph{Dokl. Akad. Nauk SSSR} {\bfseries 248} (1979) 1307--1309}].
%
% 
\bibitem{mir-HD}
B.~R.~Greene, D.~R.~Morrison and M.~R.~Plesser,
``Mirror manifolds in higher dimension,''
Commun. Math. Phys. \textbf{173} (1995), 559-598
% doi:10.1007/BF02101657
[arXiv:hep-th/9402119 [hep-th]].
%
\bibitem{flux-Sakura}
%
A.~P.~Braun, A.~Collinucci and R.~Valandro,
``G-flux in F-theory and algebraic cycles,''
Nucl. Phys. B \textbf{856} (2012), 129-179
% doi:10.1016/j.nuclphysb.2011.10.034
[arXiv:1107.5337 [hep-th]].
%
J.~Marsano and S.~Schafer-Nameki,
``Yukawas, G-flux, and Spectral Covers from Resolved Calabi-Yau's,''
JHEP \textbf{11} (2011), 098
% doi:10.1007/JHEP11(2011)098
[arXiv:1108.1794 [hep-th]].
%
S.~Krause, C.~Mayrhofer and T.~Weigand,
``$G_4$ flux, chiral matter and singularity resolution in F-theory compactifications,''
Nucl. Phys. B \textbf{858} (2012), 1-47
% doi:10.1016/j.nuclphysb.2011.12.013
[arXiv:1109.3454 [hep-th]].
%
\bibitem{BW-vhr}
%
A.~P.~Braun and T.~Watari,
``The Vertical, the Horizontal and the Rest: anatomy of the middle cohomology of Calabi-Yau fourfolds and F-theory applications,''
JHEP \textbf{01} (2015), 047
% doi:10.1007/JHEP01(2015)047
[arXiv:1408.6167 [hep-th]].
% 
\bibitem{DeWolfe}
%
O.~DeWolfe,
``Enhanced symmetries in multiparameter flux vacua,''
JHEP \textbf{10} (2005), 066
% doi:10.1088/1126-6708/2005/10/066
[arXiv:hep-th/0506245 [hep-th]].

\bibitem{ST}
%
G. Shimura and Y. Taniyama, 
``{\it Complex multiplication of abelian varieties and its applications to number theory},'' vol. 6 of {\it Publications of the Matheatical Society of Japan}.
Math. Soc. Japan, 1961. Large fraction of this book is contained as a part 
of another book: \\
%
G. Shimura, {\it Abelian varieties with complex multiplication and modular functions}, vol. 46 of Princeton Math Series, Princeton U. Press, 1998.
%
\bibitem{Fujisaki}
%
G. Fujisaki, {\it Field and Galois Theory}. Iwanami Publ. Co., 1991, written in Japanese.
%
\bibitem{Roman}
%
S. Roman, {\it Field theory}, vol. 158 of GTM. 
Springer Science and Business Media, 2005. 
%
%
\bibitem{TsushimaNagao}
%
Y. Tsushima and H. Nagao, {\it Representation theory of finite groups} 
(written in Japanese), Shoka-bo Publ. Co., 1987. 
%
\bibitem{ssAlg}
%
P. Gille and T. Szamuelly, {\it Central simple algebras and Galois cohomology},
Cambridge U. Press, 2006. 
%
\bibitem{K3-EndHdg}
%
Y. G. Zarhin, ``Hodge groups of K3 surfaces,'' 
J. Reine Angew. Math. {\bf 341} (1983) 193--220. \\
%
B. van Geemen, ``Real multiplication on K3 surfaces and Kuga--Satake varieties,'' Michigan Math. J. {\bf 58} (2008) 375--399. 
[arXiv:math/0609839 [math.AG]]
%
\bibitem{huybrechts2016lectures}
%
D.~Huybrechts, \emph{Lectures on K3 surfaces}, vol.~158 of \emph{Cambridge
  studies Adv. Math.}
\newblock Cambridge U. Press, 2016.
%
\bibitem{AK}
P.~S. Aspinwall and R.~Kallosh, ``{Fixing all moduli for M-theory on
  K3$\times$K3},''{\emph{JHEP} {\bfseries 10} (2005) 001},
  [arXiv:hep-th/0506014].
%
%
\bibitem{BKW-phys}
%
A.~P.~Braun, Y.~Kimura and T.~Watari,
``The Noether-Lefschetz problem and gauge-group-resolved landscapes: F-theory on K3  $\times$  K3 as a test case,''
JHEP \textbf{04} (2014), 050
% doi:10.1007/JHEP04(2014)050
[arXiv:1401.5908 [hep-th]].
%
%
\bibitem{DS}
%
M.~Dine, D.~O'Neil and Z.~Sun,
``Branches of the landscape,''
JHEP \textbf{07} (2005), 014
% doi:10.1088/1126-6708/2005/07/014
[arXiv:hep-th/0501214 [hep-th]].
%
Z. Sun, 
``Low energy supersymmetry from R-symmetries,''
Phys. Lett. B \textbf{712} (2012), 442-444
% doi:10.1016/j.physletb.2012.05.013
[arXiv:1109.6421 [hep-th]].
%
\bibitem{vMHS}
%
M. Kerr, ``Algebraic and Arithmetic Properties of Period Maps,'' 
in {\it Calabi--Yau varieties: Arithmetic, Geometry and Physics} 
(R. Laza, M. Sch\"{u}tt and N. Yui, eds), pp. 173--208 Springer 2015. \\
%
M. Green, P. A. Griffiths and M. Kerr, 
``Mumford--Tate Groups and Domains: Their Geometry and Arithmetic,'' 
vol. 183 of Annals Math Studies. Princeton  U. Press, 212. 
%
%
\bibitem{EIKY}
%
J.~L.~Evans, M.~Ibe, J.~Kehayias and T.T.~Yanagida,
``Non-Anomalous Discrete R-symmetry Decrees Three Generations,''
Phys. Rev. Lett. \textbf{109} (2012), 181801 
% doi:10.1103/PhysRevLett.109.181801
[arXiv:1111.2481 [hep-ph]].
%
\bibitem{Generalized-BV}
%
S. Cynk, and K. Hulek,  
``Higher-Dimensional Modular Calabi-Yau Manifolds,'' 
Canadian Mathematical Bulletin, {\bf 50}(4), (2007) 486-503. 
% doi:10.4153/CMB-2007-049-9
%
J.~Dillies, 
``Generalized Borcea--Voisin Construction,'' 
{\emph{Letters in Mathematical Physics} {\bfseries 100} (2012) 77--96}.
%
%
\bibitem{BKW-math}
%
A.~P.~Braun, Y.~Kimura and T.~Watari,
``On the Classification of Elliptic Fibrations modulo Isomorphism on K3 Surfaces with large Picard Number,''
[arXiv:1312.4421 [math.AG]].
%
\bibitem{Asp-K3}
%
P.~S.~Aspinwall,
``K3 surfaces and string duality,''
[arXiv:hep-th/9611137 [hep-th]].
%
%
\bibitem{nikulin1980finite}
V.~V. Nikulin, ``Finite automorphism groups of K{\"a}hler K3
  surfaces,''{\emph{Trans. Moscow Math. Soc} {\bfseries 38} (1980) 71--135}.
%
\bibitem{MO}
N.~Machida and K.~Oguiso, ``On {$K3$} surfaces admitting finite non-symplectic
  group actions,''{\emph{J. Math. Sci. Univ. Tokyo} {\bfseries 5} (1998)
  273--297}.
%
\bibitem{Sterk}
H.~Sterk, ``Finiteness results for algebraic {$K3$}
  surfaces,''{\emph{Math. Z.} {\bfseries 189} (1985) 507--513}.
%
% 
\bibitem{Mukai88}
S.~Mukai, ``Finite groups of automorphisms of {$K3$} surfaces and the {M}athieu
  group,'' {\emph{Invent. Math.}{\bfseries 94} (1988) 183--221}.
%
\bibitem{XiaoGal}
G.~Xiao, ``Galois covers between {$K3$} surfaces,''{\emph{Ann. Inst. Fourier
  (Grenoble)} {\bfseries 46} (1996) 73--88}.
%
%
\bibitem{Kondo89-1}
S.~Kond\=o, ``Algebraic {$K3$} surfaces with finite automorphism
  groups,''{\emph{Nagoya Math. J.} {\bfseries 116} (1989) 1--15}.

\bibitem{Kondo89-2}
S.~Kond\=o, ``On algebraic {$K3$} surfaces with finite automorphism
  groups,''{\emph{Proc. Japan Acad. Ser. A Math. Sci.} {\bfseries 62} (1986) 353--355}.
%
%
\bibitem{Keumorder}
J.~Keum, ``Orders of automorphisms of {K}3
  surfaces,''{\emph{Adv. Math.} {\bfseries 303} (2016) 39--87}.
%
%
\bibitem{KeumMax}
%
J.~Keum, 
``K3 surfaces with an automorphisms of order 66, the maximum possible,''
J. Alg. {\bf 426} (2015) 273--287.
%
%
\bibitem{Zhang}
D.-Q. Zhang, ``Automorphisms of K3 surfaces,''
  arXiv:math/0506612 [math.AG].
%
%
\bibitem{OZ99}
K.~Oguiso and D.-Q. Zhang, ``On {V}orontsov's theorem on {$K3$} surfaces with
  non-symplectic group
  actions,''{\emph{Proc. Amer. Math. Soc.} {\bfseries 128} (2000) 1571--1580}.
%
\bibitem{Kondo92}
S.~Kond\=o, ``Automorphisms of algebraic K3 susfaces which act trivially on
  Picard groups,''{\emph{J. Math. Soc. Japan} {\bfseries 44} (1992) 75--98}.
%
%
\bibitem{Taki12}
S.~Taki, ``Classification of non-symplectic automorphisms on {$K3$} surfaces
  which act trivially on the {N}\'eron-{S}everi
  lattice,''{\emph{J. Algebra} {\bfseries 358} (2012) 16--26}.
%
\bibitem{Schuett10}
M.~Sch\"utt, ``{$K3$} surfaces with non-symplectic automorphisms of 2-power
  order,''{\emph{J. Algebra} {\bfseries 323} (2010) 206--223}.
%
\bibitem{TST14}
D.~A. Tabbaa, A.~Sarti and S.~Taki, ``Classification of order sixteen
  non-symplectic automorphisms on K3 surfaces,''
  arXiv:1409.5803 [math.AG].
%
\bibitem{Vorontsov83}
S.~P. Vorontsov, ``Automorphisms of even lattices arising in connection with
  automorphisms of algebraic {$K3$}-surfaces,''{\emph{Vestnik Moskov. Univ.
  Ser. I Mat. Mekh.} (1983) 19--21}.
%

%
\bibitem{AS08}
M.~Artebani and A.~Sarti, ``Non-symplectic automorphisms of order 3 on {$K3$}
  surfaces,''{\emph{Math.Ann.} {\bfseries 342} (2008) 903--921}.
%
\bibitem{Taki11}
S.~Taki, ``Classification of non-symplectic automorphisms of order 3 on {$K3$}
  surfaces,''{\emph{Math. Nachr.} {\bfseries 284} (2011) 124--135}.
%
\bibitem{Taki10}
S.~Taki, ``Non-symplectic automorphisms of 3-power order on {$K3$}
  surfaces,''{\emph{Proc. Japan Acad. Ser. A Math. Sci.} {\bfseries 86} (2010)
  125--130}.
%
\bibitem{Artebani2011}
M.~Artebani, A.~Sarti and S.~Taki, ``K3 surfaces with non-symplectic
  automorphisms of prime
  order,''{\emph{Math. Z.}{\bfseries 268} (2011) 507--533},
  [arXiv:0903.3481 [math.AG]].
%
\bibitem{Oguiso93}
K.~Oguiso, ``A remark on the global indices of {${\bf Q}$}-{C}alabi-{Y}au
  {$3$}-folds,''{\emph{Math. Proc. Cambridge Philos. Soc.} {\bfseries 114} (1993) 427--429}.
%
\bibitem{TakionO}
S.~Taki, ``On {O}guiso's {$K3$}
  surface,''{\emph{J. Pure Appl. Algebra} {\bfseries 218} (2014) 391--394}.
%
\bibitem{Tabbaa-thesis}
D. Tabbaa, 
``Non-symplecic automorphisms of 2-power order on K3 surfaces,''
Ph. D thesis U. Poitiers. 
%
\bibitem{TS}
D.~A. Tabbaa and A.~Sarti, ``Order eight non-symplectic automorphisms on
  elliptic K3 surfaces,''  arXiv:1612.01184 [math.AG].
%
\bibitem{AS15}
M.~Artebani and A.~Sarti, ``Symmetries of order four on {K}3
  surfaces,''
  {\emph{J. Math. Soc. Japan} {\bfseries 67} (2015) 503--533}.
%
%
\bibitem{Hashimoto}
K.~Hashimoto, ``Finite symplectic actions on the {$K3$}
  lattice,''{\emph{Nagoya Math. J.} {\bfseries 206} (2012) 99--153}.
%
\bibitem{GS07}
A.~Garbagnati and A.~Sarti, ``Symplectic automorphisms of prime order on {$K3$}
  surfaces,''{\emph{J. Algebra} {\bfseries 318} (2007) 323--350}.
%
\bibitem{GS09}
A.~Garbagnati and A.~Sarti, ``Elliptic fibrations and symplectic automorphisms
  on {$K3$}
  surfaces,''{\emph{Comm.Algebra} {\bfseries 37} (2009) 3601--3631}.
%
\bibitem{Kondomax}
S.~Kond\=o, ``The maximum order of finite groups of automorphisms of {$K3$}
  surfaces,''{\emph{Amer. J. Math.} {\bfseries 121} (1999) 1245--1252}.
%
\bibitem{OZ168}
K.~Oguiso and D.-Q. Zhang, ``The simple group of order 168 and {$K3$}
  surfaces,'' in \emph{Complex geometry ({G}\"ottingen, 2000)}, pp.~165--184.
\newblock Springer, Berlin, 2002.
% 

%
\bibitem{Dolgachev2007}
I.~V. Dolgachev and S.~Kond{\={o}}, ``Moduli of K3 Surfaces and Complex Ball
  Quotients,'' in \emph{Arithmetic and Geometry Around Hypergeometric
  Functions: Lecture Notes of a CIMPA Summer School held at Galatasaray
  University, Istanbul, 2005} (R.-P. Holzapfel, A.~M. Uluda{\u{g}} and
  M.~Yoshida, eds.), pp.~43--100.
\newblock Birkh{\"a}user Basel, Basel, 2007.
%
\bibitem{0904.1922}
R.~Livn\'e, M.~Sch\"utt and N.~Yui, ``The modularity of K3 surfaces with
  non-symplectic group
  actions,''{\emph{Math. Ann.} {\bfseries 348} (2010) 333--355},
  [arXiv:0904.1922[math.AG]].
%
%
\bibitem{Taelman}
%
L. Taelman, ``K3 surfaces over finite fields with given L-function,''
Alg. and Numb. Theory {\bf 10} (2010) 1133--1146 
[arXiv:1507.08547 [math.AG]].
%
%
\bibitem{1006.1604}
A.~Garbagnati and A.~Sarti, ``On symplectic and non-symplectic automorphisms of
  {K}3 surfaces,''{\emph{Rev. Mat.
  Iberoam.} {\bfseries 29} (2013) 135--162},
  [arXiv:1006.1604[math.AG]].
%
\bibitem{1010.3904}
V.~V. Nikulin, ``Elliptic fibrations on {$\rm K3$}
  surfaces,''{\emph{Proc. Edinb. Math. Soc. (2)} {\bfseries 57} (2014) 253--267},
  [arXiv:1010.3904 [math.AG]].
% 
\bibitem{CR}
%
W.~m.~Chen and Y.~b.~Ruan,
``A New cohomology theory for orbifold,''
Commun. Math. Phys. \textbf{248} (2004), 1-31
% doi:10.1007/s00220-004-1089-4
[arXiv:math/0004129 [math.AG]].
%
\bibitem{DRS}
%
K.~Dasgupta, G.~Rajesh and S.~Sethi,
``M theory, orientifolds and G - flux,''
JHEP \textbf{08} (1999), 023
% doi:10.1088/1126-6708/1999/08/023
[arXiv:hep-th/9908088 [hep-th]].
%
%
\bibitem{Cmp-Garbag}
%
P. Comparin and A. Garbagnati, 
``van Geemen--Sarti involutions and elliptic fibrations on K3 surfaces 
double cover of P-2,''
J. Math. Soc. Japan {\bf 66} (2014) 479--522.
[arXiv:1110.6380 [math.AG]].
%
\bibitem{1806.03097}
A.~Garbagnati and C.~Salgado, ``Elliptic fibrations on K3 surfaces with a
  non-symplectic involution fixing rational curves and a curve of positive
  genus,'' arXiv:1806.03097 [math.AG].
%
%
\bibitem{P-SS}
I.~I. Pjatecki\u{i}-\v{S}apiro and I.~R. \v{S}afarevi\v{c}, ``A Torelli theorem
  for algebraic surfaces of type K3,''{\emph{Math. USSR-Izv.} {\bfseries 5}
  (1971) 547}.
%
\bibitem{BM}
%
V.~Braun and D.~R.~Morrison,
``F-theory on Genus-One Fibrations,''
JHEP \textbf{08} (2014), 132
% doi:10.1007/JHEP08(2014)132
[arXiv:1401.7844 [hep-th]].
%
\bibitem{MT}
%
D. Morrison and W. Taylor
D.~R.~Morrison and W.~Taylor,
``Sections, multisections, and U(1) fields in F-theory,''
[arXiv:1404.1527 [hep-th]].
%
\bibitem{singlrty}
%
P.~Arras, A.~Grassi and T.~Weigand,
``Terminal Singularities, Milnor Numbers, and Matter in F-theory,''
J. Geom. Phys. \textbf{123} (2018), 71-97
% doi:10.1016/j.geomphys.2017.09.001
[arXiv:1612.05646 [hep-th]].
%
D.~Klevers, D.~R.~Morrison, N.~Raghuram and W.~Taylor,
``Exotic matter on singular divisors in F-theory,''
JHEP \textbf{11} (2017), 124
% doi:10.1007/JHEP11(2017)124
[arXiv:1706.08194 [hep-th]].
%
W.~Taylor and A.~P.~Turner,
``Generic matter representations in 6D supergravity theories,''
JHEP \textbf{05} (2019), 081
% doi:10.1007/JHEP05(2019)081
[arXiv:1901.02012 [hep-th]].
%
\bibitem{CY4BV-invF}
%
A. Cattaneo, A. Garbagnati and M. Penegini, 
``Calabi--Yau 4-folds of Borcea--Voisin type from F-theory,''
Pacific J. of Math. {\bf 299} (2019) 1--31, [arXiv:1706.01689 [math.AG]].
%
%
\bibitem{Og-Km}
%
K. Oguiso, 
``On Jacobian fibrations on the Kummer surfaces of the product of non-isogenous elliptic curves,'' J. Math. Soc. Japan, {\bf 41} (1989) 651--680.
%
%
\bibitem{6auth}
%
M.~Bershadsky, K.~A.~Intriligator, S.~Kachru, D.~R.~Morrison, V.~Sadov and C.~Vafa,
``Geometric singularities and enhanced gauge symmetries,''
Nucl. Phys. B \textbf{481} (1996), 215-252
% doi:10.1016/S0550-3213(96)90131-5
[arXiv:hep-th/9605200 [hep-th]].
%
%
\bibitem{hep-th/0002012}
P.~S. Aspinwall, S.~H. Katz and D.~R. Morrison, ``{Lie groups, Calabi-Yau
  threefolds, and F
  theory},''{\emph{Adv. Theor. Math. Phys.} {\bfseries 4} (2000) 95--126},
  [arXiv:hep-th/0002012].
%
\bibitem{Miranda}
R.~Miranda, ``{Smooth models for elliptic threefolds},'' in \emph{{The
  Birational Geometry of Degenerations}}, pp.~85--133, 1983.
%
\bibitem{collision}
M.~Bershadsky and A.~Johansen, ``{Colliding singularities in F theory and phase
  transitions},''{\emph{Nucl.
  Phys.} {\bfseries B489} (1997) 122--138},
  [arXiv:hep-th/9610111].
%
%
\bibitem{KV}
%
S.~H.~Katz and C.~Vafa,
``Matter from geometry,''
Nucl. Phys. B \textbf{497} (1997), 146-154
% doi:10.1016/S0550-3213(97)00280-0
[arXiv:hep-th/9606086 [hep-th]].
%
%
\bibitem{MV-2}
D.~R. Morrison and C.~Vafa, ``{Compactifications of F theory on Calabi-Yau
  threefolds.
  2.},''{\emph{Nucl. Phys.} {\bfseries B476} (1996) 437--469},
  [arXiv:hep-th/9603161].
%
\bibitem{TW-06}
%
R.~Tatar and T.~Watari,
``Proton decay, Yukawa couplings and underlying gauge symmetry in string theory,''
Nucl. Phys. B \textbf{747} (2006), 212-265
% doi:10.1016/j.nuclphysb.2006.04.025
[arXiv:hep-th/0602238 [hep-th]].
%
\bibitem{F-flavor}
%
H.~Hayashi, T.~Kawano, Y.~Tsuchiya and T.~Watari,
``Flavor Structure in F-theory Compactifications,''
JHEP \textbf{08} (2010), 036
% doi:10.1007/JHEP08(2010)036
[arXiv:0910.2762 [hep-th]].
%
\bibitem{Kl}
%
R. Kloosterman, 
``Classification of all Jacobian elliptic fibrations on certain K3 surfaces,'' 
J. Math. Soc. Japan, {\bf 58} (2006) 665--680 [arXiv:math/0502070 [math.AG]].
%
\bibitem{BHV}
%
A.~P.~Braun, A.~Hebecker, C.~Ludeling and R.~Valandro,
``Fixing D7 Brane Positions by F-Theory Fluxes,''
Nucl. Phys. B \textbf{815} (2009), 256-287
% doi:10.1016/j.nuclphysb.2009.02.025
[arXiv:0811.2416 [hep-th]].


\end{thebibliography}
\end{document}